\documentclass[prl,twocolumn,amsmath,amssymb,superscriptaddress,showpacs,,floatfix] {revtex4-2}
\usepackage{amsmath}
\usepackage{hyperref}
\usepackage[export]{adjustbox}
\usepackage{xcolor}
\usepackage{booktabs}
\usepackage{graphicx,psfrag,color}
\usepackage{dcolumn}
\usepackage{bm}
\usepackage{mathbbol, appendix}
\usepackage{printlen}
\usepackage{subcaption}
\usepackage[justification=Justified]{caption}
\usepackage{diagbox}
\usepackage{amsmath}
\usepackage{comment}
\usepackage[normalem]{ulem}
\usepackage{float}
\usepackage[ruled,vlined]{algorithm2e}
\usepackage{xr-hyper}
\usepackage{overpic}

\newcolumntype{t}[1]{D{.}{.}{#1}}

\newcommand{\JC}{J_{\mathrm{c},ij}}
\newcommand{\SISJNOT}{\sigma_{i_0}\sigma_{j_0}}
\newcommand{\JCsub}{J_{\mathrm{c},ij}^{\rm sub}}

\begin{document}
\title{
The Physics of Local Optimization in Complex Disordered Systems}
\ifdefined\HCode\else
\externaldocument[][nocite]{final_SI_xr}
\fi
\author{Mutian Shen}
\affiliation{Department of Physics, Washington University, St.
Louis, MO 63160, USA}
\affiliation{Department of Biomedical Engineering, Washington University, St.
Louis, MO 63160, USA}

\author{Gerardo Ortiz}
\affiliation{Department of Physics, Indiana University, Bloomington, IN 47405, USA}
\affiliation{Institute for Advanced Study, Princeton, NJ 08540, USA}
\author{Zhiqiao Dong}
\affiliation{Department of Mechanical Science \& Engineering, University of Illinois Urbana-Champaign, IL 61801, USA}
\author{Martin Weigel}
\affiliation{Institut für Physik, Technische Universität Chemnitz, 09107 Chemnitz, Germany}
\affiliation{Physics Department, Emory University, Atlanta, GA, U.S.A.}

\author{Zohar Nussinov}
\email{corresponding author: zohar@wustl.edu}
\affiliation{Department of Physics, Washington University, St.
Louis, MO 63160, USA}
\affiliation{Institut für Physik, Technische Universität Chemnitz, 09107 Chemnitz, Germany}
\affiliation{Rudolf Peierls Centre for Theoretical Physics, University of Oxford, Oxford OX1 3PU, United Kingdom}
\affiliation{LPTMC, CNRS-UMR 7600, Sorbonne Universit\'e, 4 Place Jussieu, 75252 Paris cedex 05, France}

\date{\today}
\begin{abstract}
Limited resources motivate decomposing large-scale problems into smaller, ``local'' subsystems and stitching together the so-found solutions.  We explore the physics underlying this approach and discuss the concept of ``local hardness", i.e.,
  the complexity of predicting local properties of the solution from local information, for the ground-state problem of
  both P- and NP-hard spin-glasses and related frustrated spin systems. Depending on the model considered, we observe varying scaling behaviors in how errors associated with local predictions decay as a function of the size of the solved subsystem. These errors are intimately connected to global critical threshold instabilities, characterized by gapless, avalanche-like excitations that follow scale-invariant size distributions. Away from criticality, local solvers quickly achieve high accuracy, aligning closely with the results of the computationally much more expensive global minimization. We leverage these findings to introduce a heuristic contraction-based algorithm for globally studying spin-glass ground states. The local solvers further display sharp imprints of the phase transition from the spin-glass to the ferromagnetic phase as the distribution of spin-glass couplings is shifted, as well as characteristic differences for the infinite-range model, implying the existence of specific classes of local hardness. Our findings shed light on how Nature may operate solely through local actions at her disposal.
\end{abstract}
\maketitle

A prevalent, ``divide-and-conquer'' type, approach to analyzing large-scale problems involves segmenting them into smaller ``local'' components which are then solved, followed by assembling the obtained local optimal solutions in a manner akin to a jigsaw puzzle, thereby yielding meaningful results for the original large-scale problem. This \textit{modus operandi} is natural when constrained by limited time and computational resources. It is especially common when one assumes --- as often happens in problems with an underlying geometric structure (e.g., images of physical systems \cite{Imagenet} or proteins \cite{Alphafold}) --- that these local solutions provide correct and useful information about the larger system. Such local few-body solvers may accurately capture key features of general many-body systems governed by local Hamiltonians, a result known in certain contexts as {\it nearsightedness} \cite{Kohn-near,Cluster1,Cluster2,Cluster3,Cluster4}. Here, we explore physical aspects of this approach for predicting local as well as global aspects of ground states (GSs) of both short- and long-range classical spin-glass systems, a known difficult problem \cite{cipra2000ising,barahona_computational_1982} using simple local algorithms.

Similar to 
other complex systems
\cite{SG1,Steinbook,Fortunatobook,Newmanbook,Bryngelson}, the challenge posed by spin-glasses results from the combination of disorder and frustration leading to a complex
free-energy landscape~\cite{janke:07}. According to recent results these systems
generically have very low-lying excited states that are radically different from each
other and from the GSs ~\cite{Lamarcq,krzakala_disorder_2005,shen2023universal}.  This phenomenon of
system-spanning lowest-energy excitations implies that optimal (i.e., GS) 
subsystem solutions might not necessarily match the optimal full system solution. The
occurrence of such excitations is independent of the classification of the corresponding 
GS problem in the P or NP classes \cite{papadimitriou:94}. Indeed, spin-glasses with GSs of 
polynomial (P) complexity may also exhibit
system-spanning lowest-energy droplets~\cite{shen2023universal}.  
This is the case for GSs of two-dimensional (2D) Ising spin-glasses on
planar graphs, which can be found in polynomial time \cite{bieche:80a}.

Here, we investigate to which extent local (i.e., subsystem) spin-glass solvers
approximate global spin-glass GSs. Understanding such features will help
uncover the principles that underlie the effectiveness of local solvers. Our
numerical analysis (based on exact GS algorithms) concentrates on Ising
spin-glasses with Hamiltonian
\begin{eqnarray}
\label{HSG}
    H = - \sum_{\langle i j \rangle} J_{ij} \sigma_i\sigma_j,\;\;\;\sigma_i = \pm 1, 
\end{eqnarray}
with couplings $J_{ij}$ drawn from a 
Gaussian distribution of mean $\overline{J}$
and unit variance. Unless stated otherwise, we set $\overline{J}=0$. Here, the $N$ spins
are placed on 2D or 3D lattices with open boundaries and nearest-neighbor
interactions (Edwards-Anderson (EA)  model~\cite{EA}) or form a 
Sherrington-Kirkpatrick (SK) \cite{SK} model featuring all-to-all interactions (in which case we rescale $J_{ij}$ by $1/\sqrt{N}$ to guarantee a meaningful
thermodynamic limit).

\begin{figure}[htbp]
    \centering
    \includegraphics[width = 0.30\textwidth]{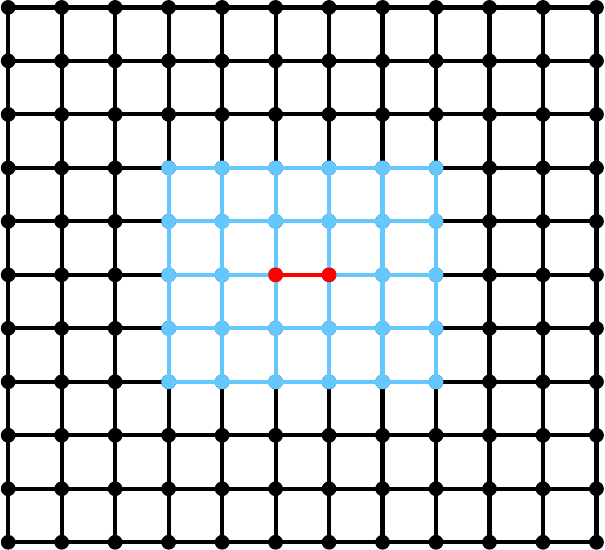}
    \caption{Computational setup for determining the relative spin orientation $\sigma_{i_{0}}\sigma_{j_{0}}$ across the central bond (shown in red).
      The correct value of
      $\sigma_{i_{0}}\sigma_{j_{0}}$ is that within the GS of the entire
      system (comprised of black, light blue, and red bonds). The LSBS computes
      $\sigma_{i_{0}}\sigma_{j_{0}}$ within the GS of the local subsystem
      (light blue and red bonds).}
    \label{fig:solver_diagram}
\end{figure}

Previous related studies attempted to combine multiple local solvers 
to address global optimization problems \cite{TSM}, fuse local GSs for the construction of global spin-glass instances with planted solutions
\cite{wang_patch-planting_2017}, and develop local spin-glass solvers using gauge-transformation-based deep reinforcement
learning \cite{RL_local,RL_local1}. Here, we take
on a different perspective and directly investigate to which extent one can predict (parts of)
the global GSs based on optimization of local subsystems. As a minimal realization of this task we study pairs of
neighboring spins (bonds) of the model and introduce a ``local single bond solver''
(LSBS) that examines 
neighboring spins within a local subsystem. We first focus on
the GS products $\SISJNOT$ for bonds $\langle i_0 j_0 \rangle$ which, in
the simplest case, are located at the center of the system, cf.\
Fig.~\ref{fig:solver_diagram}. 

We are interested to see whether the value of this spin product in the global GS matches that obtained by the LSBS for a subsystem centered on this
single bond. Recent studies \cite{hartmann_metastate_2023} examined correlations
between adjacent spins at the system center after resampling its ``shell.'' By 
contrast, here we examine whether this product as determined by (i) GS 
computations for small {\it subsystems} containing this bond and using open boundaries agrees with that in (ii) the {\it global system} GS. In this way, the LSBS solves a maximally local problem and hence is a diagnostic tool, leading to our definition of ``local hardness''. As we shall show below, however, these local solutions can be meaningfully stitched together to candidate solutions for the full problems by a contraction-based scheme. 
We will find that system-spanning correlated changes to spin-glass GSs
appear only at special critical interaction strengths~\cite{shen2023universal}. When the spin-glass couplings
are far removed from their critical threshold values, at which global avalanche-like
changes occur, the GSs are robust to small variations of the coupling
constants. This robustness enables the LSBS to become accurate whenever coupling
constants are far from their critical values. We find that for large lattice systems, the
LSBS error rate becomes independent of the global system size and
exhibits a power-law decay with subsystem size. Several
other {\it distance-dependent}  quantities also exhibit {\it power-law} behaviors (Ref.~\cite{SM}, \ref{sec:scaling_subsytem_jc}). By contrast, the dependence of the LSBS error rate on the deviations of
the coupling from its critical value 
is of an exponential form.

\begin{figure}[htb!]
    \centering
    \includegraphics[width = 0.42\textwidth]{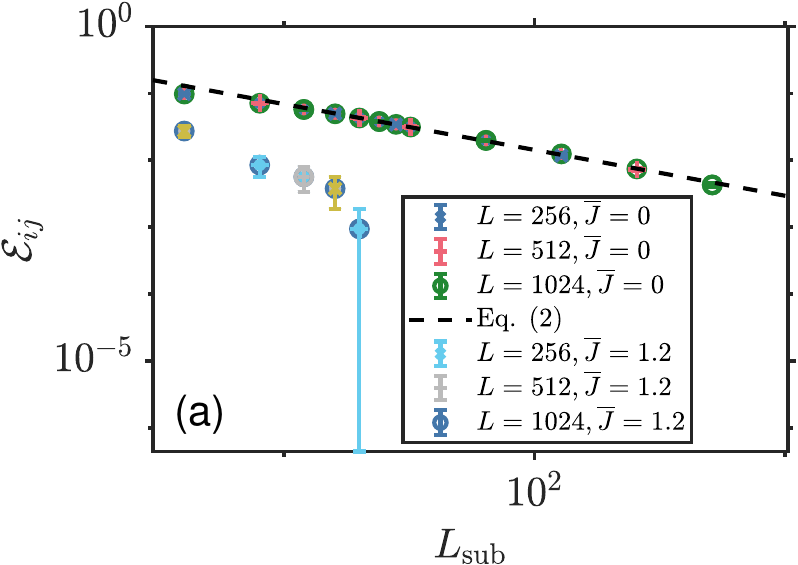}
    \includegraphics[width = 0.42\textwidth]{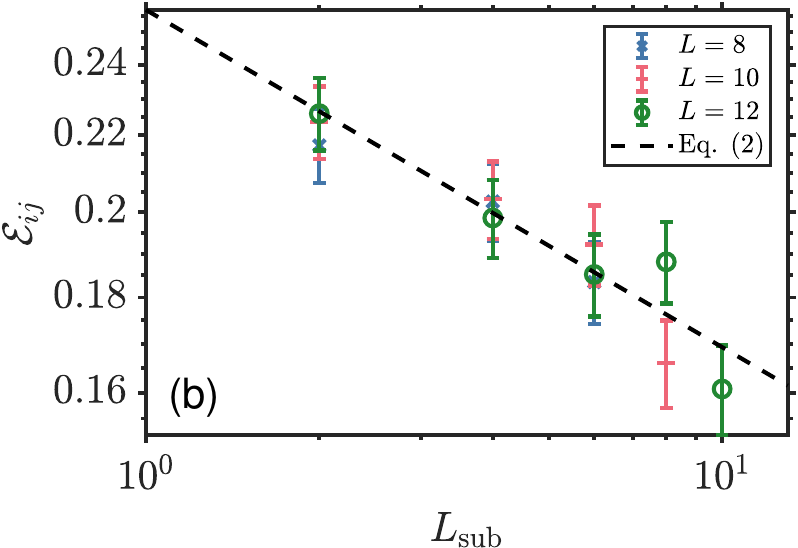}
    \includegraphics[width = 0.42\textwidth]{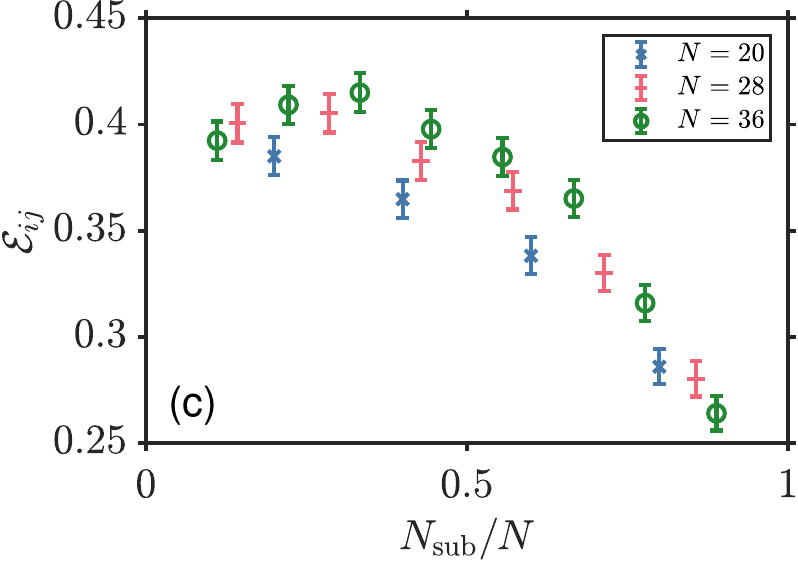}
    \caption{The disorder-averaged LSBS error rate $\mathcal{E}_{ij}$ as a function
      of the subsystem size $L_{\rm sub}$ (resp.\ $N_{\rm sub}$) for (a) the
      square-lattice (2D) EA model, (b) the cubic lattice (3D) EA model, and (c) the
      SK model. For each plot, the results for varying system sizes $L$ ($N$) are
      shown. For (a), both cases $\overline{J}=0$ and $\overline{J}=1.2$ are shown,
      whereas in (b) and (c) $\overline{J} = 0$. Note the nearly perfect {\it
        collapse} of different system-size data, indicating that the error of the LSBS
      becomes asymptotically {\it independent} of system size for (a) and (b), while this is only the case if one scales $N_{\rm sub}$ with $N$ for the SK model in (c). For the lattice
      systems in (a)--(b) and $\overline{J}=0$, the error rate is fit by 
      Eq. (\ref{eq:solver}), cf.\ the straight dashed lines.
    }
    \label{fig:solver_performance}
\end{figure}

To systematically study the relation of local and global solutions in spin systems,
we rely on exact (combinatorial) GS algorithms, using a minimum-weight
perfect matching (MWPM) method \cite{fractal2d} based on Blossom V
\cite{kolmogorov_blossom_2009} for the polynomial 2D problem and the general-purpose
Gurobi software \cite{gurobi_optimization_llc_gurobi_2022} for 3D systems and the SK
model. Figure ~\ref{fig:solver_diagram} schematically depicts the setup: we
computed GSs for the entire ($N$-spin) systems as well as for subsystems with
$N_\mathrm{sub}$ lattice sites centered on the considered bond $\langle i,j\rangle$ (corresponding
to the LSBS), comparing the predicted values of the product $\SISJNOT$ (open boundary
conditions are applied in both cases). If the results agree, the bond product was
predicted correctly (with zero error). For $L \times L$ square lattices of sizes
$L=256$, $512$, $1024$, we examined $L_{\rm sub} \times L_{\rm sub}$ subsystems with
$4 \le L_{\rm sub} \le 512$, using $\sim60\,000$ disorder realizations. For cubic
lattices of side lengths $L=10$ and $12$, the
$L_{\rm sub} \times L_{\rm sub} \times L_{\rm sub}$ subsystems were of size
$4 \le L_{\rm sub} \le 10$, and we used $\sim 2\,000$ realizations. From these
calculations we estimate the LSBS error rate $\mathcal{E}$. Figure 
\ref{fig:solver_performance} shows the dependence of $\mathcal{E}$ on the subsystem
size for several cases.  For the  $\overline{J} = 0$ lattice systems we find that,
although the 2D and 3D spin-glass GS problems belong to different
complexity classes (P and NP, respectively), in both cases the error rate is well
described by a power-law decay in the subsystem size \footnote{While we will
  largely focus on the central nearest neighbor bond (with $i=i_{0}$ and $j=j_{0}$),
  in the End Matter we discuss more general situations in which the sites
  $i$ and $j$ are a distance ${\sf r}_{ij} >1$ apart. To that end and in order to
  avoid cumbersome notation, we will often dispense with ``0'' subscripts.}
\footnote{In 2D, for the largest values of $L_{\rm sub} = {\cal{O}}(L)$, the
  right-hand side of Eq.~(\ref{eq:solver}) constitutes an upper bound on the 
  error $\mathcal{E}_{ij}$. Clearly, when $L_{\rm sub} = L$ (when the subsystem is
  the entire system), the error $\mathcal{E}_{ij}=0$ by definition.}, viz.\
\begin{eqnarray}\label{eq:solver} 
  \mathcal{E}_{ij}\equiv \frac{1-\left[\sigma_i\sigma_j\sigma^{\rm sub}_i\sigma^{\rm sub}_j\right]}{2} \sim (\ell_{\mathcal{E}}/ L_{\rm sub})^{\kappa}.
\end{eqnarray}
Here, $\left[\sigma_i\sigma_j\sigma^{\rm sub}_i\sigma^{\rm sub}_j\right]$ denotes the
pair-overlap correlation function, where $\left[ ...\right]$ represents the disorder
average, and $\ell_\mathcal{E}$ is an effective length scale.  Importantly, this
algebraic decay is {\it independent of the linear system size} $L$, cf.\ the data
collapse in Figs.~\ref{fig:solver_performance}(a) and (b). Comparing the respective
exponents $\kappa$ that are collected in Table \ref{tab:fit_params}, we see that
errors in the 2D systems drop more rapidly than in the (harder) 3D systems. When
expressed in terms of the total volume $N_{\rm sub} = L_{\rm sub}^d$, where $d$ is the lattice dimension, this
sharp change in the algebraic decays of errors between the 2D and 3D cases becomes yet more acute. In \cite{SM} (Sec. \ref{sec:scaling_kappa}), we present scaling arguments that yield values of $\kappa$ close to those that we observe numerically. 

\begin{table}[tb!]
    
  \begin{tabular}{ccccc}
    \toprule
    $d$ & $\ell_{\mathcal{E}}$ & $\kappa$ &$k_{J}$ & $\sf{J}$ \\
    \midrule
    2D & 0.20(1) & 0.685(8) &0.198(8) & 0.169(7) \\
    3D & 0.0006(10) & 0.18(3) &0.47(3) & 0.71(5) \\
    \toprule
    $d$ &  $\beta$& $\ell_{c}$ & $a_{\mathrm{c}}$ \\
    \midrule
    2D &  1.35(5)  & 1.01(3) & 1.26(1)\\
    3D &  1 & 1.40(9) & 1.3(1)\\
    \bottomrule
  \end{tabular}
  \caption{Parameters of the fits of the functional forms of Eqs.~\eqref{eq:solver},
    \eqref{eq:erregy}, and \eqref{eq:deltajc} to our data. In 2D, $\chi^2/\rm{d.o.f}$
    is found to be $0.821$, $0.477$, and $1.067$ for Eqs.~\eqref{eq:solver},
    \eqref{eq:erregy}, \eqref{eq:deltajc}, respectively. In 3D, these values are
    $1.265$, $0.794$, $0.656$. Note that for the fit of the form
    \eqref{eq:erregy} in 3D, we fixed $\beta = 1$. For the 2D systems, the form
    (\ref{eq:solver}) was fitted with a cutoff $L_{\rm sub}\geq 16$ in order to
    minimize finite-size effects.
  }
  \label{tab:fit_params}
\end{table}

While the precise 
power-law decay differs between 2D and 3D, this variability is much less dramatic than 
the P vs.\ NP computational complexity contrast %
%
between these cases. This observation suggests  a ``local hardness''
computational complexity descriptor beyond the P-NP classification
\cite{papadimitriou:94} (and related categories \cite{LSA'}) shedding further light
on the intrinsic difficulty of spin-glass GS computations. By {\it local
  hardness}, we allude to {\it how large the subsystem} on which computations are
done must be so that, when averaged over many instances, the local solver provides a
correct answer up to a fixed {\it error rate}.  Arguably, the local complexity is a relevant descriptor for physical local measurements (the microscopic state of
the full many-body system cannot, in general, be probed).

\begin{figure}[htb]
    \centering
\includegraphics[width=0.40\textwidth]{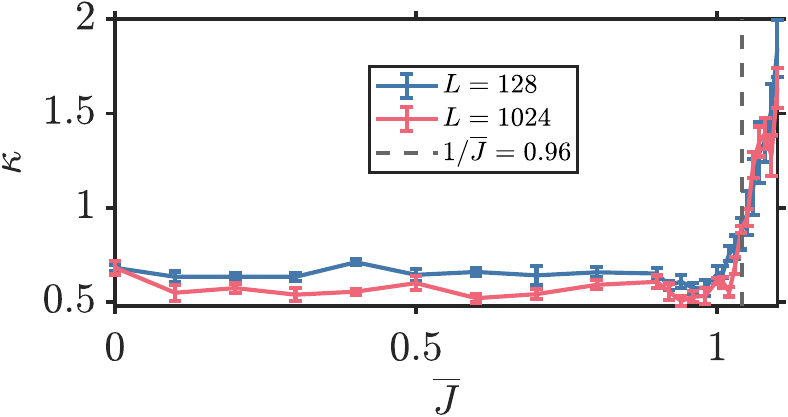}
    \caption{For an asymmetric bond distribution, as $\overline{J}$ increases the system transitions from a spin-glass phase to a disordered ferromagnetic phase  \cite{melchert_scaling_2009}. We observe how the fitting coefficient $\kappa$ changes as we adjust $\overline{J}$. A larger value of $\kappa$ implies that the error rate of the subsystem solver decays faster, with a crossover to an exponential decay beyond the threshold $1/\overline{J}=0.96$, where the systems enters the ferromagnetic phase \cite{melchert_scaling_2009}.}
    \label{fig:jbarphase}
\end{figure}

For the case of non-zero average coupling $\overline{J}$, we find the same power-law
decay as long as we remain in the (zero-temperature) spin-glass phase, thus
suggesting some degree of universality in the behavior of local hardness. As soon as
one enters the disordered ferromagnetic phase, however, there is a much faster, exponential
decay of the error rate, cf.\ the data for $\overline{J} = 1.2$ in
Fig.~\ref{fig:solver_performance}(a), Fig.~\ref{fig:jbarphase} (End Matter) and further details in Ref.~\cite{SM}, Sec.~\ref{sec:shifted_gaussian}. Similarly, we find identical 2D-type scaling and exponents for
planar antiferromagnets having faint random perturbations of their
couplings from an otherwise constant value (Ref.~\cite{SM}, Sec.~\ref{sec:fully_frustrated}). For the all-to-all SK model, on the other
hand, the data in Fig.~\ref{fig:solver_performance}(c) reveal that the local hardness
\emph{does} depend on the total system size $N$ and not only on 
the $N_\mathrm{sub}$ randomly chosen spins forming the ``local''  subsystem, thus
indicating a higher degree of local hardness of the SK model. This is consistent
with the fact that there is {\it no local geometry} in the SK system, and that it consequently features a
more complex free-energy landscape with infinitely many thermodynamic states
\cite{parisi_infinite_1979}. As a result, one needs to scale $N_\mathrm{sub}$
proportionally to $N$ to achieve low error rate for the LSBS in the SK model.

\begin{figure}[t]
    \centering
      \includegraphics[width = 0.40\textwidth]{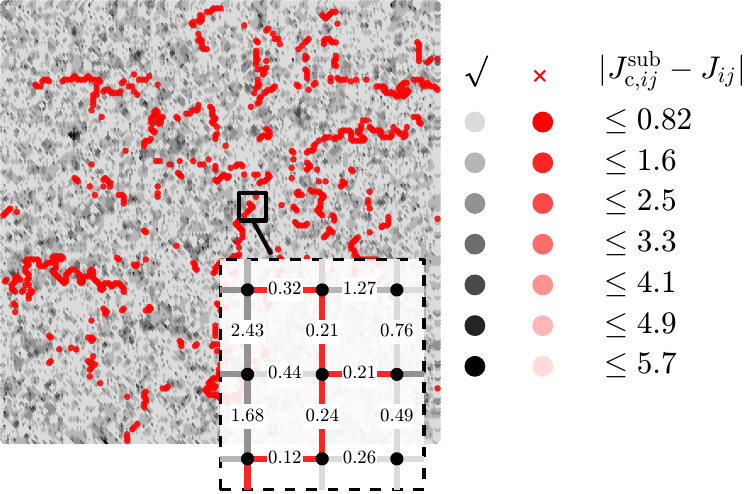}
      \caption{LSBS solutions for 
      nearest-neighbor bonds $\langle ij\rangle$ in an $L=128$ system. An $L_{\rm sub}=40$ subsystem is chosen to be centered about any such bond $\langle i j \rangle$ (near the boundaries, the
     subsystem becomes correspondingly smaller). $\JCsub$ is the critical threshold of bond $\langle ij\rangle$. $\color{red} \times$ denotes an error wherein 
     the local solver does not match the global solution, while $\checkmark$ refers to correct LSBS predictions.
     Errors 
     arise more readily 
     for 
     smaller 
      $|\JCsub-J_{ij}|$. It is noteworthy that correct LSBS predictions also
appear for small values of $|\JCsub- J_{ij} |$. Zoomed in 
fragment:  
individual bond $|\JCsub-J_{ij}|$ values are overlayed on color coded correct/incorrect LSBS predictions for these bonds.}
      \label{fig:solver_diagram'}
\end{figure}

We next turn to our central endeavor --- that of understanding the physics underlying
local solvers. Our approach relies on the concept of ``critical thresholds'' $\JC$
\cite{newman_ground_2021,shen2023universal}, defined as follows. Across any bond, the
spin product $\sigma_i\sigma_j$ is either $+1$ or $-1$. Consider now continuously
varying $J_{ij}$ from $- \infty$ to $+ \infty$ leaving the strengths of all other
bonds unchanged. As we have shown elsewhere \cite{shen2023universal}, there is a
\emph{unique} value $J_{ij} = \JC$ where the relative orientation of $\sigma_i$ and
$\sigma_j$ changes from $\sigma_i = -\sigma_j$ ($J_{ij} < \JC$) to
$\sigma_i = \sigma_j$ ($J_{ij} > \JC$) and, hence, two GSs become
degenerate. A similar definition applies for the subsystem --- changing the coupling
$J_{ij}$ beyond $\JCsub$ leads to a change of the subsystem GS. For the
LSBS we have the following important observation: If $\JCsub$ and $\JC$ are either
(i) both smaller or (ii) both larger than $J_{ij}$, then the LSBS (determining the
value of $\sigma_i \sigma_j$ as calculated within the smaller subsystem GS)
will yield the correct $\sigma_i\sigma_j$ value (i.e., that computed within the
global system GS). 
The opposite case of $J_{ij}$ being sandwiched
between $\JC$ and $\JCsub$, leading to a failure of LSBS, is more likely to occur if
$J_{ij}$ is close to the critical thresholds. This is illustrated in Fig.~\ref{fig:solver_diagram'} correlating the error rate of bond predictions with the couplings' deviation from criticality.  
Thus, the (subsystem) \emph{stability} of a bond, captured by the distance $|\JCsub-J_{ij}|$ between the bare coupling and its subsystem critical threshold
can 
serve as a predictor of LSBS reliability
\footnote{A very similar behavior is observed for the distance $|\JC-J_{ij}|$, see
  the discussions in Supplemental Material Sec. \ref{sec:scaling_subsytem_jc}, but this quantity is not accessible from the
  perspective of the local solver.}. When $\JCsub$ and $J_{ij}$ are far apart, then
$\JCsub$ and $\JC$ may indeed easily be both larger or both smaller than
$J_{ij}$. Conversely, when $\JCsub$ and $J_{ij}$ are close, slight miscalculations
can lead to $\JCsub$ and $\JC$ to appear on opposite sides of $J_{ij}$, thus causing
the LSBS to 
yield incorrect $\sigma_i\sigma_j$ values (see also Ref. \cite{SM}, Sec. \ref{sec:core_concept}).

At the critical threshold $\JC$, the GS responds sensitively to
infinitesimal single-bond changes. Similar to avalanches in the random-field Ising
model \cite{sethna:01}, neural networks \cite{Avalanche1}, systems featuring Highly
Optimized Tolerance \cite{HOT}, and other problems \cite{Avalanche2}, the
GS spins that are flipped as a result of such changes form what we call
Zero Energy Droplets (ZEDs) featuring fractal boundaries and sizes conforming to power-law distributions
\cite{shen2023universal}.
In \cite{SM} (Sec. \ref{sec:scaling_kappa}) we suggest a simple relation, which appears to be consistent with our data, that links ZED boundary Hausdorff dimensions to the exponent $\kappa$ of Eq. (\ref{eq:solver}).
The ZED distributions and the heterogeneous structure of these rare events, characterized by highly non-uniform spatial distributions, bears resemblance to the temperature 
\cite{fernandez2013temperature,baity2021temperature,katzgraber_temperature_2007} and disorder chaos \cite{krzakala_disorder_2005,katzgraber_temperature_2007} phenomena in spin glasses, where rare events 
drive system-spanning changes; 
these distributions may be tied to complex critical spatial correlations (Ref.~\cite{SM}, Sec.~\ref{sec:sg_criticality}). 
Configurations associated with
general excitations may be constructed as composites of (single bond) ZEDs
\cite{shen2023universal}. These excitations are, by construction, of
vanishing energy. Right at the critical threshold of a given bond, the global GS is extremely sensitive to infinitesimal
changes of that local bond coupling, and the LSBS 
attains its lowest accuracy. However, when the coupling constants are sufficiently removed from their
(measure zero) critical thresholds, the local solver becomes progressively
more accurate. As is seen in Fig.~\ref{fig:solver_diagram'} as well as in Fig.~\ref{fig:error_jc_diff}, End Matter, the 
error rate decays with
increasing $|\JCsub-J_{ij}|$. 
This decay is well described by   
\begin{eqnarray}\label{eq:erregy}
\mathcal{E}_{ij} \sim k_J \, e^{-\left (\frac{|\JCsub-J_{ij}|}{\sf{J}}\right )^\beta}, 
\end{eqnarray}
with $\beta>1$ (2D) and $\beta =1$ (3D), respectively, see Table \ref{tab:fit_params}. 
As for the dependence on $L_{\rm sub}$ expressed in Eq.~(\ref{eq:solver}), the error rate $\mathcal{E}_{ij}$ exhibits the same functional dependence on the 
ility in 2D and 3D (Ref.~\cite{SM}, Sec.~\ref{sec:2Dvs3D}). The modulus 
$|\JCsub-J_{ij}|$ is equal to twice the bond  excitation energy $\Delta E$ 
within the subsystem, cf.\
Eq.~(\ref{eq:jccompute}) in Ref.~\cite{SM}, Sec.~\ref{sec:compute_jc}. Thus, the LSBS accuracy 
is determined by this 
excitation energy. 

We further tuned the central bonds of the $L=128$ square and $L=12$ cubic lattices to
their critical values and 
computed 
the respective changes in critical threshold
($\Delta \JC$) of all other bonds. We find that the average difference $|\Delta \JC|$
decays algebraically with 
distance from the modified central bond (see
Fig.~\ref{fig:jclocality}, End Matter),
\begin{eqnarray}\label{eq:deltajc}
  [ |\Delta \JC| ] \sim (\ell_{c}/r)^{a_c}. 
\end{eqnarray}
In 2D and 3D systems, both the length $\ell_c$ and the exponent $a_c$ are of
order unity, the fit parameters are provided in Table \ref{tab:fit_params}. For
sufficiently large $r$, the central bond coupling negligibly impacts the $\JC$ of
distant bonds. In other words, the critical coupling of a given bond is almost
completely determined only by the couplings on bonds that are close to it. This
exemplifies the ``locality of the critical threshold $\JC$.'' 
Since
the ZED volume distribution decays algebraically \cite{shen2023universal}, for
large distances $r$ between the tuned and observed bonds, the probability
of ZED excitations connecting these bonds becomes small. 
When resampling the
system boundary, $[ |\Delta \JC| ]$ decays similar to Eq.~(\ref{eq:deltajc}), see
Ref.~\cite{SM}, Sec. \ref{sec:scaling_subsytem_jc}.  Thus, generating a subsystem by removing peripheral bonds is of limited
impact on the critical coupling of the central bond thus explaining the proximity of
$\JCsub$ and $\JC$.

In summary, using {\it exact GS algorithms}, we study the effectiveness of divide-and-conquer type local subsystem solvers and relate it to critical threshold ZED physics \cite{shen2023universal} of system-spanning avalanche-like vanishing-energy excitations. In short-range spin-glasses, away from critical thresholds, the response of the system GS to local perturbations decays rapidly. Generally, the difference between local and global minimization only becomes significant near critical coupling threshold values. Away from these critical values, local GS solvers achieve increased accuracy. Importantly, the associated error rate of the local solvers becomes {\it independent of the system size} [see Fig.~\ref{fig:solver_performance} (a)] for large systems and decreases rapidly with the subsystem size. Local solvers may thus become {\it increasingly effective} for large $L$. 
Our analysis reveals that various length-dependent quantities 
display algebraic scaling just like the scale-free distributions of avalanche size \cite{shen2023universal}, while for ferromagnetic phases we observe an exponential decay of errors with subsystem size (see also End Matter and \cite{SM}, Sec.~\ref{sec:shifted_gaussian}). These behaviors may rationalize the success of more sophisticated 
solvers in spin-glasses \cite{wang_patch-planting_2017,RL_local,RL_local1} and other systems \cite{Imagenet, Alphafold}, and might be broadly applicable in systems not constrained by the ``Overlap Gap Property'' \cite{OGP}. For the all-to-all SK model, on the other hand, the error rates always depend on the size of the full system. These differences lead us to conjecture a classification of possible levels of local hardness (Ref.~\cite{SM}, Sec.~\ref{sec:level_hardness}). Local solvers are found to be efficient for the studied lattices irrespective of whether the global problems are of P (2D) or NP (3D) complexity. The occurrence of rare events of bonds being close to their critical couplings explains why local solvers cannot, in general, guarantee to find exact GSs of the full systems. This mechanism which is reminiscent of chaos in spin glasses \cite{fernandez2013temperature,baity2021temperature,katzgraber_temperature_2007} is generic for the classes of systems we study.
It is instructive to 
briefly relate the 
local-hardness perspective 
to the metastate and chaotic-size-dependence framework~\cite{Newmanbook}: there, one asks whether local observables have an effectively ``concentrated'' or more ``dispersed'' distribution over thermodynamic states. In our 
setting, 
large stability $|\JCsub-J_{ij}|$ 
corresponds to a computable proxy 
for 
a robust metastate-like concentration of local observables 
of low LSBS error while small stability reflects 
competition between nearly degenerate states. 

While these considerations were focused on the hyper-local task of predicting the spins on both ends of a single bond, it is possible to stitch together the results of such calculations to candidate ground states of the full system.
In the End Matter and Ref. \cite{SM} (Sec.~\ref{sec:contraction_Solver}), we 
outline an LSBS-based solver that uses the subsystem critical thresholds $\JCsub$ to identify ``well-determined'' bonds, i.e., bonds for which $|\JCsub-J_{ij}|$ is large and the LSBS error rate is exponentially small, cf.~Eq.~(\ref{eq:erregy}). This identification  enables the contraction of the system into a smaller number of cohesive large-scale elements (or ``communities'' 
\cite{Girvan_Newman2002,Fortunato2010,Comm_detection_local,Comm_detection_local'}) before solving for the GS. This extension is here presented as proof-of-concept demonstration of how local hardness can guide algorithm design, and further detailed study is required to gauge its full potential. Generalized to 
settings that include external fields and multi-body interactions 
(Ref.~\cite{SM}, Sec.~\ref{sec:multibody}), 
our framework becomes applicable to other important NP-hard optimization problems, such as the Minimum Vertex Cover problem (see End Matter and Ref. \cite{SM}, Sec.~\ref{sec:mvc_ising}). 
Locality 
further 
underlies numerous many-body physics problems \cite{Kohn-near,Cluster1,Cluster2,Cluster3,Cluster4} where 
degrees of freedom may also be {\it continuous}; replacing Eq. (\ref{HSG}) by $H = - \sum_{ij} J_{ij} s_i s_j + u \sum_{i} (s_i^2-1)^2$ with $u \gg 1$ (Eq. (\ref{HSG}) corresponds to $u \to \infty$) for continuous $s_{i} \in \mathbb{R}$ suggests that our 
findings 
extend to continuous 
systems of sufficiently large quartic interaction strength $u$. In a 
continuous spin variant defined by a {\it global constraint}-- the spherical model~\cite{Berlin+Kac,kosterlitz_spherical_1977}-- the locality that we described here is altered (Ref.~\cite{SM}, Sec.~\ref{sec:solvable}).
Another interesting application area for the ideas outlined here lies in machine learning, with the potential of replacing slow global gradient methods by fast sequential local stochastic gradient descent-type approaches 
\cite{Harju1997,Comm_detection_local,Comm_detection_local',Ruder2017,review_ML_SGD,MatDis}. Local plastic variants of the backpropagation algorithm used in machine learning might emulate efficient biological neural computing \cite{Song2020,Goodfellow}. Other notable aspects of local minimization and optimized structure appear in biological and artificial neural networks 
\cite{NN_modular,modular_brain}.
We view our 
LSBS-based contraction 
solver as a starting point for future investigations of local-hardness–guided heuristics. 
Dividing large scale problems into smaller 
easier-to-solve fragments may also be beneficial for quantum computing schemes for which smaller scale systems might suffer less from decoherence. 
Extensions of the GS problem analyzed here, including products between distant spins (see Ref. \cite{SM}, Sec. \ref{sec:distant_spins}) to 
finite temperature correlation functions 
may provide illuminating  
connections between our results and finite size scaling  \cite{FS1,FS2,FS3,FS4}. 
In future work, we will further 
detail relations to previously unreported criticality (Ref.~\cite{SM}, Sec.~\ref{sec:sg_criticality}). 
The physics of local optimization may underlie Nature's 
efficiency, shaping computation 
across diverse realms.

{\bf Acknowledgments.} We thank Mark Goh for a discussion in which he alerted us to \cite{OGP}. Z.N. and M.W. are grateful for support from TU Chemnitz through the Visiting Scholar Program. Z.N. is further grateful for support through the Washington University Seeding Projects for Enabling Excellence \& Distinction (SPEED) and the Leverhulme-Peierls
Senior Researcher Professorship at Oxford via a
Leverhulme Trust International Professorship Grant No. LIP2020-014 (S. L. Sondhi). G.O. gratefully acknowledges support from the Institute for Advanced Study.

\bibliography{ref.bib}
\bibliographystyle{my}

\bigskip





\pagebreak
\newpage
\newpage

\clearpage

\section*{End Matter}



Here, we provide numerical evidence for Eqs.~(\ref{eq:erregy}) and (\ref{eq:deltajc}), the existence of a transition in the LSBS accuracy as the distribution of couplings is shifted (so as make the system more ferromagnetic and easier to solve), and we discuss beyond nearest-neighbor correlations found by the LSBS. 

\subsection*{Error rates for  deviations from the critical threshold}
 
\begin{figure}[htb]
    \centering
        \includegraphics[width = 0.45\textwidth]{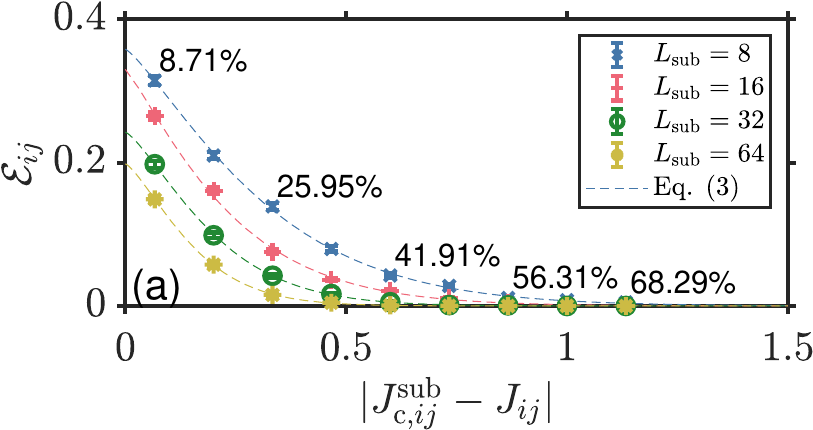}
    \includegraphics[width = 0.45\textwidth]{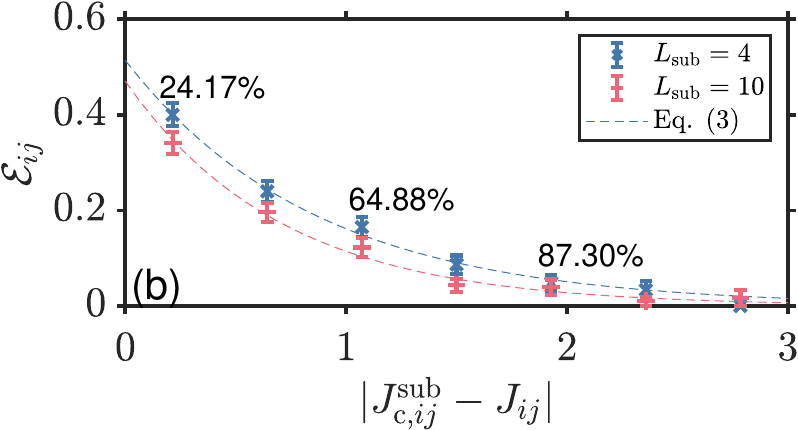}
    \caption{Error rate $\mathcal{E}_{ij}$ as a 
    function of $|\JCsub-J_{ij}|$ for (a) a 2D (square lattice), $L=1024$ system with $L_{\rm sub}=8$, $16$, $32$, $64$ and (b) a 3D (cubic) system of $L=12$ with $L_{\rm sub}=4$, $10$. Point percentages denote cumulative probabilities  (i.e., fraction of instances) for values of $|\JCsub-J_{ij}|$ lower than their abscissa.}  
    \label{fig:error_jc_diff}
\end{figure}

Illustrating the applicability of the analysis in the main text, Fig. \ref{fig:error_jc_diff} demonstrates
that the LSBS error
indeed 
decays 
with increasing $|\JCsub-J_{ij}|$ in a manner that adheres to the stretched exponential drop of Eq. (\ref{eq:erregy}). 



\begin{figure}[htb]
    \centering
    \includegraphics[width = 0.35\textwidth]{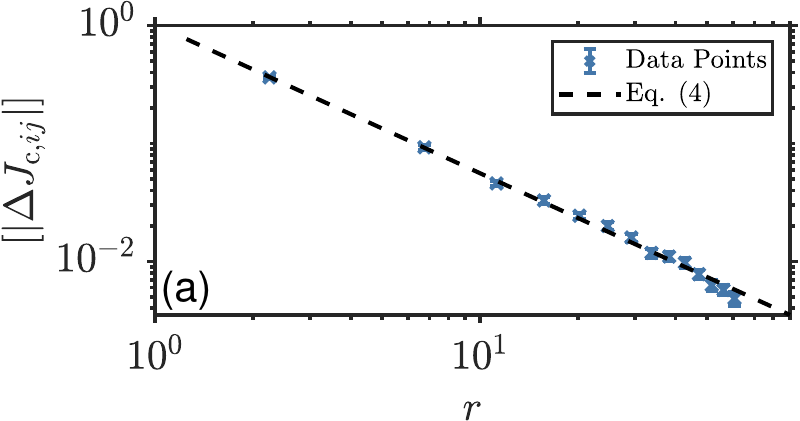}
    \includegraphics[width = 0.35\textwidth]{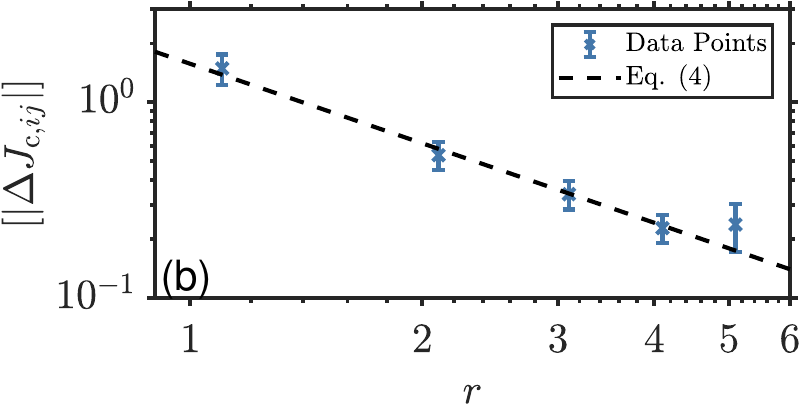}
    \caption{
    Average change of critical threshold values  ($[ |\Delta {\JC}|]$) of bonds a distance of $r$ along a Cartesian direction from $-\infty$ to $+\infty$ in (a) 2D ($L=128$) and (b) 3D ($L=12$) systems. Dashed: Eq. (\ref{eq:deltajc}) with parameters in Table  \ref{tab:fit_params}. As is more evident in 2D, when $r= {\cal{O}}(L)$, the average $[ |\Delta \JC|]$ is bounded from above by these ultra-local forms with the fitted $\ell_{c}$ being of the order of the lattice constant.}
    \label{fig:jclocality}
\end{figure}

\subsection{Distance dependence of critical threshold values}

As discussed in the main text, the critical threshold of a given bond is largely determined only by the couplings of other bonds that are very close to it. When we tuned the central bonds of $L=128$ square and $L=12$ cubic lattices to
their critical values, the respective critical threshold changes
($\Delta \JC$) of all other bonds at a distance $r$ away conformed to Eq.~(\ref{eq:deltajc}) with $\ell_{c}$ of the order of the lattice constant, see Table  \ref{tab:fit_params} (with the fit of Eq.~(\ref{eq:deltajc}) becoming an upper bound for the largest distances in the 2D system). This is illustrated in Fig. \ref{fig:jclocality}. In Ref. \cite{SM}, Sec. \ref{sec:correl_critical_couplings} (see Fig. \ref{fig:visualization_Jc_change} therein), we further provide a calculated real-space visualization of such a typical rapid drop of the change in the critical coupling away from the origin. We further observe a smaller effect on distant bonds for smaller ZEDs.

\begin{figure}[htb]
    \hspace*{-3cm}\begin{overpic}[width = 0.15\textwidth]{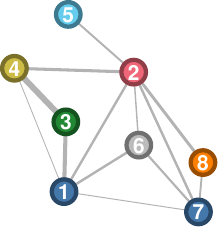}
        \put(-50,4){\fontsize{12}{16}\fontfamily{phv}\selectfont{(a)}}
    \end{overpic}\\[0.3cm]
    \begin{overpic}[width = 0.40\textwidth]{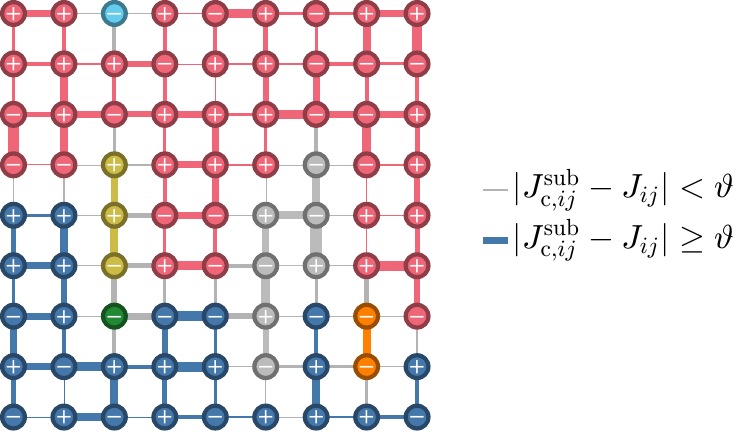}
        \put(-8,-2){\fontsize{12}{16}\fontfamily{phv}\selectfont{(b)}}
    \end{overpic}
    \caption{Schematic of the contraction solver. (a) Well-determined bonds ($|J_{ij} - \JC| \geq \vartheta$, where $\vartheta$ is a manually chosen threshold) form connected components that are contracted to effective sites (shown as larger nodes), inheriting all interactions between the component and the rest of the system. (b) The relative spin orientations (indicated by $+$ and $-$ signs) of sites within each connected component are locked in by the well-determined bonds. The GS of the full system can then be reconstructed from the GS of the contracted system by propagating the component orientations. Different colors represent distinct connected components.}
\label{fig:contraction_solver_schematic}
\end{figure}

\subsection*{Contraction solver based on LSBS}
\label{sec:contraction_Solver}

The LSBS provides local predictions for individual bond spin products. Here we outline how these local predictions can be combined into a \emph{contraction solver} that approximates the full GS spin configuration. For each bond, we compute its subsystem critical threshold $\JC$ and identify \emph{well-determined bonds} satisfying $|J_{ij} - \JC| \geq \vartheta$, where $\vartheta$ is a chosen threshold. These bonds, being far from their critical thresholds, have high local prediction accuracy. Well-determined bonds form connected components (or ``communities'' \cite{Girvan_Newman2002,Fortunato2010,Comm_detection_local,Comm_detection_local'}) within which relative spin orientations are reliably fixed. Each such component is contracted to a single effective site, inheriting all interactions between the component and the rest of the system. The contracted system is solved exactly, and the original spin configuration is reconstructed by propagating the component orientations. This contraction procedure is illustrated schematically in Fig.~\ref{fig:contraction_solver_schematic}. For appropriate choices of $\vartheta$, this approach yields configurations very close to the exact GS while substantially reducing the effective problem size. Detailed schematics, algorithmic steps, and performance analysis are provided in the Supplemental Material (Ref.~\cite{SM}, Sec.~\ref{sec:contraction_Solver}). \bigskip

\subsection*{Minimum Vertex Cover as an Ising problem}

\begin{figure}[htb]
    \centering
\includegraphics[width=0.48\textwidth]{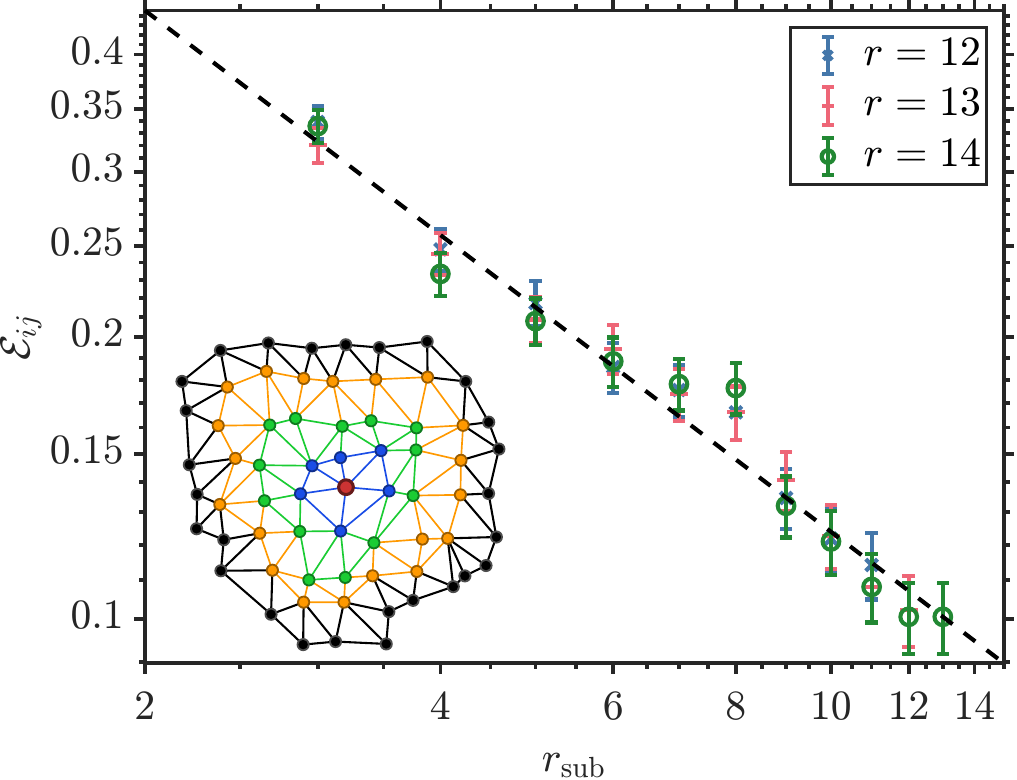}\\
    \caption{Minimum Vertex Cover (MVC) as an Ising problem on random planar graphs. LSBS error rate for the central spin in the MVC mapping, of the random planar graph shown in the inset, as a function of subsystem size, showing a power-law-like decay with exponent $\kappa \approx 0.8$. (Lower left inset) Random planar graph generated from $n=10^4$ points by Delaunay triangulation, with sites colored by topological distance $r$ from the geometric center.}
    \label{fig:mvc_main}
\end{figure}

As a representative NP-hard optimization problem beyond spin glasses, we also examined the Minimum Vertex Cover (MVC) on random planar graphs generated by Delaunay triangulation. Mapping a vertex $i$ being in the cover to $\sigma_i=+1$ and otherwise $\sigma_i=-1$, the cost function can be written as
\begin{equation}
\label{eq:mvc_hamiltonian_main}
H_{\text{MVC}} = A \sum_{i\in V}\frac{1+\sigma_i}{2}+B\sum_{\langle ij\rangle\in E}\frac{1-\sigma_i}{2}\frac{1-\sigma_j}{2},
\end{equation}
with couplings $J_{ij}=-B/4$ on edges $\langle ij\rangle\in E$ and fields $h_i=-A/2+(B/4)\deg(i)$ (for $A=1$, $B=2$). Using LSBS on subsystems of topological radius $r$ around a central site, we predict its relative spin orientation in the MVC GS and compare to the exact solution, finding a power-law-like decay of the LSBS error rate with subsystem size (Fig.~\ref{fig:mvc_main}) with exponent $\kappa \approx 0.8$, close to the spin-glass case. This suggests that our locality-based hardness picture applies more broadly to combinatorial optimization problems.

\end{document}


\title{
Supplemental Material for:\\ ``The Physics of Local Optimization in Complex Disordered Systems''}

\ifdefined\HCode\else
\externaldocument[][nocite]{final_no_red_xr}
\fi 
\author{Mutian Shen}
\affiliation{Department of Physics, Washington University, St.
Louis, MO 63160, USA}
\affiliation{Department of Biomedical Engineering, Washington University, St.
Louis, MO 63160, USA}

\author{Gerardo Ortiz}
\affiliation{Department of Physics, Indiana University, Bloomington, IN 47405, USA}
\affiliation{Institute for Advanced Study, Princeton, NJ 08540, USA}
\author{Zhiqiao Dong}
\affiliation{Department of Mechanical Science \& Engineering, University of Illinois Urbana-Champaign, IL 61801, USA}
\author{Martin Weigel}
\affiliation{Institut für Physik, Technische Universität Chemnitz, 09107 Chemnitz, Germany}

\author{Zohar Nussinov}
\email{corresponding author: zohar@wustl.edu}
\affiliation{Department of Physics, Washington University, St.
Louis, MO 63160, USA}
\affiliation{Institut für Physik, Technische Universität Chemnitz, 09107 Chemnitz, Germany}
\affiliation{Rudolf Peierls Centre for Theoretical Physics, University of Oxford, Oxford OX1 3PU, United Kingdom}
\affiliation{LPTMC, CNRS-UMR 7600, Sorbonne Universit\'e, 4 Place Jussieu, 75252 Paris cedex 05, France}
\maketitle

\renewcommand{\thesection}{S\arabic{section}}

\renewcommand{\thefigure}{S\arabic{figure}}
\renewcommand{\thetable}{S\arabic{table}}
\renewcommand{\theequation}{S\arabic{equation}}
\renewcommand{\thealgocf}{S\arabic{algocf}}

\setcounter{figure}{0}
\setcounter{table}{0}
\setcounter{equation}{0}

\setcounter{secnumdepth}{3}


\section{Computation of critical threshold coupling}\label{sec:compute_jc}
To determine 
the critical threshold $\JC$, 
we first compute the GS of the system and extract $\underline{\sigma_i\sigma_j}$, 
the value of this nearest-neighbor spin product 
in the GS with the original coupling on the link $\langle i j \rangle$ which we here explicitly  denote by $\underline{J_{ij}}$. 
Then, we modify 
the coupling $J_{ij}$ so that it 
has a sign opposite to that of  
$\underline{\sigma_i\sigma_j}$ with 
$|J_{ij}|$ sufficiently large  
so as to force the product of spins associated with this bond ($\sigma_i\sigma_j$) to change its sign in the new resulting GS. 
Next, this 
GS 
for the 
now modified coupling is determined; this new state constitutes an excited state 
(of energy $\Delta E>0$) 
for the original system (in which 
$J_{ij}$ is 
not changed). 
Once the energy $\Delta E$ is numerically computed
the critical coupling may then be determined via the equality 
\begin{equation}\label{eq:jccompute}
    \JC = \underline{J_{ij}}-\underline{\sigma_i\sigma_j}\Delta E/2.
\end{equation}
On general lattices and graphs, the probability density of critical couplings depends mainly on the 
(inherently {\it local}) coordination number \cite{shen2023universal}. This suggests that, on average, general functions $f$ depending on $\JC$ may be 
local (this includes the Heaviside function $f= \Theta(J_{ij} - \JC)$ which provides the aforementioned sign of
the bond $\langle i j \rangle$ (i.e., the product $\sigma_i \sigma_j$) that the LSBS aims to find). 
Applied to any subsystem 
containing the link $\langle i j \rangle$, we may similarly compute the 
{\it subsystem critical coupling}  
$\JCsub$. 

\section{Critical avalanches at transitions between global GSs}\label{sec:critical_avalanche}
\begin{figure}
    \centering
    \includegraphics[width=0.95\linewidth]{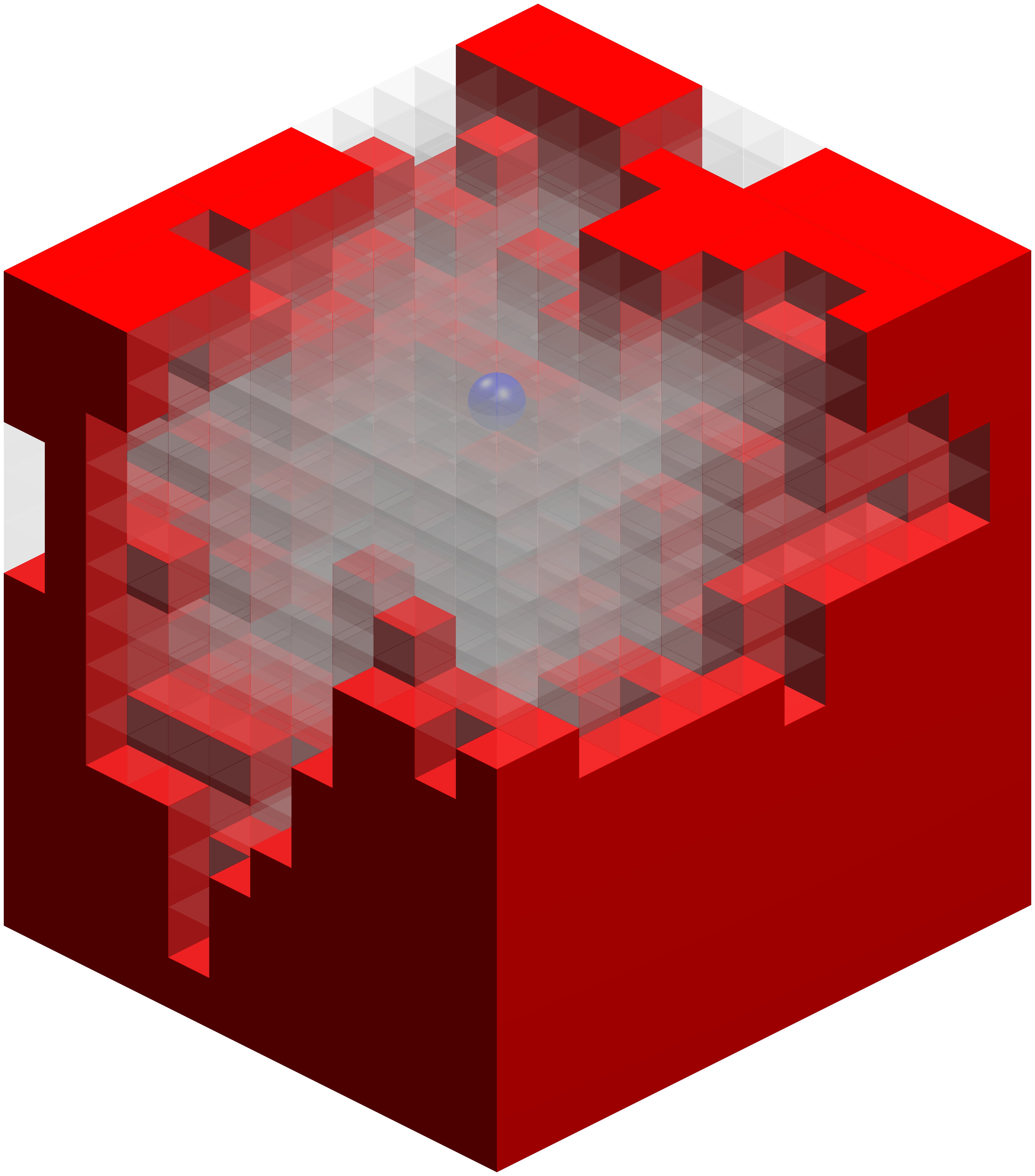}
    \caption{A computed critical avalanche associated with a transition (see text) between optimal global solutions- different GSs of the EA model. When the local spin-glass solution provided by the LSBS (Fig. \ref{fig:solver_diagram}) for the central  bond $\sigma_{i_{0}} \sigma_{j_{0}}$ (blue sphere) differs from the global one, the flipped spins form bubbles (above ``cloud'' surrounding the central bond) that occur at all scales (governed by critical power law distributions) \cite{shen2023universal}. Shown here is a typical critical avalanche for an $L=12 $ cubic lattice.}
    \label{fig:magnetprincess}
\end{figure}

Ref.~\cite{shen2023universal} examined, in disparate lattices, the transitions between different global solutions when a single nearest neighbor coupling traversed its critical value $\JC$ to universally find 
scale invariant power law distributions for ZED volume and area \cite{shen2023universal}. In all studied 2D and 3D lattices, critical system spanning avalanches appear at $\JC$ in which a divergent number of spins change their values. These avalanches are characterized by nontrivial exponents that (within obtainable numerical accuracy) are largely determined only by the spatial dimensionality and insensitive to specific lattice details. As explained in the main text, the rarity of these transitions --- the fact that they only occur at special $\JC$ values --- underlies the accuracy of the local optimization problem solved by LSBS (as compared to the exact full system GS). The highly heterogeneous structure of these rare avalanche events, where correct and incorrect LSBS predictions coexist for bonds with similar $|\JCsub-J_{ij}|$ values, bears similarity to temperature chaos \cite{fernandez2013temperature,baity2021temperature,katzgraber_temperature_2007} and disorder chaos \cite{krzakala_disorder_2005,katzgraber_temperature_2007} phenomena in spin glasses, where rare events also exhibit heterogeneous spatial distributions. In Fig. \ref{fig:magnetprincess}, we provide an example of a typical critical avalanche that occurs at $\JC$. Various details of these avalanches (which constitute zero energy droplet (ZED) excitations at $\JC$) are given in Ref.~\cite{shen2023universal}.

\section{More on a core concept underlying the LSBS}
\label{sec:core_concept}
\begin{figure}[htb!]
    \centering
    \includegraphics[width = 0.45\textwidth]{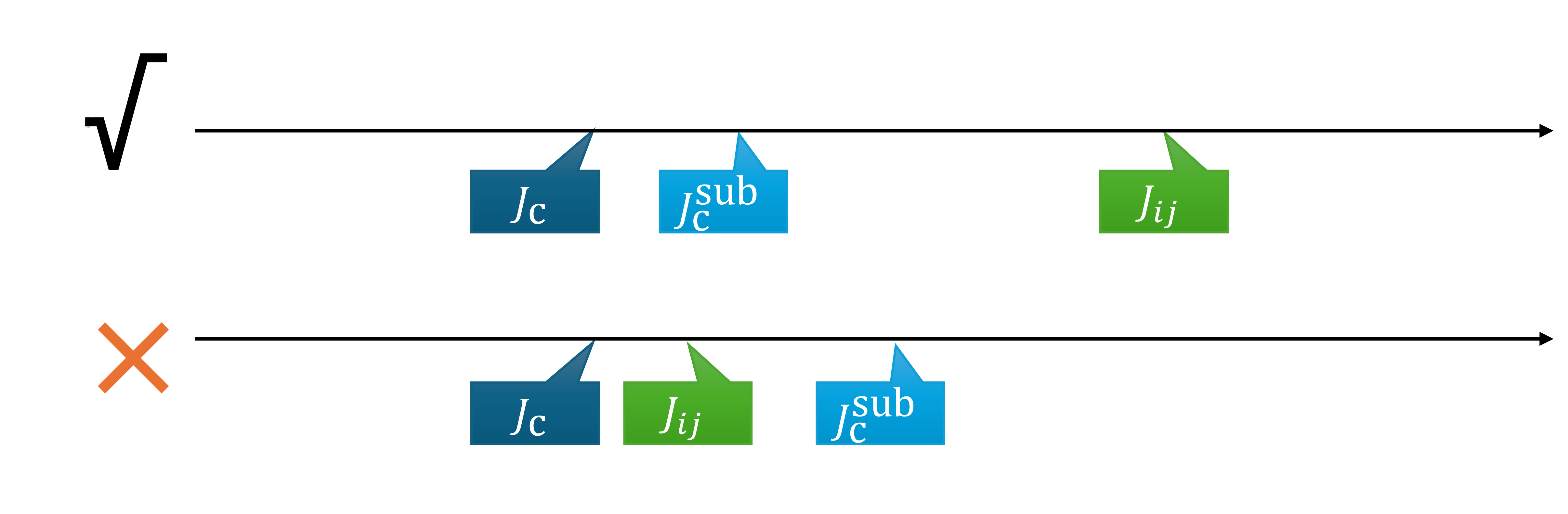}
    \caption{A schematic of the relative sizes of  the (i) the bond coupling constant $J_{ij}$ of a bond $\langle i j \rangle$ (green), (ii) the exact critical threshold $\JC$ (navy blue) of this bond 
    within the entire system
    and (iii) the critical threshold $\JCsub$ (light blue) of this bond as computed within a local patch or subsystem 
 that includes this bond at its center. 
 The exact GS spin product $\sigma_i\sigma_j=\pm 1$ when, respectively, the coupling 
 $J_{ij}\gtrless \JC$. 
 The value of $\sigma_i\sigma_j$ as determined  by the approximate local (subsystem) bond solver coincides with the exact full system GS answer 
 when both $(\JCsub-J_{ij})$ and $(\JC-J_{ij})$ are of the same sign ($\checkmark$ at top panel). When the signs of $(\JCsub-J_{ij})$ and $(\JC-J_{ij})$ are opposite, the local bond solver yields an incorrect prediction 
 ($\color{red} \times$ at bottom panel). As we will later detail (Eq. (\ref{Jcvars}) and Fig. \ref{fig:Jscub-Jx_vs_L_sub}), we indeed find that with increasing subsystem size, the deviation between the global and subsystem critical threshold value decreases, thus further consistently explaining the increasing accuracy of the LSBS with subsystem size.}
    \label{fig:jc_diagram}
\end{figure}

As discussed in the main text (and illustrated in Fig. \ref{fig:solver_diagram'}), a central theme lying at the heart of the LSBS concerns the proximity of the critical threshold value of the same bond $(ij)$ in subsystem and in the full system. If the full system ($J_c$) and subsystem ($J_{c}^{\rm sub}$) critical values for the bond $(ij)$ are close to each other then the local solver may still remain accurate  for
couplings $J_{ij}$ that are not far the critical threshold value in the subsystem $J_{c}^{\rm sub}$. Specifically, if the coupling $J_{ij} > \max (J_{c}^{\rm sub}, J_{c})$ or $J_{ij} < \min (J_{c}^{\rm sub}, J_{c})$ then the LSBS will provide the exact answer. The only ``dangerous'' situation in which the LSBS will provide the incorrect result is when the coupling for the bond $(ij)$ will lie between its global critical $\JC$ value and its subsystem critical threshold value $J_{c}^{\rm sub}$. If the latter threshold values are very close to each other then the probability this ``dangerous'' situation will occur can become exceedingly low. The graphic of Fig. \ref{fig:jc_diagram} further depicts this key notion. As illustrated in Fig.~\ref{fig:solver_diagram'} of the main text, as the coupling $J_{ij}$ becomes far removed from the local critical coupling $J_{c}^{\rm sub}$ for the bond $(ij)$, the LSBS indeed achieves higher accuracy.

\section{Extension of the LSBS to distant spins}
\label{sec:distant_spins}
\begin{figure}[H]
    \centering
    \includegraphics[width = 0.35\textwidth]{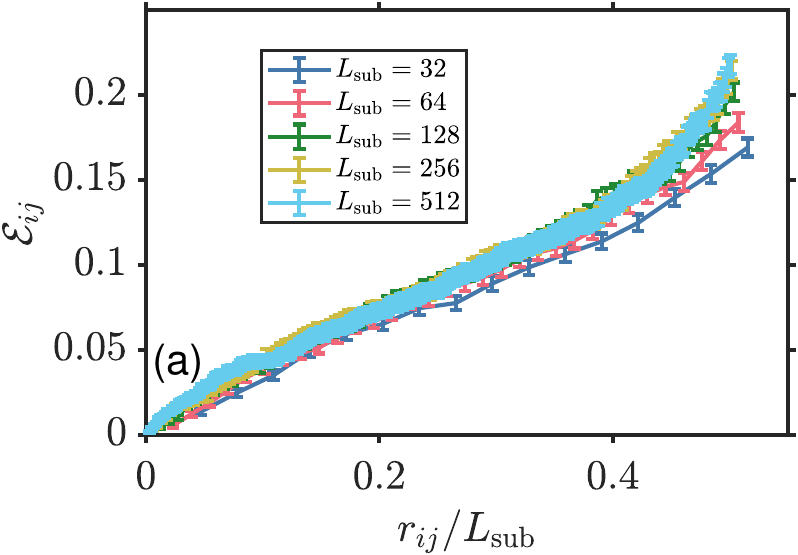}
    \includegraphics[width = 0.35\textwidth]{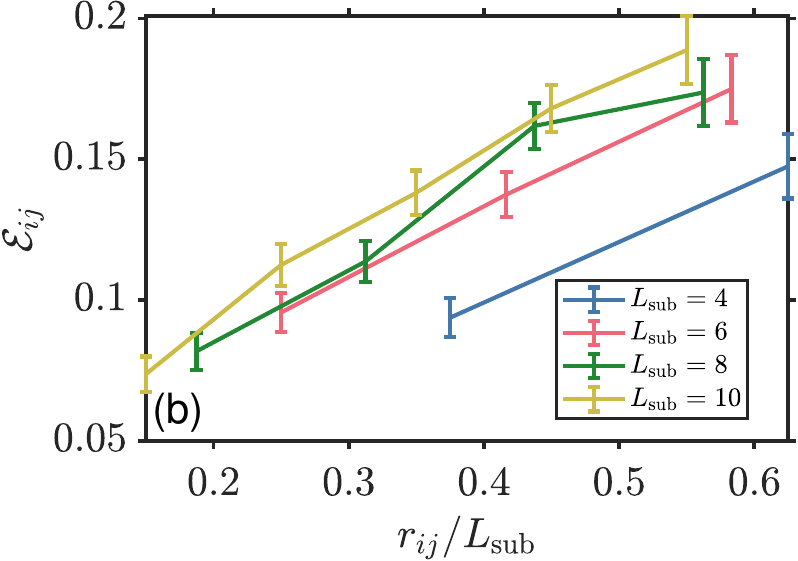} 
    \caption{LSBS error rate for GS bond products $\sigma_i \sigma_{j}$ when sites $i, j$ are a {\it distance ${r}_{ij}$ apart} along a Cartesian axis. 
    Results for (a) 2D and (b) 3D systems. The abscissa is normalized by the subsystem size, ${r}_{ij} /L_{\rm sub}$. As $L_{\rm sub}$ increases, ${\mathcal{E}}_{ij}$ 
    gradually collapses onto a single curve (as for the ${ r}_{ij} =1$ case of the main text).}
    \label{fig:corrdist}
\end{figure}

In the main text, we examined the accuracy of local solvers in determining nearest-neighbor spin products (or correlations) within the GS. In Fig. \ref{fig:corrdist}, we illustrate the result of our approach when the spins $\sigma_i$ and $\sigma_j$ are no longer nearest neighbors but are rather a distance ${\sf r}_{ij}$ away from each other along the Cartesian axes. Again, we compute the spin product within the GS of a local subsystem that contains these two spins. As may be expected, as the distance ${\sf r}_{ij}$ becomes larger relative to the subsystem size the error rate of the local solver increases.

\section{Visualization of the spatial correlations of critical couplings}
\label{sec:correl_critical_couplings}
\begin{figure}[htb]
    \centering
    \includegraphics[width=0.95\linewidth]{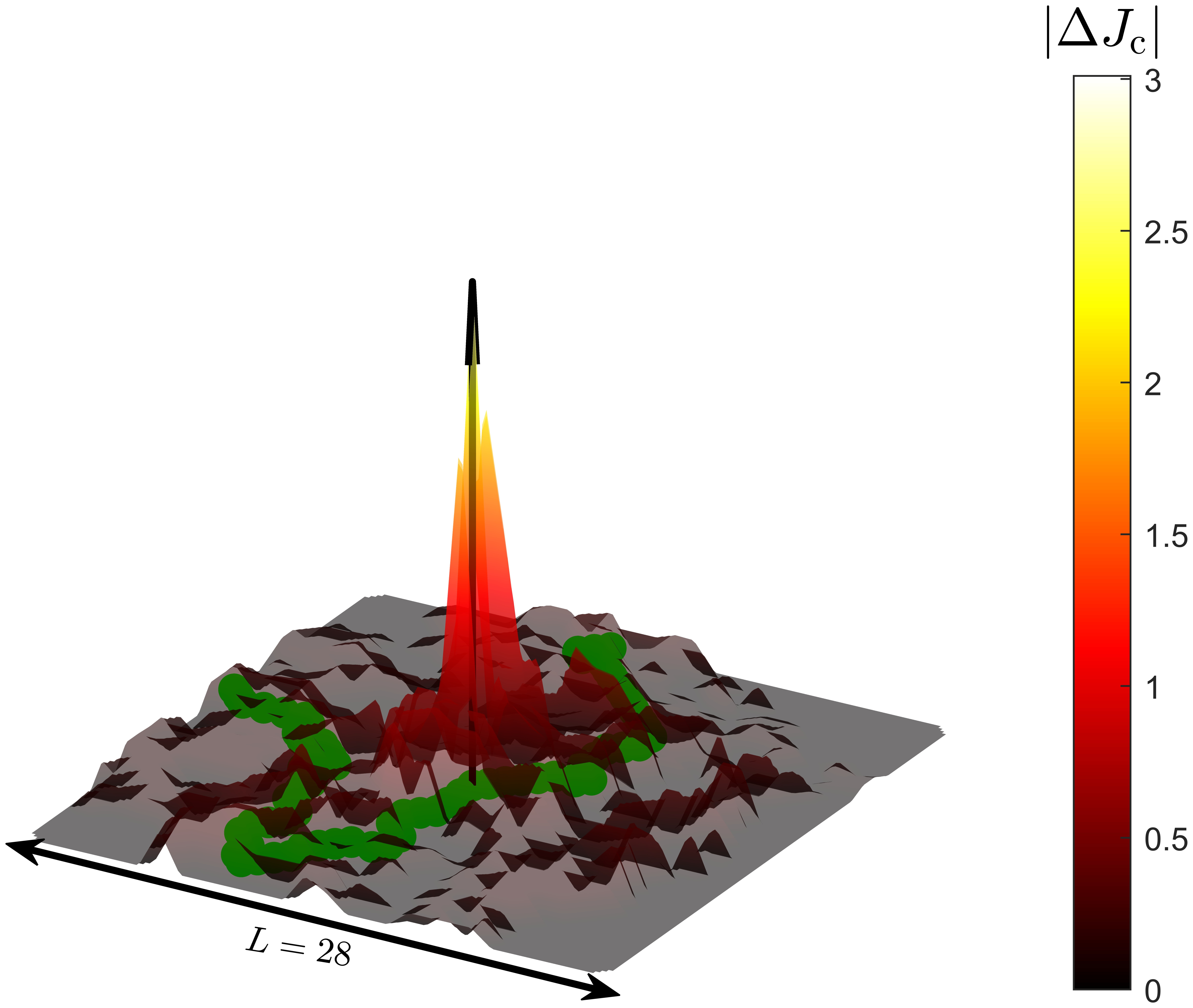}
    \caption{Numerically determined typical real-space realization of the data plotted in Fig.~\ref{fig:jclocality}. On an $L=28$ square lattice, altering the central bond coupling $J_{i_{0} j_{0}}$ from $-\infty$ to $+\infty$ generally induces the change of critical thresholds $\JC$'s of all other bonds in the system. The surface color map visualizes the amplitude of $\JC$'s change for a typical realization of the system. The green contour represents the flipped bonds between the GSs before and after $J_{ij}$ crosses its own $\JC$ as it is monotonically varied.}
    \label{fig:visualization_Jc_change}
\end{figure}

In Fig.~\ref{fig:jclocality}, we illustrated that the ultra-local influence of the average variation of $\JC$ at the origin may have on critical couplings at a distance $r$ away. These changes were given (and more generally bounded) by Eq.~(\ref{eq:deltajc}) with $\ell_J$ of the order of the lattice constant, see Table~\ref{tab:fit_params}. In Fig.~\ref{fig:visualization_Jc_change}, we provide a calculated real-space visualization of such a typical rapid drop of the couplings away from the origin by setting the nearest neighbor at the origin to its critical value.  

\section{Levels of Local Hardness}
\label{sec:level_hardness}

Before proceeding with the enumeration of viable formal levels of local hardness, we must underscore that the conventional P/NP classification relates to exact solutions of the full global problem (a vanishing error rate)- not that for a finite error rate as discussed in the current work which may naturally arise when only solving for a local fragment of the problem.

We next couch our spin-glass GS problem in a far broader context and ask how hard it generally may be for local solvers to solve global optimization problems. When examining systems (problems) composed of a large number of spins (or elements) $N \gg 1$, different levels of local hardness are conceivable:  
\begin{itemize}
    \item [I.] There exists an $N^*$ that is independent of the entire system size (total number of spins) $N$, such that as long as the size of the subsystem is larger than $N^*$, the error rate of the subsystem GS solver is strictly always $0$.
    \item [II.] For any maximally tolerable error rate $\epsilon$, there exists a subsystem size $N^*(\epsilon)<N$ that does not depend on the entire system size $N$, such that solving for the GSs in  subsystems $N_{\rm sub} \ge N^{*}(\epsilon)$ yields an error rate that is $\le \epsilon$.
    \item [III.] For any maximally tolerable   error rate $\epsilon$, there exists a subsystem size $N^*(\epsilon,N)<N$ that depends on the entire system size $N$ (and may diverge as $N\to \infty$), such that solving this subsystem can yield an error rate $\le \epsilon$.
    \item [IV.] Even if the subsystem size is made arbitrarily large so long as it smaller than that of the full system (i.e., $N_{\rm sub} < N$), it is impossible to guarantee that the error obtained when comparing the exact GS of the subsystem with that of the full system can be made lower than a finite error rate $\epsilon_{*})$. 

\end{itemize}

\begin{figure*}[htb]
    \centering
    \includegraphics[width = 0.45\textwidth]{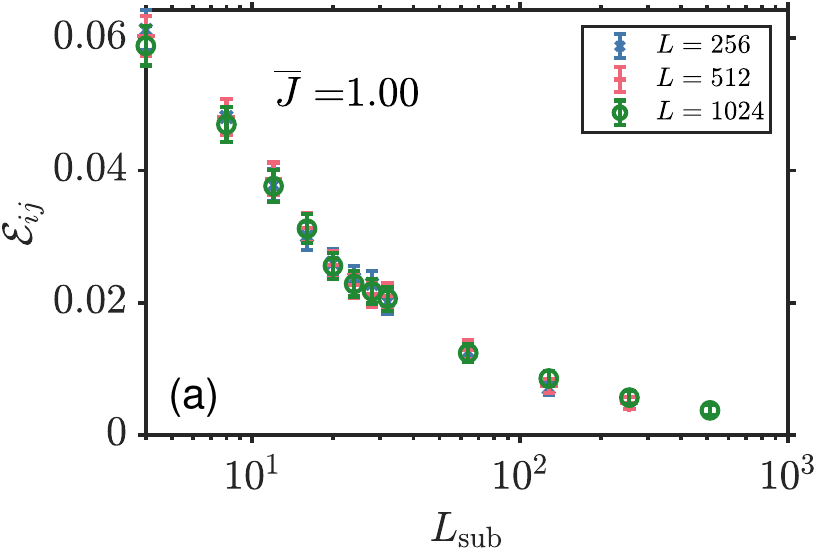}
    \includegraphics[width = 0.45\textwidth]{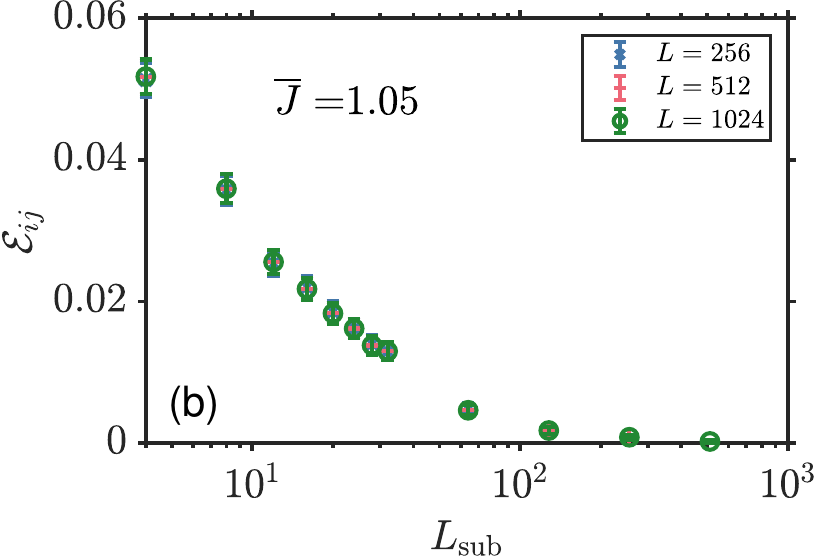}
    \includegraphics[width = 0.45\textwidth]{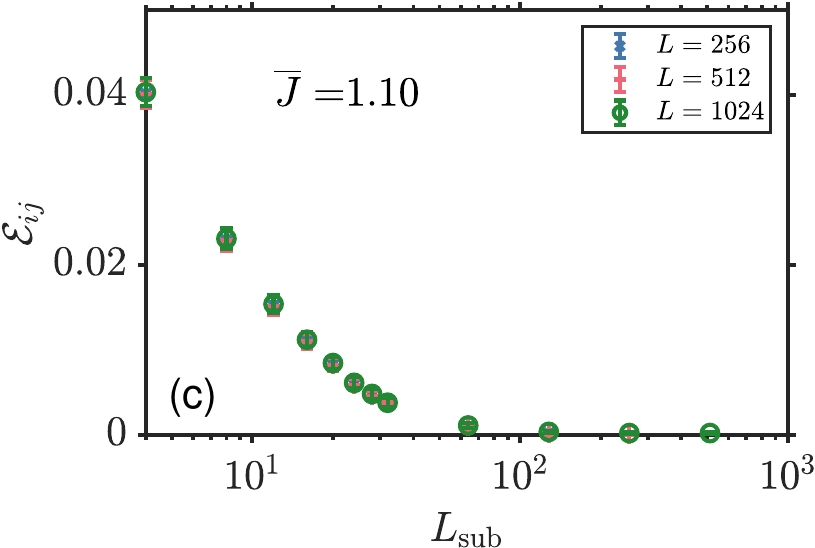}
    \includegraphics[width = 0.45\textwidth]{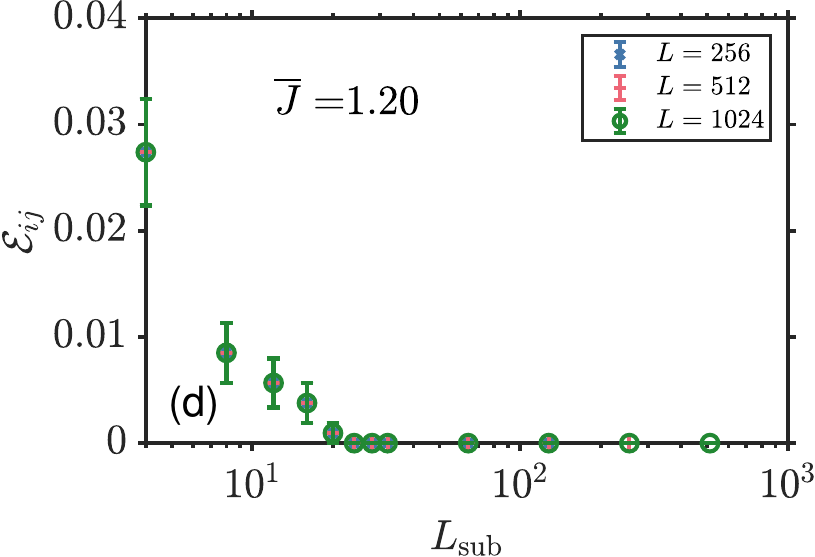}
    \includegraphics[width = 0.45\textwidth]{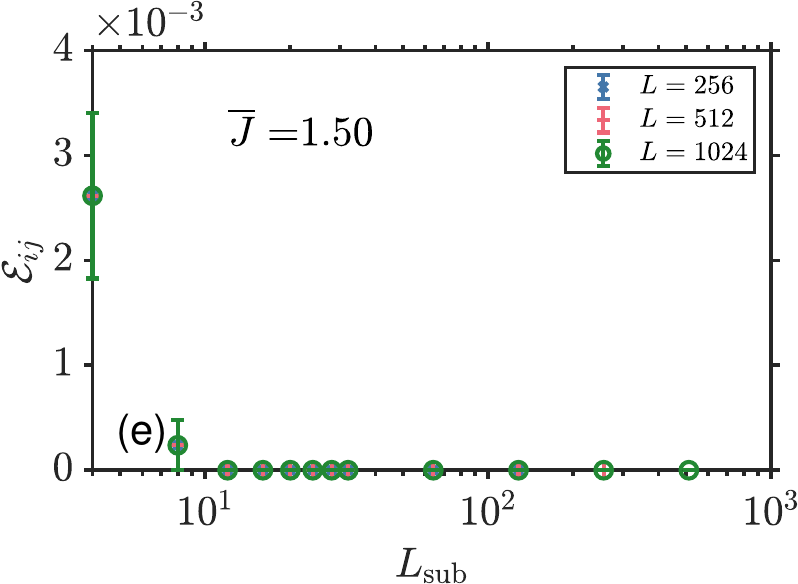}
    \includegraphics[width = 0.45\textwidth]{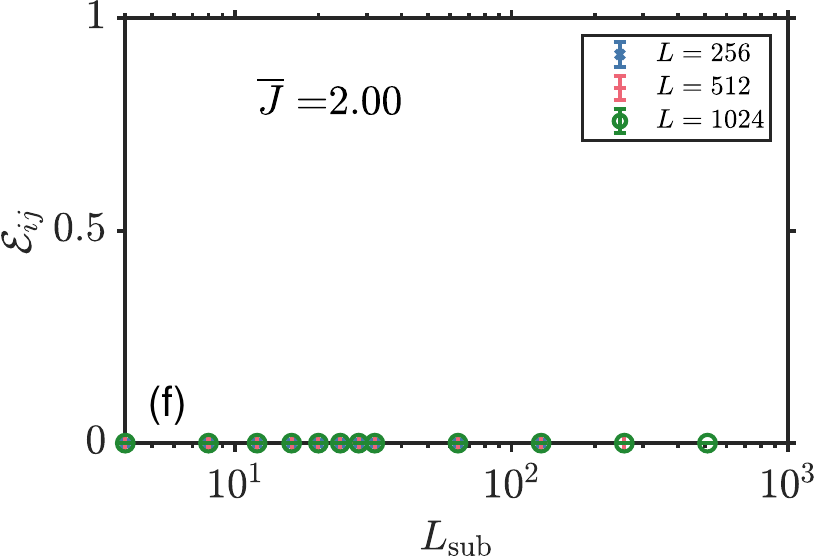}
    \caption{(a)-(f): The LSBS error rate ($\mathcal{E}_{ij}$) as a function of subsystem size ($L_{\rm sub}$) for 2D EA spin-glasses having a shifted Gaussian distribution of their couplings (mean $\overline{J} \neq 0$). 
    In panels (a)-(f), the respective values are 
$\overline{J}=1.0,1.05,1.10,1.20,1.50,$ and $2.0$. The setting is similar to that of Fig. \ref{fig:solver_performance}(a) of the main text. As $\overline{J}$ is elevated for the shown values, the error rate rapidly tends to zero with increasing subsystem length $L_{\rm sub}$.}
    \label{fig:shift}
\end{figure*}

The diametrically opposite extremes of Type I and Type IV hardness are known from numerous simple contexts. For instance, 

$\bullet$ Most conventional textbook type systems (e.g., the classical ferromagnet or antiferromagnet) correspond to type I systems. (In the classical ferromagnet and antiferromagnet, small finite patch  subsystem (respectively, uniform or  staggered magnetization) GSs capture the exact GSs of the full system.)  

$\bullet$ Type IV hardness may, e.g., appear for problems exhibiting the Overlap Gap Property \cite{OGP} as well as bona fide random systems. Indeed, if we were to guess the product of $N$ randomly sampled numbers $\pm 1$ that are chosen with equal probability, the estimates on products of numbers in smaller subsystems of any size $N_{\rm sub} < N$ would always have a $50\%$ error rate.  

The intermediate levels of hardness that we find suggest an underlying richer fine structure of the local complexity. Inasmuch as finite-size numerical results can be extrapolated to the thermodynamic limit, our calculations indicate that spin-glasses (and thus other general complex problems) straddle and are connected (when control parameters are further varied) to the above extreme limits of I and IV. In particular, 

$\bullet$ Our computations (see Fig. \ref{fig:solver_performance}) 
suggest that the local hardness of computing both  2D and 3D spin-glass GSs is 
that of level II. 

$\bullet$ The 2D spin glass having its couplings sampled from a `shifted' univariate Gaussian distribution for the coupling constants having a mean of $\overline{J}=1.2$ has level I hardness (see Sec. \ref{sec:shifted_gaussian} for the other shift values). Interestingly, in the latter case, we found a hardness transition point $\overline{J}\simeq 1.1$, which is marginally consistent with $1/r_c=1.04(1)$, as the ferromagnet-to-spin-glass transition value of $\overline{J}$ reported in the literature~\cite{melchert_scaling_2009}.   

$\bullet$ The local hardness associated with computing the GSs for the fully connected SK model \cite{SK} is (at least) that of level III.  
\section{Comparing the LSBS error rate between the 
2D and 
3D systems}
\label{sec:2Dvs3D}
\begin{figure}
    \centering
    \includegraphics[width=0.45\textwidth]{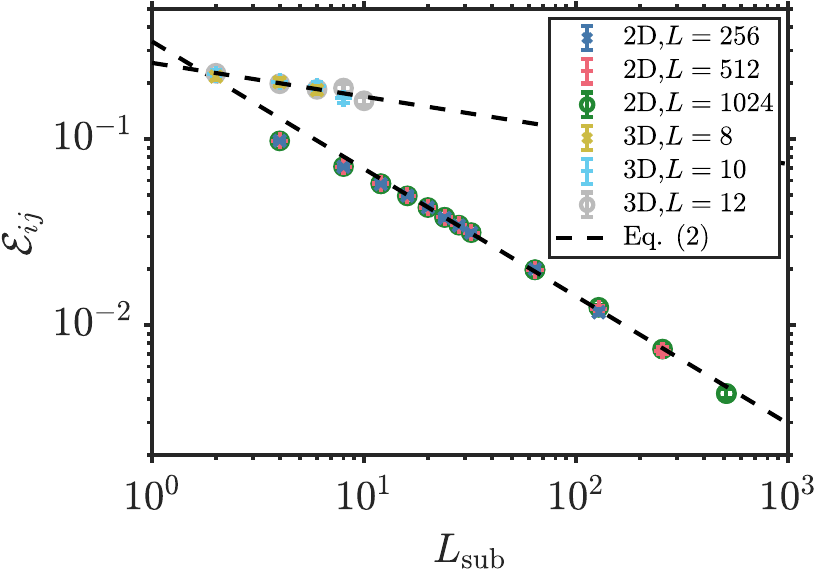}
    \caption{A comparison between the subsystem size ($L_{\rm sub}$)  dependence of the LSBS error rate ($\mathcal{E}_{ij}$) for the  EA models in $d=2,3$ dimensions. Here, we collate the results of Figs. \ref{fig:solver_performance}(a) and  (b) with the dashed line fit of Eq. (\ref{eq:solver}) (and associated parameters $\kappa$ and $\ell_{\mathcal{E}}$ of Table \ref{tab:fit_params}). The drop of the error rate with increasing  subsystem size is far more pronounced in the easier to solve (P) 2D systems than it is in the NP-hard 3D systems. This difference is  made yet more pronounced if the error rate is examined as a function of number of spins in the subsystem $N_{\rm sub} = L_{\rm sub}^{d}$ instead of the linear system scale  $L_{\rm sub}$.}
    \label{fig:errratemerged}
\end{figure}

\begin{figure}
    \centering
    \includegraphics[width=0.45\textwidth]{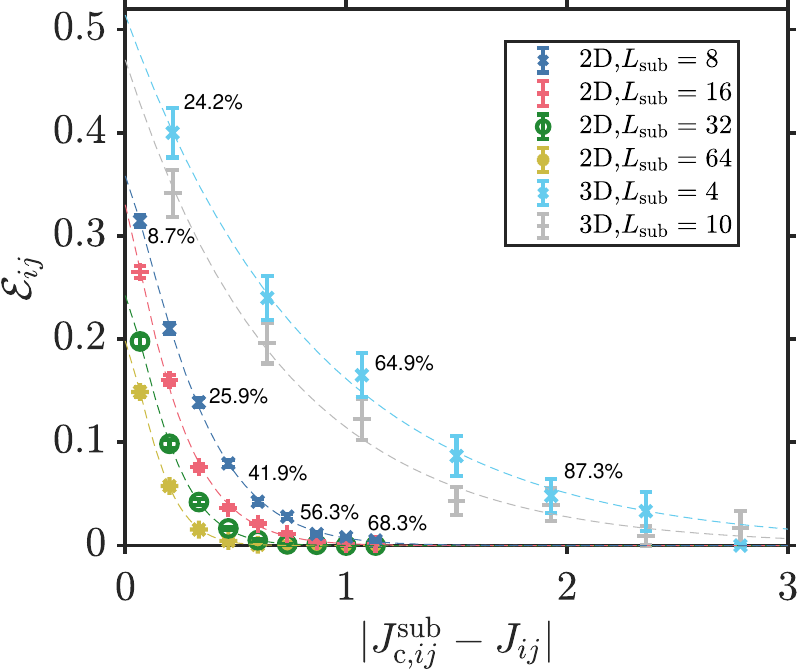}
    \caption{The dependence of the LSBS  error rate ($\mathcal{E}_{ij}$) on the deviation of the link strength from its critical threshold value $|\JCsub-J_{ij}|$. Shown is a hybrid of Figs. \ref{fig:error_jc_diff}(a)(b) allowing for a comparison between the 2D and 3D EA systems. As in Figs. \ref{fig:error_jc_diff}(a)  and (b), the point percentages denote cumulative probabilities, i.e., the fraction of instances with $|\JCsub - J_{ij}|$ less than or equal to the corresponding abscissa value. The dashed lines are given by Eq. (\ref{eq:erregy}) with the parameters of Table \ref{tab:fit_params}. As this figure makes clear, by comparison to the 2D systems, in the harder 3D systems, the error rate decays far more slowly with increasing deviations from the critical threshold values.  
    }
    \label{fig:jcprofilemerged}
\end{figure}

As we underscored in the main text, finding the ground states of the 2D EA is a problem of Polynomial complexity while determining the ground states of the 3D EA spin-glass systems is an NP-hard problem. In order to more readily quantify the precise differences between these different spin-glass type problems, in Figs.~\ref{fig:errratemerged} and ~\ref{fig:jcprofilemerged}, we present a composite contrasting, on the same set of axes, the decay of the LSBS error rate with subsystem size $L_{\rm sub}$ (Fig.~\ref{fig:errratemerged}) and deviation $|\JCsub-J_{ij}|$ of the coupling of the bond investigated by the LSBS from its critical threshold value (Fig. \ref{fig:jcprofilemerged}) in 2D and 3D EA systems. Even though the general form of the functional dependence of the LSBS error rate on the {\it is similar in both the 2D and 3D systems} (see Eqs. (\ref{eq:solver},\ref{eq:erregy}) of the main text), the precise values of the LSBS error rates are notably larger in the harder-to-solve 3D EA spin glasses.

\section{LSBS Examples for Shifted Gaussian Distributions}
\label{sec:shifted_gaussian}
With the exception of Fig.~\ref{fig:solver_performance} (a) in the main text, our analysis there largely focused on a symmetric Gaussian distribution of coupling constants (centered about $\overline{J} =0$). As $\overline{J}$ becomes larger, the system becomes less frustrated; in the trivial large $\overline{J}$ limit, the GS becomes ferromagnetic. Naturally, everything else being kept fixed, for large $\overline{J}$, the error $\mathcal{E}_{ij}$ of Eq.~(\ref{eq:solver}) must decay to zero more quickly. In the End Matter (Fig.~\ref{fig:jbarphase}), we illustrated the existence of a transition in the error rate as the mean of the Gaussian distribution of couplings was shifted by an amount $\overline{J}$. Here, we provide schematics for the results obtained for different $\overline{J}$. Fig. \ref{fig:shift} shows sample plots of the associated rapid drop of the error rate in subsystem size with increasing $\overline{J}$ from which such values of $\kappa$ are deduced. In the taxonomy that we introduced above when discussing the levels of local hardness, as the Gaussian distribution of the coupling constants $J_{ij}$ is progressively biased towards larger $\overline{J}$ values, the performance of the LSBS improves. These changes (see also Fig. \ref{fig:jbarphase}) are suggestive of a transition from level II to level I hardness at  $\overline{J} \approx 1$.  

\begin{figure}[htb]
    \centering
    \begin{subfigure}[b]{0.22\textwidth}
    \includegraphics[width = \textwidth]{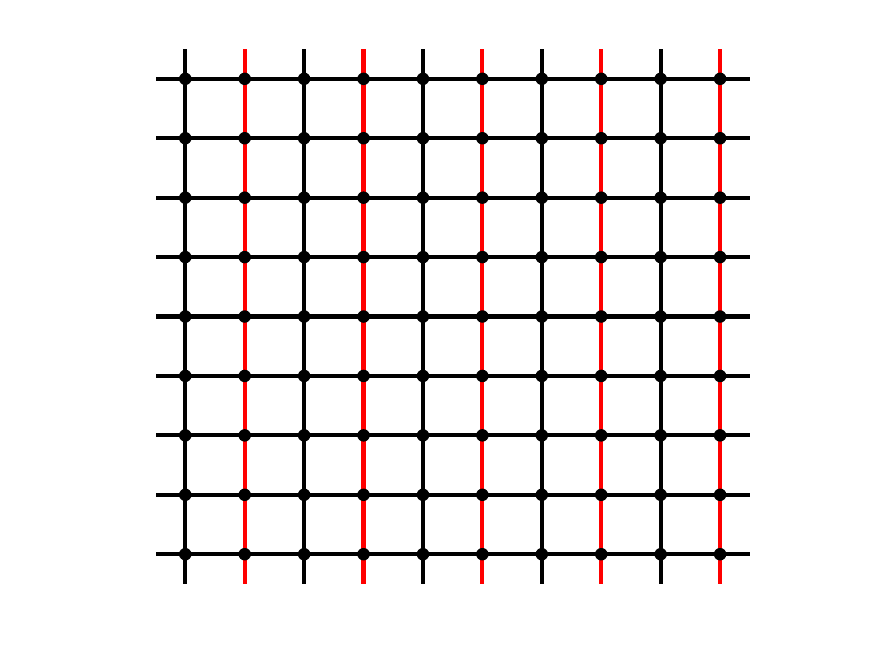}
    \caption{}
    \end{subfigure}
    \begin{subfigure}[b]{0.22\textwidth}
    \includegraphics[width = \textwidth]{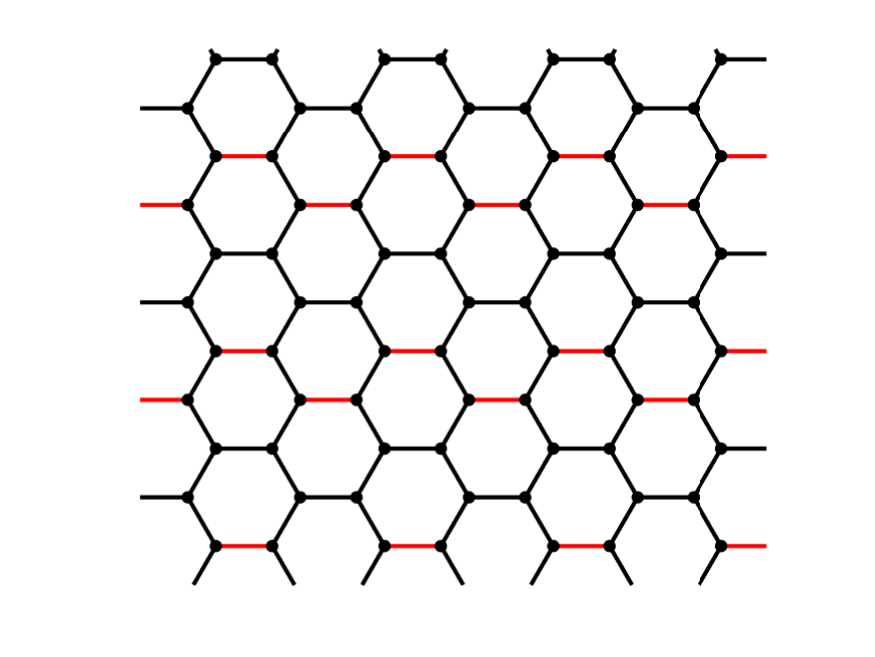}
    \caption{}
    \end{subfigure}
    \begin{subfigure}[b]{0.22\textwidth}
    \includegraphics[width = \textwidth]{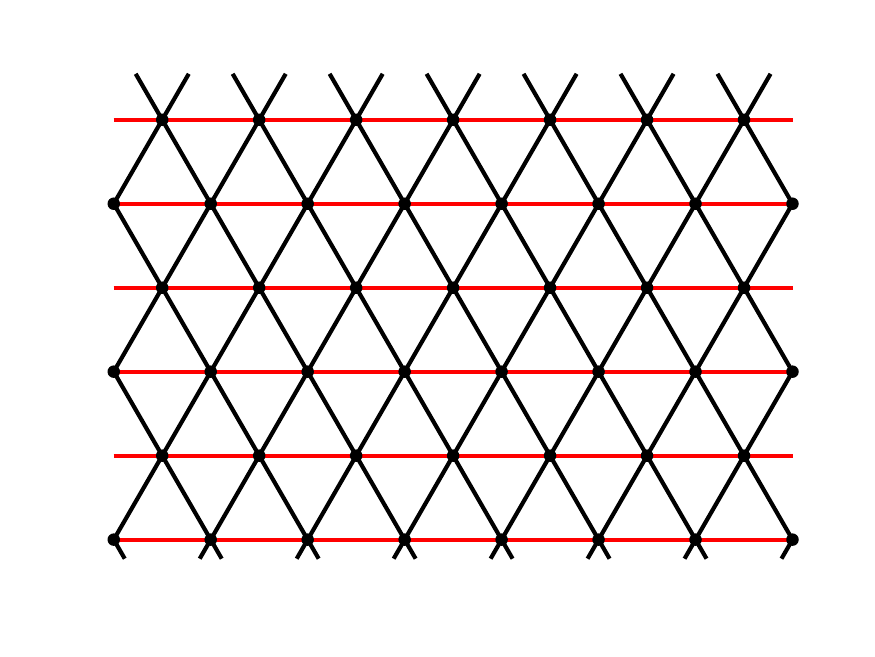}
    \caption{}
    \end{subfigure}
    \caption{Specific nearest-neighbor coupling realizations in our studied fully-frustrated square (a), honeycomb (b), and triangular (c) lattice spin systems. The red and black lines respectively denote equal strength bonds with ferromagnetic and antiferromagnetic couplings. The product of the coupling constants around any minimal (respectively, square, hexagonal, or triangular) plaquette is negative. As described in the text, these bonds were weakly perturbed to lift an otherwise exponential GS degeneracy which would render the problem determining the sign of the central bond in the GS meaningless.}
    \label{fig:ff}
\end{figure}

\begin{figure*}[htb]
    \centering
    \includegraphics[width = 0.45\textwidth]{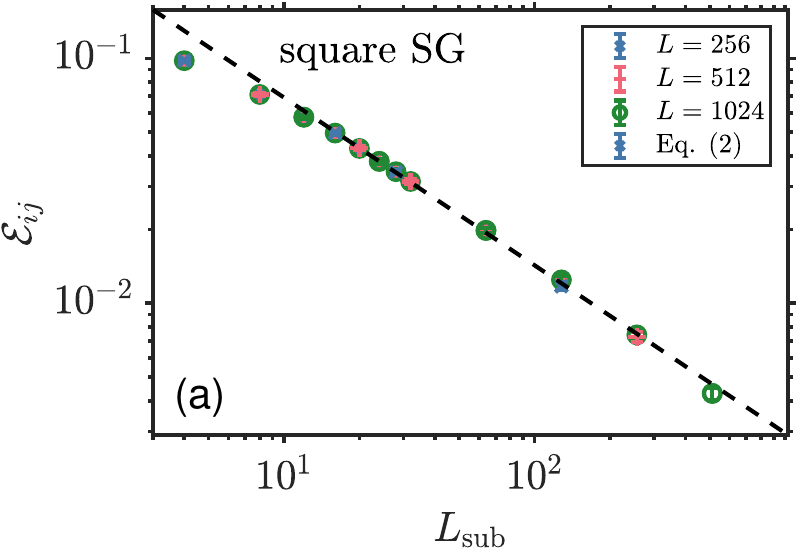}
    \includegraphics[width = 0.45\textwidth]{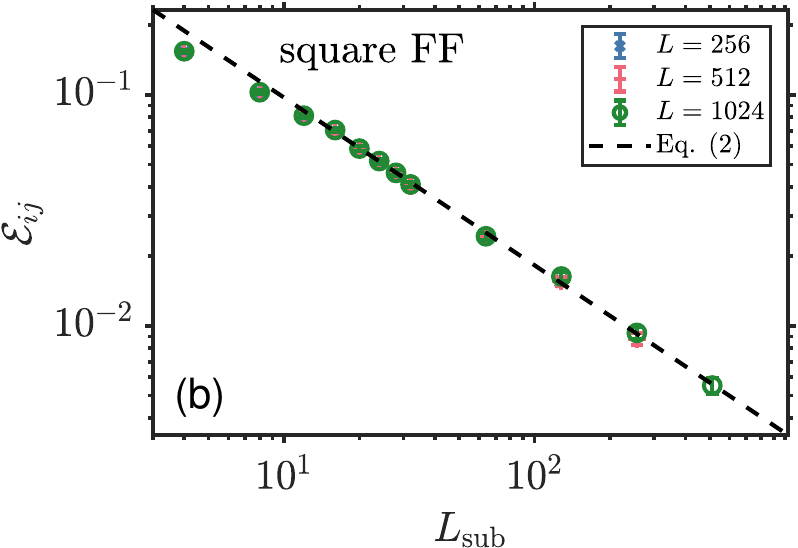}
    \includegraphics[width = 0.45\textwidth]{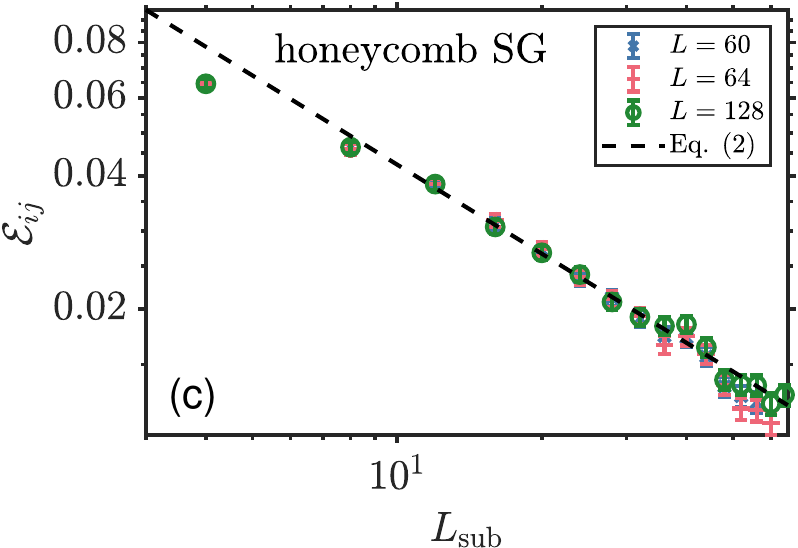}
    \includegraphics[width = 0.45\textwidth]{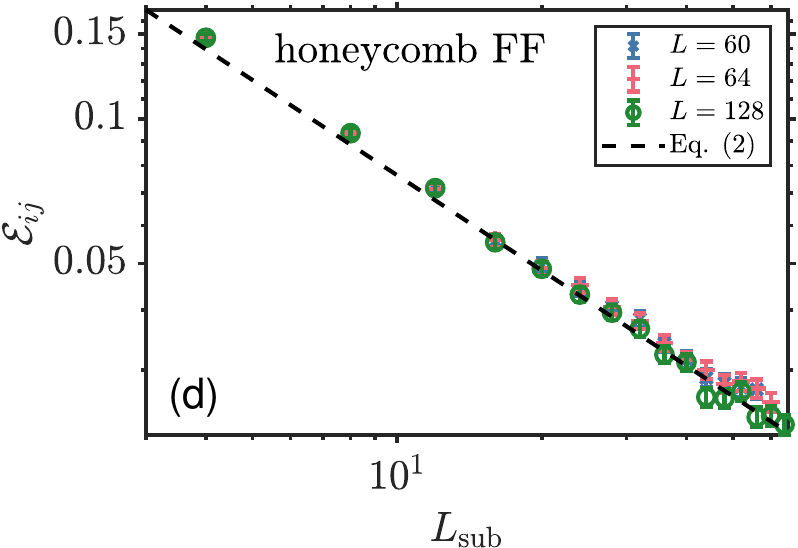}
    \includegraphics[width = 0.45\textwidth]{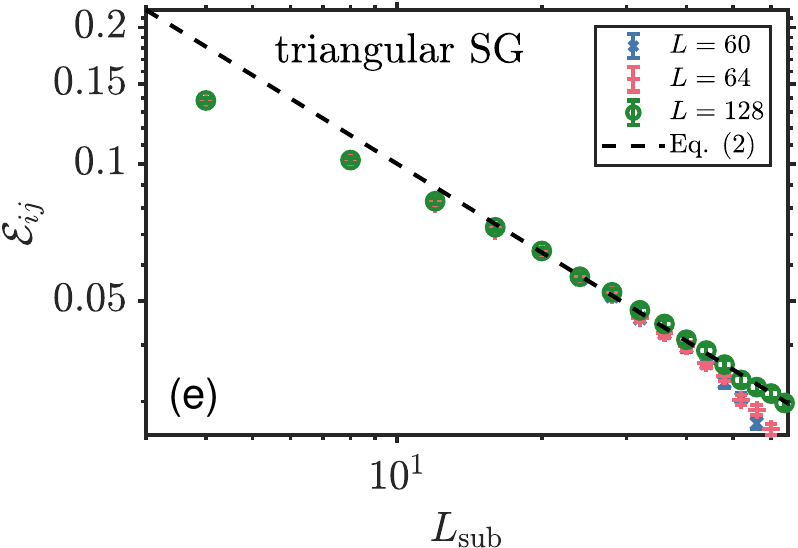}
    \includegraphics[width = 0.45\textwidth]{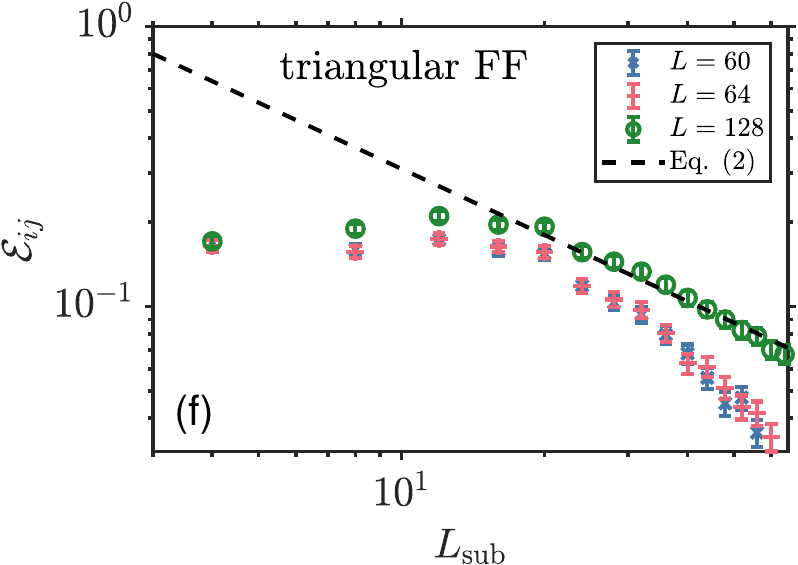}
    \caption{A side by side comparison of the performance of LSBS on 2D spin-glass systems (left) and their weakly perturbed ($\sigma_p=0.05$) fully-frustrated counterparts (right). Shown are the results for square lattice spin-glasses and the  perturbed fully frustrated square lattice systems (panels (a) and (b) respectively), honeycomb (c,d), and triangular lattice (e,f) systems. For this weak perturbation $\sigma_p$, except for the fully-frustrated triangular lattice systems, the error rate decays as a power of $L_{\rm sub}$. (Results for stronger perturbations of the fully-frustrated triangular lattice system are given in Fig. \ref{fig:ff_result_triangle}.) $\kappa_{\rm square}=0.685(8),\kappa_{\rm square,FF}=0.73(1),\kappa_{\rm honeycomb}=0.67(3),\kappa_{\rm honeycomb,FF}=0.66(2),\kappa_{\rm triangle}=0.648(9),\kappa_{\rm triangle,FF}=0.79(4)$.}
    \label{fig:ff_result}
\end{figure*}

\section{LSBS on 2D Fully-Frustrated Systems}
\label{sec:fully_frustrated}
A natural question concerns the extension of local solvability that we studied in the main text for spin-glass systems to several fully-frustrated systems. Specifically, towards that end, we apply the LSBS to fully-frustrated 2D spin systems on the (i) square, (ii) triangular,
and (ii) honeycomb lattices. 
In these fully-frustrated systems, within every minimal (respectively, single square, triangular and hexagonal) plaquette, at least one bond cannot be satisfied. In Fig. \ref{fig:ff}, we  provide an illustration of these systems. The specific setup of these systems is as follows: as seen in the figure, some bonds in the system are set to be ferromagnetic, while others are antiferromagnetic with the product of all bonds around a plaquette being negative. In the absence of disorder, such systems may have an exponentially large GS degeneracy 
 (in the full system size $N$) in which local bonds are not uniquely determined (as in, e.g., Wannier classical result \cite{Wannier} for the triangular lattice Ising antiferromagnet and Villain's demonstration of the exponential degeneracy of the fully frustrated Ising square lattice model \cite{Villain} invoking Fisher's asymptotic result for dimer coverings \cite{Dimer_fisher}). This renders a comparison between the LSBS to the global GSs ill-posed since both single bond ($\sigma_i \sigma_j$) values are possible in a global GS. To overcome this vexing issue and lift the exponential degeneracy, we added weak Gaussian perturbations of standard deviation of $\sigma_p=0.05$ to the otherwise uniform unit (absolute value) strength nearest neighbor couplings.
 
\begin{figure*}[htb]
    \centering
    \includegraphics[width = 0.45\textwidth]{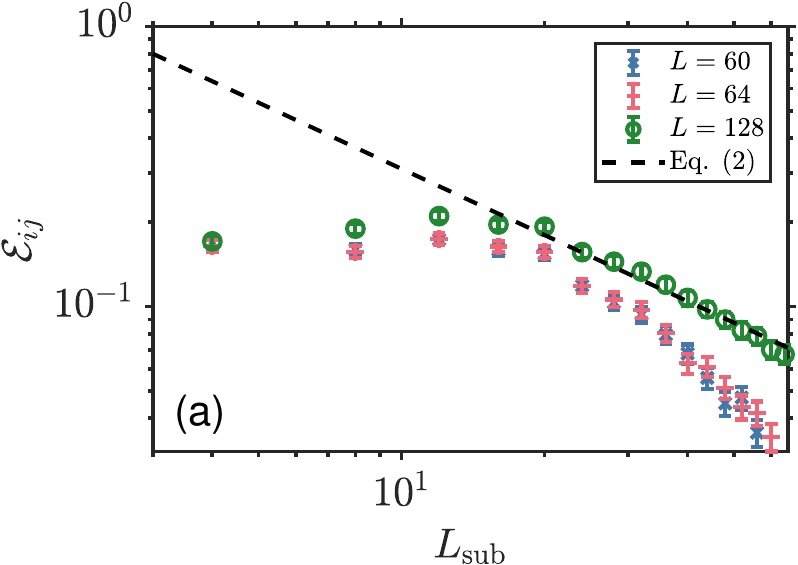}
    \includegraphics[width = 0.45\textwidth]{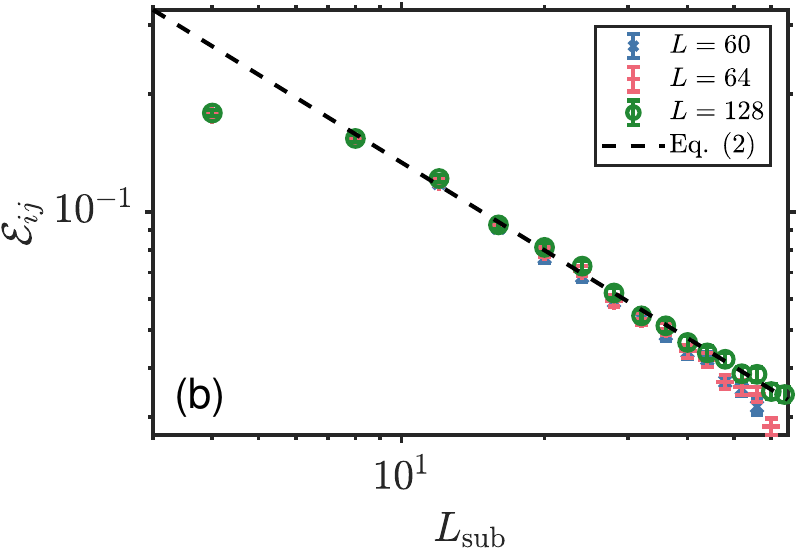}
    \includegraphics[width = 0.45\textwidth]{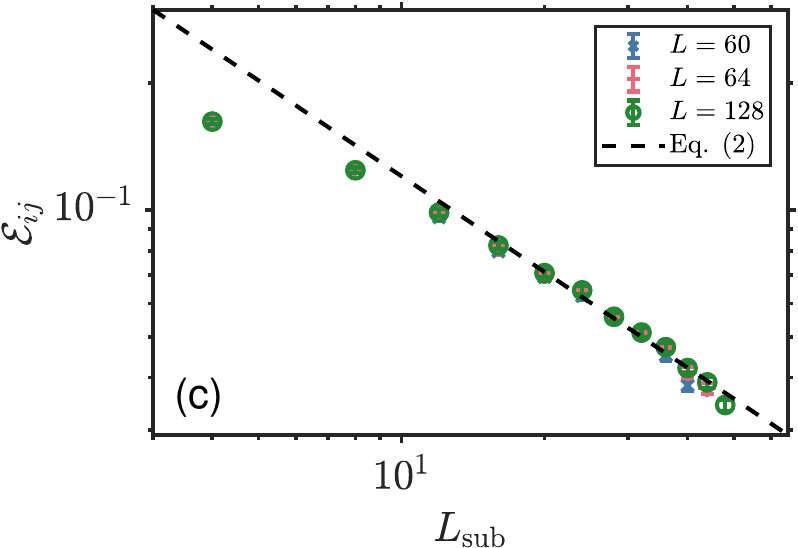}
    \caption{Fully frustrated triangular lattice systems with increasing perturbation strength $\sigma_p$ of their coupling constants. In panels (a,b,c), we  respectively show the results for perturbations of strength $\sigma_p= 0.05, 0.20, 0.40$ (in units of the uniform absolute value of the unperturbed coupling constants of the fully frustrated system). The dashed lines correspond to the fit of Eq. (\ref{eq:solver}) for triangular spin-glass systems (see bottom two panels of  Fig. \ref{fig:ff_result}). The found exponents $\kappa$ for these fittings are 0.79(4), 0.75(2), 0.76(3), with $\chi^2/{\rm d.o.f}=1.282,0.590,2.003$. }
    \label{fig:ff_result_triangle}
\end{figure*}

In Fig.~\ref{fig:ff_result}, we present the results of  the LSBS on these perturbed fully-frustrated systems. The fully-frustrated square and honeycomb lattice spin systems exhibit characteristics similar to those of the spin glass, where the error rate approximately follows a power law with respect to $L_{\rm sub}$. The right and left panels of Fig.~\ref{fig:ff_result} provide, respectively, a side by side comparison of the perturbed fully frustrated spin systems with their spin-glass counterparts. The found exponents $\kappa$ in the power law form for the error rate of Eq.~(\ref{eq:solver}) are $\kappa_{\rm square}=0.685(8),\kappa_{\rm square,FF}=0.73(1),\kappa_{\rm honeycomb}=0.67(3),\kappa_{\rm honeycomb,FF}=0.66(2)$ are nearly matching for the honeycomb and square lattices. 
For the lowest perturbation strength shown ($\sigma_p=0.05$), the fully-frustrated triangular lattice spin system does not exhibit a power law decay of the average LSBS error in $L_{\rm sub}$ while its spin-glass counterpart obeys Eq. (\ref{eq:solver}) with $\kappa_{\rm triangle}=0.55(1)$. In Fig.~\ref{fig:ff_result_triangle}, we show the different performance of LSBS for the perturbed ($\sigma_p=0.05, 0.20, 0.40$) fully frustrated triangular lattice spin system. As $\sigma_p$ increases, the difficulty encountered by the LSBS decreases accordingly. We speculate that this may be because the original (unperturbed uniform)  system's `degeneracy' is further split as $\sigma_p$ grows in size. 
For the strongest perturbation $\sigma_p=0.40$ investigated, we found $\kappa_{\rm triangle,FF}=0.79(4)$ while for 
the triangular lattice spin-glass $\kappa_{\rm triangle}=0.648(9)$ (Fig. \ref{fig:ff_result}). Note that all fittings were performed for $L_{\rm sub}\geq 16$ to minimize the finite size effect. We note that for different lattices, a fixed value of $L_{\rm sub}$ corresponds to a different number of spins $N_{\rm sub}$ the subsystem solver examines.  The $\chi^2/\rm{d.o.f}$ values for the fittings in Fig. \ref{fig:ff_result} (a)-(f) are, respectively, 0.821, 0.770, 0.960, 0.836, 0.558, and  1.282.


\section{Scaling of the subsystem critical threshold}  
\label{sec:scaling_subsytem_jc}
\subsection{Boundary Conditions}

In the main text, we considered how 
changing the value of one 
bond 
induces a change for the critical thresholds $\Delta \JC$ 
of other bonds a distance $r$ away. Here, we 
examine another question: 
If we change the boundary condition, instead of solely tuning one bond to its critical value, what will the change of the critical threshold of the central bond look like?

\begin{figure}[htb]
    \includegraphics[width = 0.45\textwidth]{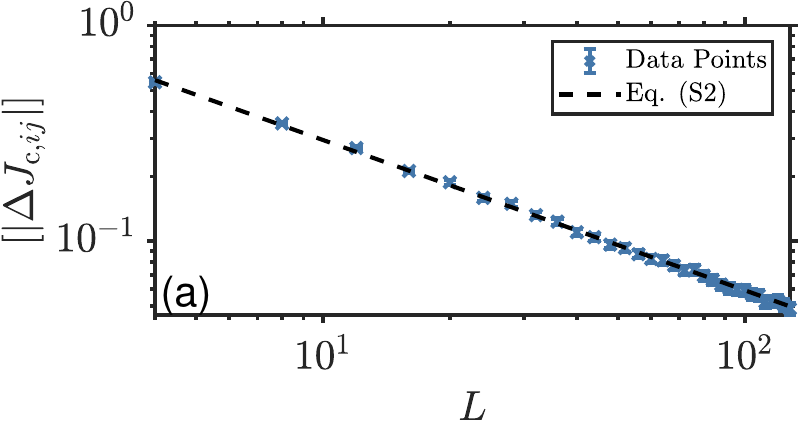}
    \includegraphics[width = 0.45\textwidth]{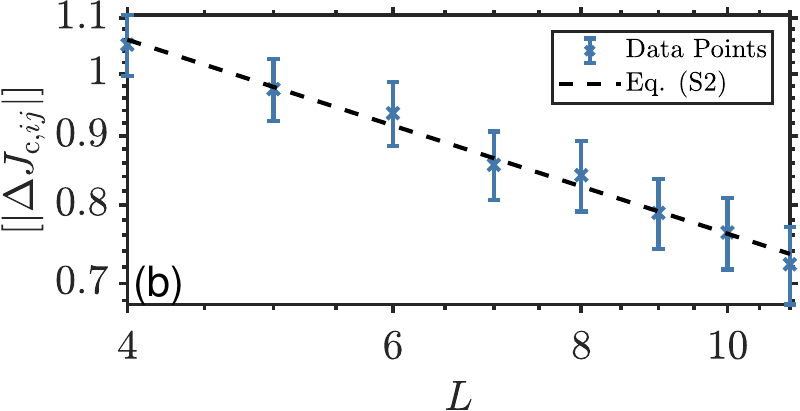}
    \caption{After randomly resampling all spins 
    on the boundaries, 
    we calculated the average change of the critical threshold $[ |\Delta J_c|]$ of the central bond as a function of system size $L$. This change is well-fitted by the power-law of Eq. (\ref{eq:deltajcboundpower}), where for (a) 2D systems- $\ell_{L,2D}=1.73(2)$ and $a_{L,2D}=0.697(4)$, and in (b) 3D systems- $\ell_{L,3D}=4.7(1)$ and $a_{L,3D}=0.36(2)$.}
    \label{fig:boundary}
\end{figure}

Towards this end, 
we studied $4 \le L\le 128$ square and $4 \le L  \le 11$ cubic lattice systems. We first calculated the critical threshold $\JC$ of the original system. Subsequently, we let all the spins $\sigma_i$ on the boundaries to assume random values of $\pm 1$ and then calculate the $\JC$ of the central bond in this new system. 
The disorder averaged magnitude of change $[ |\Delta J_c|]$ of the central bond is displayed in Fig. \ref{fig:boundary}. Somewhat similar to the change of critical couplings when only a single bond is  changed that was discussed in the main text (Eq. (\ref{eq:deltajc})) and End Matter (Fig. \ref{fig:jclocality})), we found a power-law decay 
\begin{eqnarray}\label{eq:deltajcboundpower}
   [ | \Delta \JC| ] = (L/\ell_L)^{-a_L},
\end{eqnarray} 
with $\ell_{L,2D}=1.73(2), ~a_{L,2D}=0.697(4)$ ($\chidof=0.544$) and $\ell_{L,3D}=4.7(1), ~ a_{L,3D}=0.36(2)$ ($\chidof=0.080$) for the square and cubic systems respectively (see Fig. \ref{fig:boundary}.) The 2D exponent is close to that found ($0.7 \pm 0.02)$ for the dependence of the error rate of a patch solver on its size \cite{hartmann_metastate_2023}).

\subsection{Subsystem Size}

\begin{figure}
    \centering
    \includegraphics[width=0.45\textwidth]{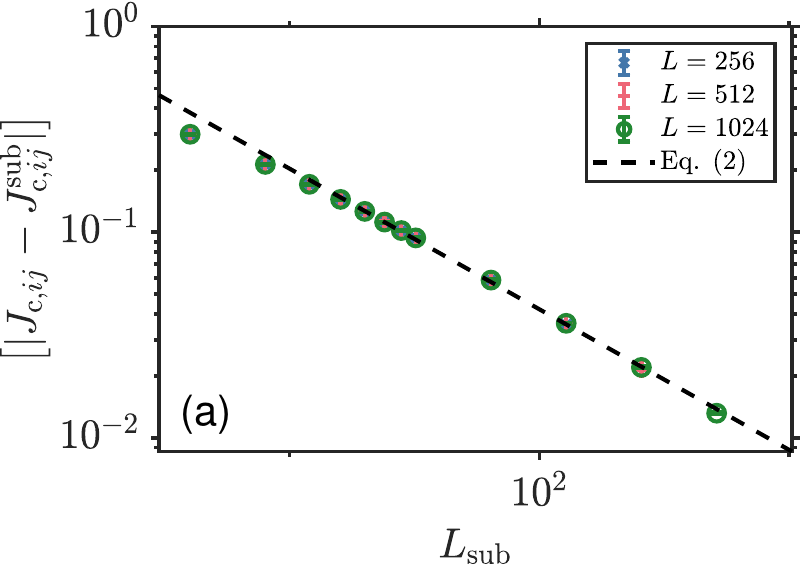}
    \includegraphics[width=0.45\textwidth]{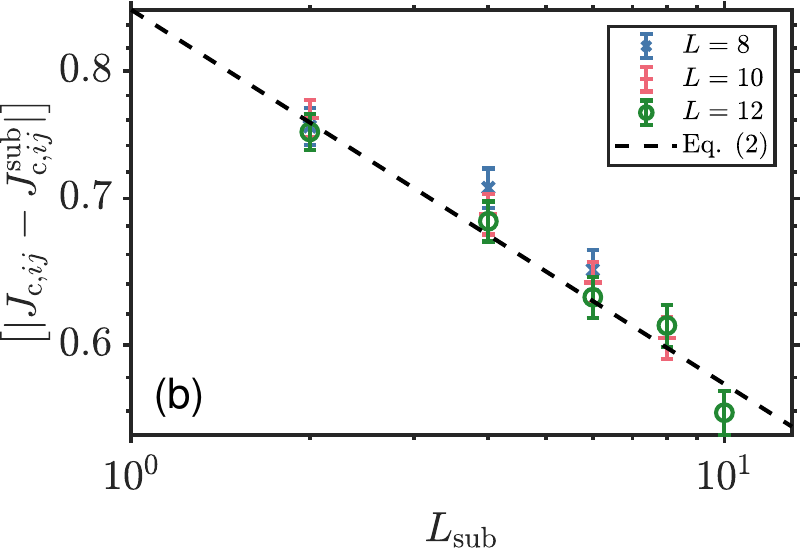}
    \includegraphics[width=0.45\textwidth]{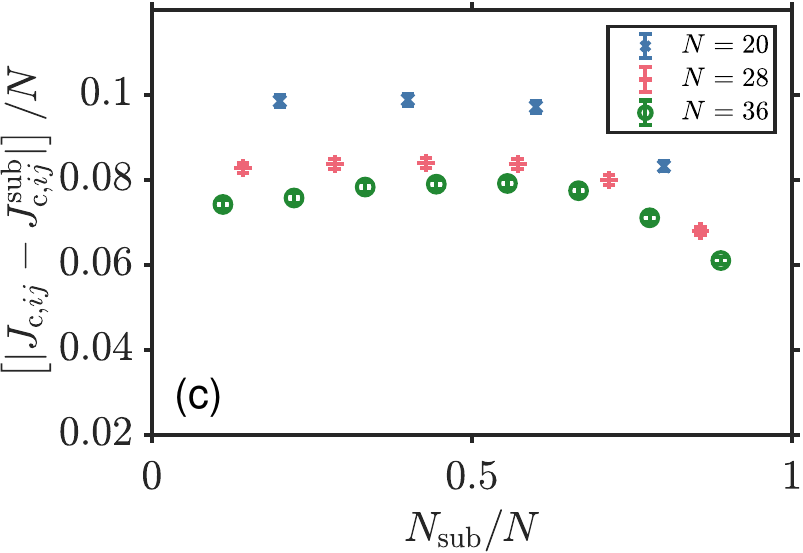}
    \caption{For (a) the square and (b) cubic lattice EA model, the relation between $\left[ |J_{\rm c}-J_{\rm c}^{\rm sub}|\right]$ and ${ L_{\rm sub}}$ also follows a power law and has a very similar fitting coefficient compared to $\mathcal{E}_{ij}$, Eq. (\ref{Jcvars}).
    In Eq. (\ref{Jcvars}), for 2D systems $\ell_J=0.98(4)$ and $\kappa_J=0.684(7)$; for 3D, $\ell_J=0.4(1)$ and $\kappa_J=0.17(2)$. The $\chi^2/{\rm d.o.f}$ values for (a) and (b) are 11.698 and 1.823 respectively. (c) The SK model. For the 2D case, although we have adopted a cutoff $L_{\rm sub} \geq 16$, we still obtained a relatively large $\chi^2/{\rm d.o.f}$ value. When increasing the $L_{\rm sub}$ cutoff to 32 and 64, the corresponding $\chi^2/{\rm d.o.f}$ value decreases to 4.276 and 2.650, with $\ell_J=1.12(4),1.20(5)$ and $\kappa_J=0.703(6),0.713(6)$.}
    \label{fig:Jscub-Jx_vs_L_sub}
\end{figure}

In Fig. \ref{fig:Jscub-Jx_vs_L_sub}, we show the deviation of the critical threshold coupling of the central bond in the subsystem relative to the true (global) critical threshold value of this bond in the full system as a function of the subsystem size. Here, too, we find an algebraic decay,
\begin{equation}
    \label{Jcvars}
    \left[ |J_{\rm c}-J_{\rm c}^{\rm sub}|\right]\sim (\ell_J/L_{\rm sub})^{\kappa_J}.
    \end{equation}
    The rapid drop of this deviation with subsystem size further rationalizes our observed decreasing error rate with increasing subsystem size of Eq. \ref{eq:solver} (see also cartoon of Fig. \ref{fig:jc_diagram} explaining how such a small deviation favors accurate LSBS outcomes).  
    Since the LSBS errors are incurred by a shift in the value of the subsystem $J_{\rm c}^{\rm sub}$ relative to its true value of $\JC$, it is natural to expect that the error rate $\cal{E} \propto |J_{\rm c} - J_{\rm c} ^{\rm sub}|$ for small $|J_{\rm c} - J_{\rm c} ^{\rm sub}|$. Indeed, within our numerical error, the exponents $\kappa_J$ found via the fit of Eq. (\ref{Jcvars}) conform with those for the error rates
    (Eq. (\ref{eq:solver})) on different lattices,
    \begin{equation}
    \label{kappa=kappaJ}
    \kappa_J = \kappa.
    \end{equation}
    Specifically, for the 2D (square lattice) case,  $\kappa_{J} = 0.684(7)$ 
    (Fig. \ref{fig:jc_diagram}) while $\kappa = 0.685(8)$ (Table \ref{tab:fit_params}). 
    Similarly, in 3D (cubic lattice) systems, $\kappa_J = 0.17(2)$ while $\kappa = 0.18(3)$. 

\section{Some Details of the Numerical Experiments}
\label{sec:numerical_details}
When we refer to simulations performed on a system with linear size $L$, it should be clarified that the system is not strictly square in shape. For instance, a two-dimensional system with $L=1024$ actually corresponds to a lattice of $1025 \times 1026$ spins, just as in Ref. \cite{shen2023universal}. This tiny adjustment is made in order to ensure that the specific central bond is placed precisely at the center of the system.

Next, we discuss the strategy for subsystem sampling. A standard procedure, which we refer to as \emph{parallel sampling}, proceeds as follows: one first randomly samples an entire system of size $L=1024$, and then selects a subsystem of size $L_{\rm sub}=512$ for analysis. This process yields one data point. To obtain a data point at a smaller subsystem size, e.g., $L_{\rm sub}=256$, one would ideally resample the entire system and then extract the smaller subsystem. However, this approach can be computationally expensive, particularly when data across multiple $L_{\rm sub}$ values are needed.

\begin{figure}[htb!]
    \centering
    \includegraphics[width=0.48 \textwidth]{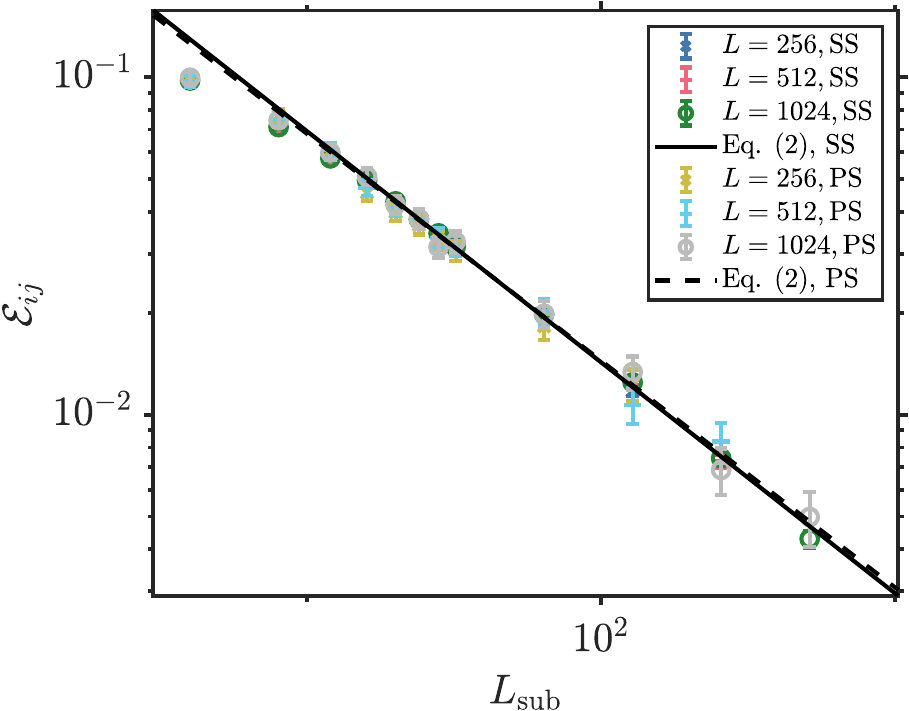}
    \caption{Comparison of the fit to Eq.~(2) for subsystem data extracted using the sequential and parallel sampling strategy respectively. Both curves correspond to systems with total size $L=1024$ and subsystem size $L_{\rm sub}=512$. SS and PS are short for Sequential Sampling and Parallel Sampling respectively.}
    \label{fig:sampling_comparison}
\end{figure}

To mitigate computational costs while still acquiring a sufficient number of data points for statistically reliable results (e.g., LSBS error rate), we adopt an alternative procedure, referred to as \emph{sequential sampling}. In this approach, a single $L=1024$ system is sampled, and multiple nested subsystems, such as $L_{\rm sub}=512$, $256$, and $128$, are subsequently extracted and analyzed from the same entire system. Although the measurements obtained from these subsystems are theoretically correlated, this correlation should have minimal impact on our final results. For example, as shown in Fig. \ref{fig:sampling_comparison}, we compare the fits to Eq. \eqref{eq:solver} obtained for $L=1024$ and $L_{\rm sub}=512$, and the results are in close agreement—both visually and in terms of the fitted parameters: $\ell_{\mathcal{E},\rm{parallel}}=0.18(3)$, $\kappa_{\rm{parallel}}=0.67(2)$, $\ell_{\mathcal{E},\rm{sequential}}=0.20(1)$, $\kappa_{\rm{sequential}}=0.685(8)$.

\section{Exactly solvable limits}
\label{sec:solvable}
In what follows, we discuss two limits in which an exact analysis is possible:
(1) 1D systems and (2) large $n$ (or spherical model) realizations. These simply exactly solvable models will yield results notably different from those that we found for the 2D and 3D Ising EA spin-glasses.

\subsection{1D Ising spin-glass} One-dimensional spin-glass systems are rather special in many regards (including their exceptionally trivial exact solvability). Here, the critical thresholds exhibit distinct behaviors for periodic boundary conditions (PBC) and free (or open) boundary conditions (FBC). For the free boundary condition, all critical thresholds $\JC=0$, ensuring the locality of $\JC$. 
For the periodic boundary condition, we define $\JC\equiv \zeta|\JC|$, with $\zeta=\pm 1$ indicating the sign of $\JC$. The value of $\zeta$ is chosen such that the system is \emph{frustrated}, i.e., contains an odd number of antiferromagnetic $J_{ij}<0$ bonds. 
In the frustrated system, the GS is determined by having the smallest absolute bond being unsatisfied. Consequently, $|\JC|=\min\limits_{(i,j)\neq(i_0,j_0)}|J_{ij}|$. Obviously, both the sign and the magnitude of $\JC$ are determined non-locally. In summary, we found that the locality holds for the FBC case, but not for the PBC case.



\subsection{Spherical Model Spin-Glasses}
Another analytically solvable limit is that of the spherical \cite{Berlin+Kac} (or large $n$) soft spin-glass model \cite{kosterlitz_spherical_1977} counterpart of the Ising spin-glass Hamiltonian of Eq. (\ref{HSG}). This model is defined by the following Hamiltonian and global constraint, 
\begin{equation}
\label{s:spherical}
    H = -\sum_{\langle i j \rangle}J_{ij}s_is_j,~~\sum_i s_i^2=N.
\end{equation}
Unlike the $N$ Ising spins $\{\sigma_{i}\}_{i=1}^{N}$ 
of Eq. (\ref{HSG}), the $N$ spins $\{s_{i}\}_{i=1}^
{N}$ are now {\it arbitrary real numbers} subjected only to the single {\it global normalization} constraint in Eq. (\ref{s:spherical}) concerning the sum of their squares. That is, only the mean (over the entire $N$ spin system) value of $s_i^{2}$ is constrained to be unity unlike the Ising system where $N$ constraints of the type $\sigma_i^2=1$ appear for each of the $N$ spins $\sigma_i$. We find that this global constraint removes the otherwise inherent locality of the critical thresholds. The Hamiltonian of Eq.  (\ref{s:spherical}) is a bilinear in the real variables $s_{i}$ and thus the GS can be found by diagonalizing the real symmetric $N \times N$ matrix whose elements are the couplings $J_{ij}$. Following the below theorem, we may readily establish that 
for any bond $\langle i j \rangle$ in the system 
(keeping all other bonds fixed),
the product $s_i s_j$ 
may vanish in a single interval of $J_{ij}$ values (including possibly only a single coupling). 

\emph{Theorem.} Given a real symmetric matrix $A_{m\times m}$, consider the perturbation $A\to A' \equiv A+\epsilon P$, with $P_{12}=P_{21}=+1$,  $P_{ij}=0$ otherwise, and $\epsilon>0$. 
The normalized principle eigenvectors $v\equiv(v_1,v_2,\cdots,v_m)^T$ of $A$ and $v'$ of $A'$ (with respective eigenvalues $\lambda$ and $\lambda'$) satisfy $v_1v_2\leq v'_1v'_2$. 

\emph{Proof}. Note that $v_1v_2=v^TPv/2$, therefore we only need to prove $v^TPv\leq v'^TPv'$. By the definition of the vector $v$ as the (normalized) principle eigenvector of $A$,  
\begin{eqnarray}
     \label{eq:pri1} v^TAv&\geq v'^TAv' 
     \\
    \label{eq:pri2} v^T(A+\epsilon P)v&\leq v'^T(A+\epsilon P)v'
\end{eqnarray}
Plugging Eq. (\ref{eq:pri1}) into Eq. (\ref{eq:pri2}),  
$v^TPv\leq v'^TPv'$. 
\qed

This theorem formally establishes (the intuitively expected behavior) that the two-spin product $s_is_j$ is monotonically non-decreasing in the coupling $J_{ij}$ between them. (In the respective asymptotic $J_{ij}\to \pm \infty$ limits, 
the monotonic behavior saturates and the product  $s_is_j=\pm \frac{N}{2}$.) The demonstrated  monotonicity ensures that 
we can define a {\it single interval} $[\JCL,\JCH]$  (that may possibly be a point of zero measure) for which the nearest-neighbor product $s_i s_j$ vanishes in the GS of Eq. (\ref{s:spherical}). Here, 
$\JCL\equiv\sup\{J_{ij}|s_is_j<0\}$ and   $\JCH\equiv\inf\{J_{ij}|s_is_j>0\}$. 
In the following, 
we will 
use the shorthand $\JC$ to denote the 
interval $[\JCL,\JCH]$.
Unlike the Ising variant studied in the main paper text that had a single point value of the critical coupling, a general interval $[\JCL,\JCH]$ will not be of vanishing measure. 

\subsubsection{Symbiosis and Competition}
\begin{figure}[htb]
    \centering
    \includegraphics[width=0.49\textwidth]{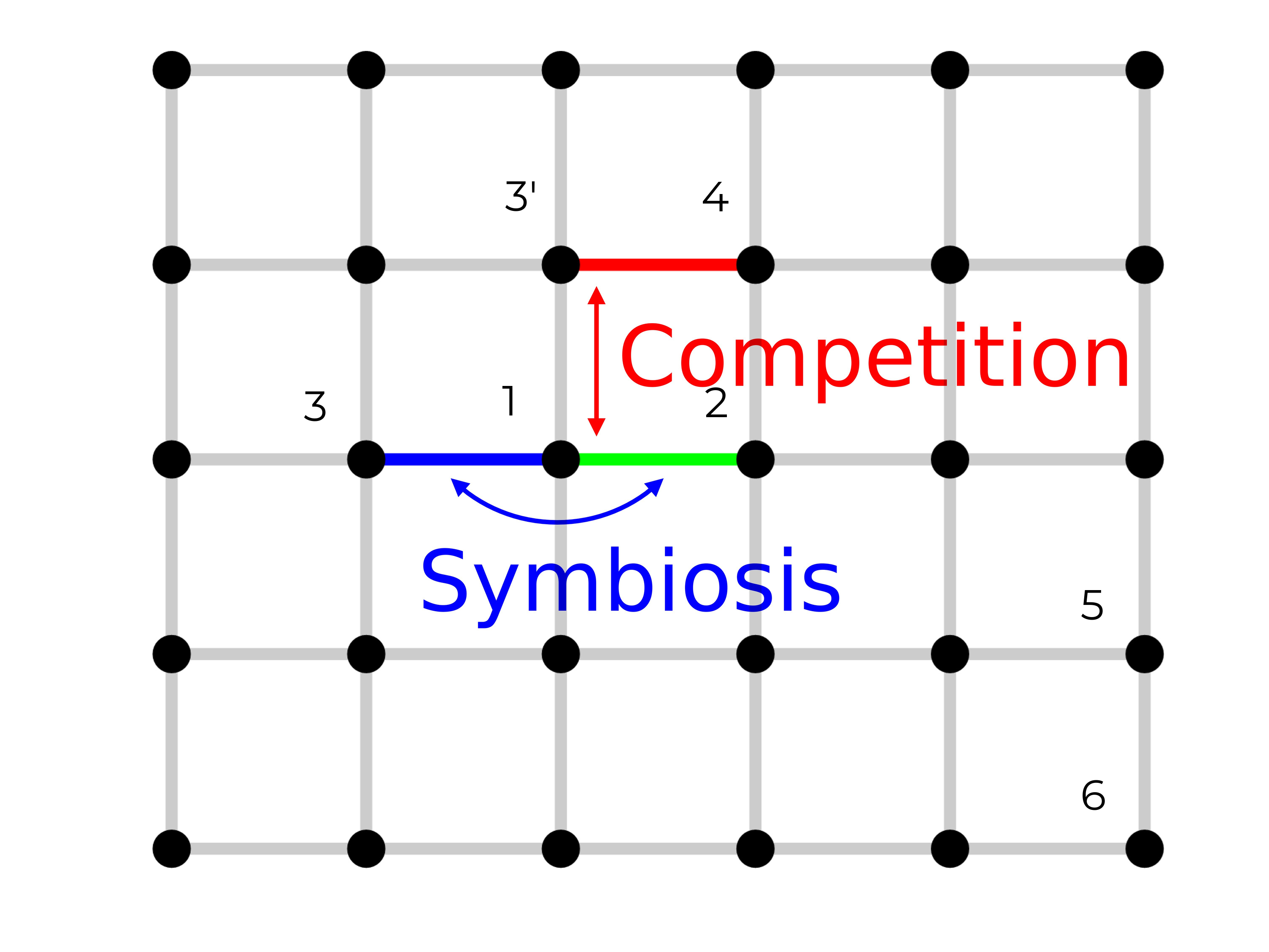}
    \caption{A sketch of symbiotic (green-blue) and the competitive (green-red)  bonds in the spherical model spin-glass, see text. We underscore that {\it irrespective of how far away they are}, all bonds (e.g., bond $(5,6)$) that do not share a common spin with the central (green) bond exhibit 
    an identical competitive relation with it.} 
    \label{fig:s_and_c}
\end{figure}

\begin{figure}
    \centering
    \includegraphics[width=0.45\textwidth]{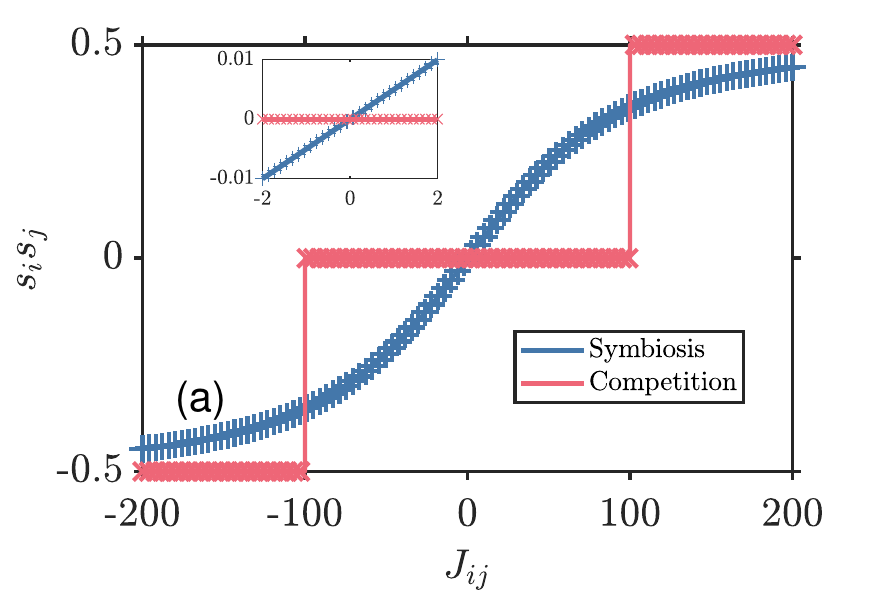}
    \includegraphics[width=0.45\textwidth]{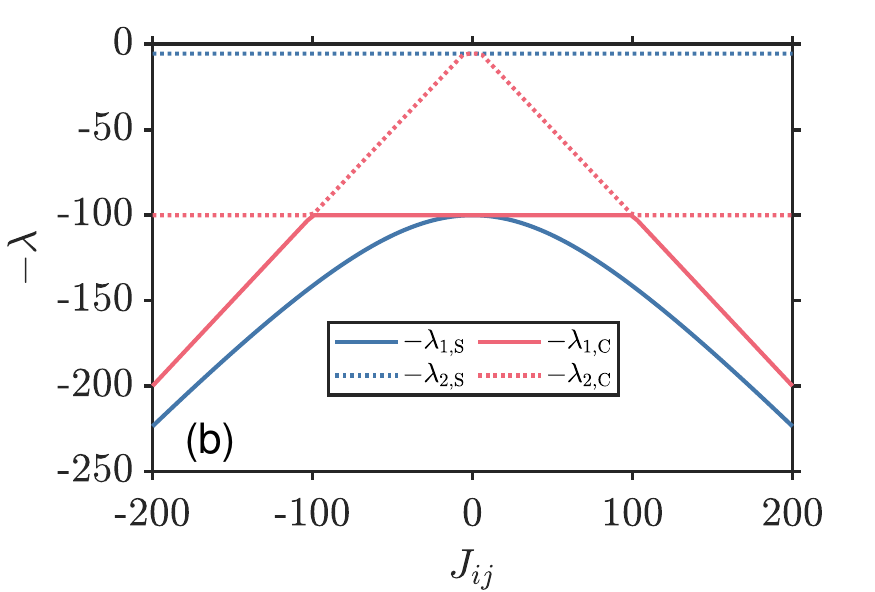}
    \caption{(a) Computations for the spherical spin-glass model on an $L=10$ cubic lattice. We calculate  nearest-neighbor spin products, namely $s_1s_3$ of the blue bond, and $s_{3'}s_4$ of the red bond, as a function of their respective couplings $J_{ij}$, while fixing the coupling of the green bond to be $J_0=+100.0$. Blue and red scattering crosses represent the actual results from numerical simulation. Blue and red continuous curves show the predictions from the symbiosis and competition models given in Eq. (\ref{eq:s_and_cmodel}). The inset shows more data points within a smaller range of $J_{ij}$ around zero. (b) On the same system, numerically we show the opposites of the first (solid) and second (dashed) largest eigenvalues $-\lambda$ of the system in symbiosis (blue, $-\lambda_{1(2),S}$) and competition (red, $-\lambda_{1(2),C}$) model respectively. Degeneracy is absent in the symbiosis model. In the competition model, a pair of degenerate points emerge when $|J_{ij}|$ of the red bond (Fig. \ref{fig:s_and_c}) is similar to $|J_{ij}|$ of the blue bond. Just like the top side, the theoretical curve and the numerical curve (points) almost completely overlap, so to prevent clutter here, we only plotted the numerical curve.}
    \label{fig:s_and_clambda}
\end{figure}

Unlike the discrete Ising model of Eq. (\ref{HSG}), within the spherical model, the influence of one bond on another bond's critical threshold $\JC$ is rather trivial. 
As we illustrate 
in Fig. \ref{fig:s_and_c}, even if two bonds share the same Euclidean distance to a modified bond, other facts are at play. 
To highlight this difference, the following extremal analysis would be very useful. In a 3D, $L=10$ system, setting the blue bond to be $J_0=+100.0$ as a `strong' bond and we tune the value of the green bond and the red bond respectively, from $-2J_0$ to $2J_0$, we see quite different behaviors of $s_is_j$, see the top side of Fig. \ref{fig:s_and_clambda}. Now, we present two models to explain this behavior.

We neglect most of the sites and interactions and then we write down the model of ``symbiosis'' (green-blue) and ``competition'' (green-red): $H\simeq 
 H_{\rm{S}}=-J_{13}s_1s_3-J_{12}s_1s_3$ and $H\simeq H_{\rm{C}}= -J_{3'4}s_{3'}s_4-J_{12}s_1s_2$. Here we label the blue, green, and red bonds as $\{1,2\},\{1,3\},\{3',4\}$, for the convenience of the matrix representations. 
 Obviously, these bilinear forms appearing in the Hamiltonian can be expressed as  $H_{\rm{S}|\rm{C}} = -\frac{1}{2}J_0s^{T}(A_{\rm{S}|\rm{C}}) s$,
\begin{equation}\label{eq:s_and_cmodel}
A_{\rm{S}} =\begin{bmatrix}
0 & 1 & \alpha \\
1 & 0 & 0 \\
\alpha & 0 & 0 ~
\end{bmatrix} ,
A_{\rm{C}}=\begin{bmatrix}
0 & 1 & 0& 0 \\
1 & 0 & 0& 0 \\
0 & 0 & 0&\alpha\\
0 & 0 & \alpha&0
\end{bmatrix}.
\end{equation}
Here, the subscript $\rm{S}|\rm{C}$ denotes 
symbiosis 
or competition. 
In the context of the symbiosis captured by $H_{\rm{S}}$, the vector  $s^T = (s_1,s_2,s_3)$ while for the competition $H_{\rm{C}}$, the vector $s^{T} =(s_1,s_2,s_{3'},s_4)$. The constant
$\alpha$ is the ratio of the $J_{ij}$ values for blue/red bonds to the green bonds, $J_{ij}/J_0$. For $A_{\rm{S}}$, the principal eigenvector  is $(\frac{1}{\sqrt{2}},\frac{1}{\sqrt{2\alpha^2+2}},\frac{\alpha}{\sqrt{2\alpha^2+2}})^T$.
For $A_{\rm{C}}$, the principle eigenstate is $(\frac{1}{\sqrt{2}},\frac{1}{\sqrt{2}},0,0)^T$ for $-1< \alpha <1$, $(\frac{1}{2},\frac{1}{2},\frac{1}{2},\frac{1}{2})^T$ for $\alpha=\pm 1$, $(0,0,\frac{1}{\sqrt{2}},-\frac{1}{\sqrt{2}})^T$ for $\alpha<-1$, and $(0,0,\frac{1}{\sqrt{2}},\frac{1}{\sqrt{2}})^T$ for $\alpha>1$.
These eigenstates can provide a fairly good explanation for the behavior of the spin product $s_1s_3$ or $s_{3'}s_4$ (see the upper part of Fig. \ref{fig:s_and_clambda}).

\begin{figure}
    \centering
    \includegraphics[width=0.49\textwidth]{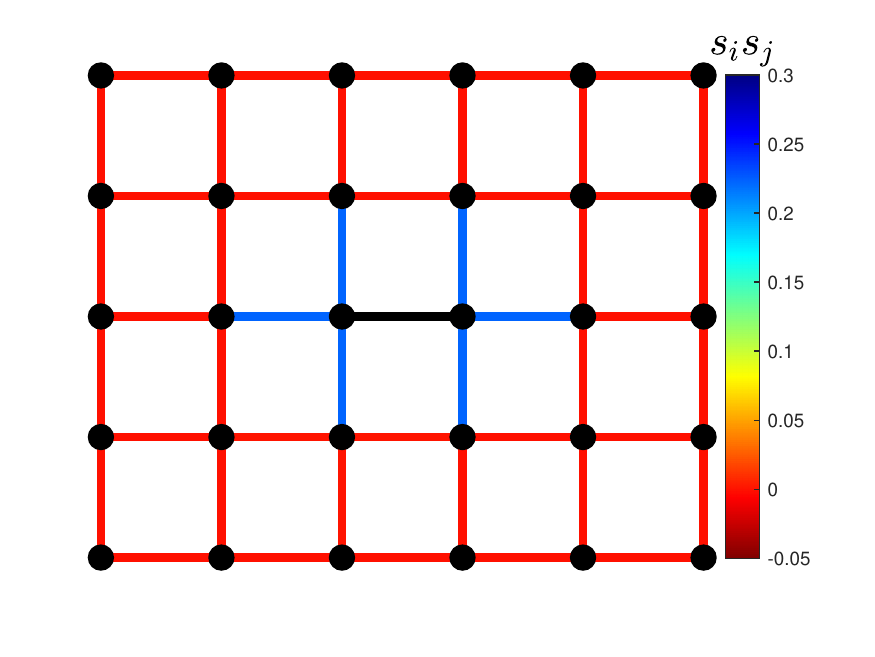}
    \caption{A spherical spin glass mode having the central black bond  coupling  $J_{i_{0}j_{0}} = +100$ with all other couplings $J_{ij}$ being $+50$. The values of the computed GS spin products $s_i s_j$ are then color coded (see legend at right).}
    \label{fig:colorsphericallocality}
\end{figure}

\begin{figure}
    \centering
    \includegraphics[width=0.49\textwidth]{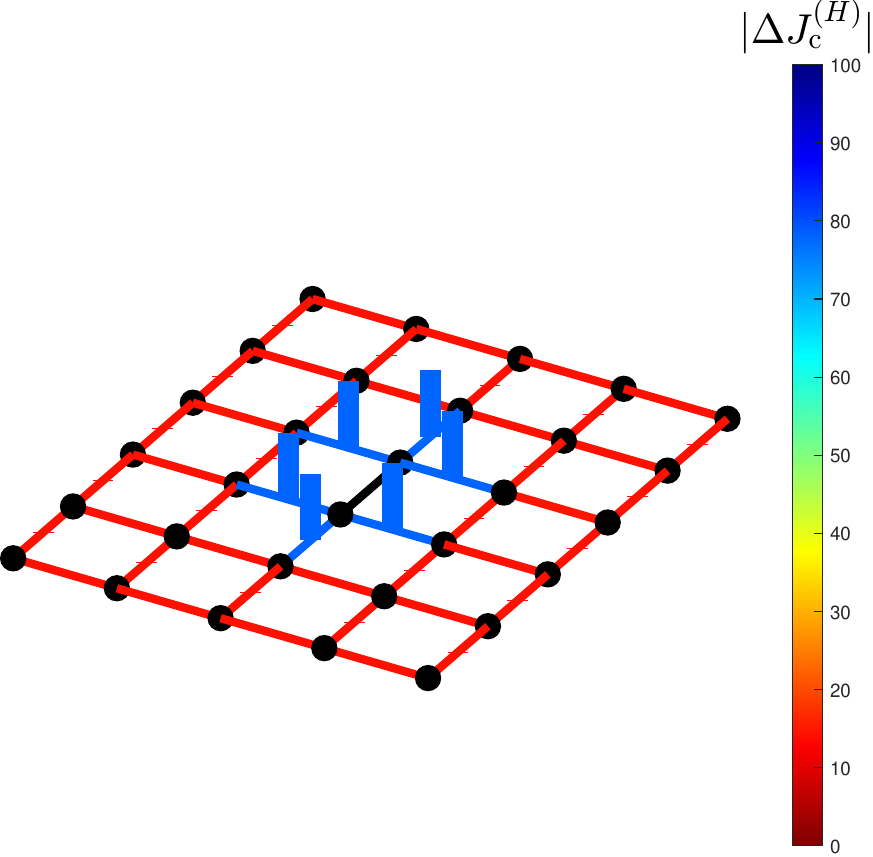}
    \caption{Calculated change of the critical threshold ($|\Delta J^{(H)}_{\rm c}|$) when tuning the central bond strength  $J_{i_{0}j_{0}}$ from $0$ to $100$. Each bond in a small $5\times 6$ square lattice  system with initial bond strengths is drawn from the symmetric Gaussian distribution (i.e., that centered about a mean of $\overline{J} =0$). We find that for the continuous spins in the spherical model (unlike the discrete spins in the Ising spin-glass model in which there is a single critical coupling), there is, generally, a {\it range of critical threshold coupling constant values}. The displayed $|\Delta J^{(H)}_{\rm c}|$ 
    are associated with minima of 
    the gap between the principle eigenvalue and the second largest eigenvalue (see text).}
    \label{fig:deltajcplotspherical}
\end{figure}

Note that for any bond that does not share a spin with the bond $(1,2)$ in Fig. \ref{fig:s_and_c}, for example, the bond $(5,6)$, no matter how far away they are, it will exhibit an identical competitive relationship as the bond $(3',4)$, with the bond $(1,2)$. This can be verified by, for example, setting $J_{12}=+100.0$ and all other bonds $J_{ij}=+50.0$ and observing $s_is_j$, see Fig. \ref{fig:colorsphericallocality}. It can also be verified by showing the distribution of $|\Delta J^{(H)}_{\rm c}|$ as $J_{i_{0}j_{0}}$ changes from $0$ to $+100$, see Fig. \ref{fig:deltajcplotspherical}.

Now, let us go back and consider the impact of the strong green bond on $\JC$ of the other bonds. For the blue bond, according to the symbiosis model, there is a nonzero slope at the origin for $s_1s_3$. Therefore, when we consider the influence of other bonds as noise on it, these influences do not significantly move its zero point. On the other hand, for the red bonds, $s_{3'}s_4$ will remain approximately $0$ over a considerable range. As a result, other bonds can easily disturb its zero point, causing $s_{3'}s_4$'s zero point to potentially move far away from zero. This implies that despite the same Euclidean distance, the impact of the blue bonds on the $\JC$ of the green and red bonds differs considerably.

We can also consider an alternative definition for $\JC$, which is to set $J_{ij}=\JC$ (which could be a point, a pair of points, or an interval) in such a way that the difference between the principal eigenvalue and the second principal eigenvalue is minimized. It is worth noting that in the limiting case illustrated in the top and bottom sides of Fig. \ref{fig:s_and_clambda}, these two definitions actually yield the same value(s) of $\JC$. In fact, according to the models in Eq. \eqref{eq:s_and_cmodel}, we could easily get $-\lambda_{1,S}=-\sqrt{\alpha^2+1},-\lambda_{2,S}=0;$ and $-\lambda_{1,C}=\min(\pm \alpha,-1),-\lambda_{1,C}=\max(\pm \alpha,-1)$. Of course, to get similar curves on the bottom side of Fig. \ref{fig:s_and_clambda}, one should replace $\alpha$ here with $J_{ij}/J_0$ and multiply the eigenvalues by $J_0$.

\begin{figure}[htb]
    \centering
    \includegraphics[width = 0.45\textwidth]{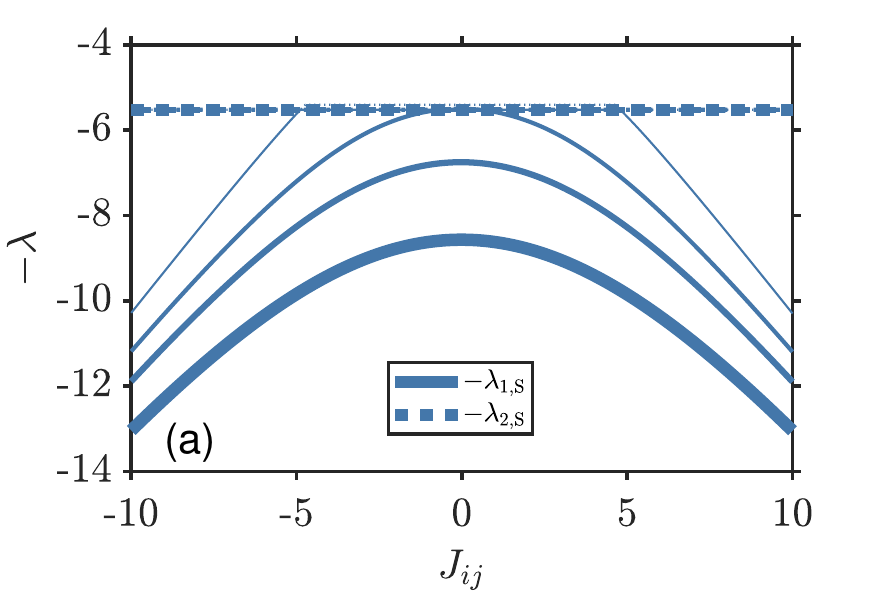}
    \includegraphics[width = 0.45\textwidth]{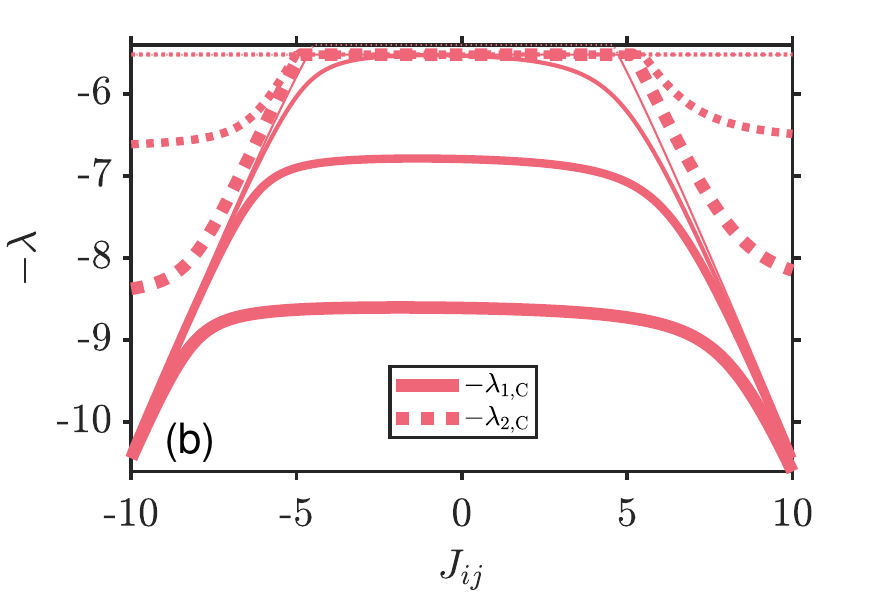}
    \caption{(a)-(b) The sign inverted largest and second-largest eigenvalues $-\lambda_1,-\lambda_2$ when tuning the $J_{ij}$ value of the symbiotic (blue) bond and the competitive (red) bond, on a $3D,L=10$ system, with different $J_{ij}$ values of the central (green) bond. Thicker curves correspond to stronger central bonds.}
    \label{fig:s_and_clambda_change}
\end{figure}

What needs to be added is that this competition model is not limited to the red bond shown in Fig. \ref{fig:s_and_c} In fact, all bonds that are not directly connected to the green bond are in competition with the green bond. To be specific, similar behaviors can be seen in Fig. \ref{fig:s_and_clambda} for all these competitive bonds. 
Note that this conclusion holds for both definitions of $\JC$. When the strength $J_{ij}$ of the modified bond (green bond) increases in absolute value, see Fig. \ref{fig:s_and_clambda_change}:
\begin{itemize}
    \item For bonds having a symbiotic  relationship, $\JC$ will be more like a single point tending zero, that is, $\JCL\to 0,~\JCH\to 0$.
    \item For bonds with a competition relationship, $\JC$ will be more like an interval and asymptotically $\JCL\to -J_0,~\JCH\to J_0$.
\end{itemize}

 Overall, in the spherical model which is subject to the above discussed global normalization constraint, the locality of the critical threshold does not exist like that in the Ising spin model. Specifically, although modifying the $J_{ij}$ of a bond will usually affect other bonds' critical thresholds globally, the effects are not all the same. Symbiotic bonds and competitive bonds can be distinguished by whether they have the minimum topological distance to the modified bond, and they exhibit completely different behaviors regarding $\JC$.



\section{Scaling argument for observed exponents}
\label{sec:scaling_kappa}
Independent of any particular theory, as rigorously established in \cite{binomial}, the Gaussian EA spin-glass system exhibits a single GS pair that are related by a global spin inversion (so long as the continuum limit of continuous coupling is taken prior to the thermodynamic large system limit). This implies that in the $L \to \infty$ limit, the bond product $ \sigma_{i} \sigma_{j}$ amongst nearest neighbor spins assumes a unique value (of either $+1$ or $-1$) in either of its GSs --- the correct value in the global GS pair. Thus, with $\langle \cdot \rangle$ denoting the GS average, the further disorder averaged $[|\langle \sigma_{i} \sigma_{j} \rangle|] =1$. Our results indicate that the LSBS accuracy is independent of the system size $L$. This implies that we may consider the LSBS in the $L \to \infty$ limit, and examine the dependence of $[| \langle \sigma_{i} \sigma_{j} \rangle| ]$ with, some abuse of notation, $\langle \cdot \rangle$ now denoting the subsystem GS average on $L_{\rm sub}$ with the external spin configurations averaged over. The variation of this disorder averaged GS spin product from unity arises solely due to the disorder average error rate $\overline{\mathcal{E}}_{ij}$ of a finite subsystem LSBS (now further averaged over external spin configurations that are consistent with the GS of the subsystem) that also samples incorrect (i.e., opposite sign values) of the bond spin product as compared to its value in the true original $L \to \infty$ rendition. That is, 
\begin{equation}
\label{Ebar_}
[|\langle \sigma_{i} \sigma_{j} \rangle| ] = 1 -\overline{\mathcal{E}}_{ij}.
\end{equation}
Interestingly, for the square lattice in \cite{hartmann_metastate_2023} the dependence of $[|\langle \sigma_{i} \sigma_{j} \rangle|]$ on a particular choice of the ratio between the system size and the subsystem size ($L/L_{\rm sub} = 2)$ was examined when resampling the ``shell'' of the subsystem (thus emulating $\overline{\mathcal{E}}_{ij}$).  Our square lattice value of $\kappa$ in Table I (0.685(8)) is close to the reported value (of $0.70 \pm 0.02$) in \cite{hartmann_metastate_2023} for resampling subsystem shells. The two are possibly related by the two considerations listed below. 

$\bullet$ (i) The proven uniqueness of the GS pair \cite{binomial}. We underscore that this rigorous result concerning uniqueness of the GSs does not hinge on any particular theory or assumption. Rather, it always holds provided that the continuum limit of the spin couplings is taken prior to the thermodynamic limit. 

and 

$\bullet$ (ii) The independence of error rate  $\mathcal{E}_{ij}$ on $L$ that we establish in the current work (thus allowing us to consider the thermodynamic $L \to \infty$ limit). 

In the current work, there are no external spins to the subsystem when the GSs are computed on the open subsystem while in \cite{hartmann_metastate_2023} spin configurations external to the subsystem are resampled and averaged over. The calculation in the current work was a direct one on a small local subsystem with no regard to what the spin configurations in the external system might be. However, intuitively, the similarity between the two (that is, with $\overline{\mathcal{E}}_{ij}$ of Eq. (\ref{Ebar_}) possibly being equal to the error rate $\mathcal{E}_{ij}$ (Eq. (\ref{eq:solver})) that we focus on in the current work) is suggestive. The decay of the error rate with subsystem size indeed points to the robustness of the subsystem GS configurations to the inclusion of additional external spins. In what briefly follows we discuss what will transpire if we may consider the fraction of incorrect nearest neighbor bonds (i.e., the error rate $\mathcal{E}_{ij}$) as the fraction of bonds that lie on the boundary of a ZED of linear scale $L_{\rm sub}$. The latter is the number of bonds flipped between two GSs and thus incorrect in the ``original'' GS relative to all nearest neighbor bonds in the subsystem. In such a situation then, extending the suggestions of \cite{hartmann_metastate_2023}, the value of $\kappa$ in Eq. (\ref{eq:solver}) may scale as
\begin{equation}
\label{kappadds}
\kappa = d-d_{s}
\end{equation}
with $d_{\mathrm{s}}$ the surface fractal dimension of the ZED. Reverting to the argument of \cite{hartmann_metastate_2023} and trivially extending it to general dimensions, this is so since the number of ``wrong'' bonds in a region of linear size  $L_{\rm sub}$ scales as $L_{\rm sub}^{d_{s}}$ whereas the total number of bonds in that region $\sim L_{\rm sub}^{d}$ leading to a fraction of wrong bonds (the error rate) scaling as $L_{\rm sub}^{d_{s}-d}$. 
Specifically, for our analyzed 2D (square lattice) systems $\kappa =0.685(8)$ (Table \ref{tab:fit_params}) that is numerically close to $(d-d_{s})= 0.72$ given the fractal dimension of the ZED surface $d_{s}= 1.275(30)$ \cite{shen2023universal}. For the cubic lattice system (for which our numerical errors may be larger given the smaller system size that we are able to examine), the ZED fractal dimension $d_\mathrm{s} = 2.76$ \cite{shen2023universal}; this suggests a value of $\kappa = 0.24$ as compared to our obtained value of $\kappa = 0.18$ (Table \ref{tab:fit_params}) for our examined cubic lattice systems.
The numerically observed Eq. (\ref{kappa=kappaJ}) suggests that the exponent $\kappa_J$ in Eq. (\ref{Jcvars}) may, similarly, be equal to $(d-d_s)$.
Furthermore, given the rather universal character of the ZED volume and area distributions \cite{shen2023universal} and Eq. \ref{kappadds}, near-universal (spatial dimension dependent yet specific lattice type independent) values of the error rate exponent $\kappa$ may be anticipated for EA spin-glass systems. 
This is consistent with the values of $\kappa$ that we found for different 2D lattices (square: 0.685(8), honeycomb: 0.67(3), and triangular (for which the largest deviation from our other investigated 2D lattice occurs): 0.648(9)).

\section{Spin-glass constrained disordered-average criticality}
\label{sec:sg_criticality}
We conclude by discussing the prospect of disorder-averaged GS criticality at the critical threshold and thus general transitions (since any transition between GSs arises from varying couplings across their critical threshold). By ``criticality'' we allude here to algebraic deviations of gauge invariant (i.e., invariance of the Hamiltonian and associated distribution of coupling constants $J_{ij}$ under the simultaneous transformations  $\sigma_{i} \to \eta_{i} \sigma_{i}, ~J_{ij} \to \eta_{i} J_{ij} \eta_{j}$ with arbitrary local $\eta_{i} = \pm 1$) correlation functions from their asymptotic long distance limit. 

In the following, $\langle \cdot \rangle$ denote averages over the set of all GSs for a fixed set of couplings when degeneracy arises- when one GS pair (i.e., two states related by the global inversion of all spins) of the system becomes degenerate with another pair. We wish to compute the two-point correlation function $G_{ij} \equiv \langle \sigma_i  \sigma_j \rangle$ at the transition between the degenerate states (in systems having an unbiased probability distribution of $J_{ij}$, the GS pair averages $\langle \sigma_i \rangle =0$). To consider gauge invariant quantities that do not vanish identically, we examine the disorder average $\Gamma_{i,j} \equiv [G^{2}_{ij}]$ \footnote{We may trivially, with no further change, implicitly average $G_{ij}$ over sites $i$ and $j$ of fixed separation $|i-j| = r$ prior to taking the disorder average to obtain $\Gamma_{ij}$.}. Here, the disorder average is that over the set of all (gauge-invariant) couplings for which GS degeneracy arises. Now, within each element of the set of possible GS transitions across the critical threshold, with the said disorder average over all couplings where degeneracy arises following an internal average over all GSs at those couplings, we have, longhand, 
\begin{eqnarray}
\label{triv_long}
&& \Gamma_{ij} = [({\langle \sigma_{i} \sigma_{j} \rangle  - \langle \sigma_i \rangle  \langle \sigma_j \rangle)^2}] =  [({\langle \sigma_{i} \sigma_{j} \rangle)^2}] =  \nonumber
\\ && =  [(\chi_{i \in {\sf ZED}} ~ \chi_{j \in {\sf ZED}}+ \chi_{i \not \in {\sf ZED}} ~ \chi_{j  \not \in {\sf ZED} })^2] \nonumber
\\ && = [(\chi_{i \in {\sf ZED}}  ~ \chi_{j \in \sf ZED})
+  (\chi_{i \not \in {\sf ZED}}  ~ \chi_{j \not \in \sf ZED})]. 
\end{eqnarray}
In Eq. (\ref{triv_long}), $\chi_{i \in {\sf ZED}}=1$ if $i$ lies in the ZED and $\chi_{i \in {\sf ZED}}=0$ otherwise (with the opposite definition  for $\chi_{i \not \in {\sf ZED}}$). In Eq. (\ref{triv_long}) we employed the trivial observation that if both sites $i$ and $j$ lie in the ZED or both of these sites lie outside the ZED, then the product of the spins at these two sites ($\sigma_i \sigma_j$) assumes the same value (either ``$1$'' in all of these states or ``$-1$ '' in all of the four degenerate states (two degenerate pairs of GSs)); these uniform sign values add coherently in the GS  average. Thus, regardless of its sign, the average $\langle \sigma_i \sigma_j \rangle$ over all GSs (the latter four states) is of unit norm. The deviation of the GS and disorder averaged $\Gamma_{ij}$ from unity is given by the probability that one of the sites $i$ or $j$ lies within the ZED with the other site ($j$ or $i$, respectively) being outside the ZED. For a sequence of ZEDs connecting degenerate GSs (the general case), we consider the probability distribution associated with the ``last'' ZED.

We now consider the specific (constrained) case in which the degenerate GSs are such that $i \in {\sf ZED}$ and
turn to the asymptotic scaling of large distance $r=|i-j|$ of $\Gamma_{ij}$. Since the (disorder averaged) cumulative ZED volume distribution function is a power law  \cite{shen2023universal},  $\Gamma_{ij}$ decays as a power law in the distance $r$. We earlier found \cite{shen2023universal} that the cumulative probability associated with ZED volume $|D|$ scaled as 
\begin{equation}
  P(|D| \ge \mathcal{V}) = 1-F(\mathcal{V}) = \frac{1}{\mathcal{V}_0^{\kappa_\mathrm{v}}} \Omega\left(\frac{\mathcal{V}}{\mathcal{V}_0}\right) \sim  k_\mathrm{v} \mathcal{V}^{-\kappa_\mathrm{v}}.
  \label{Fs}
\end{equation}

Eq. (\ref{Fs}) implies that the probability density for the volume decays algebraically with an
exponent $\kappa_\mathrm{v}+1$. Thus, denoting by $\tilde{P}_{r}$ the probability density for the ZED to be of linear size $r$, we have that $
    \tilde{P}_r ~dr \propto \mathcal{V}^{-(\kappa_\mathrm{v}+1)} d\mathcal{V} \propto r^{-d(\kappa_\mathrm{v}+1)} r^{d-1} dr = r^{-(d \kappa_\mathrm{v}+1) } dr$. To find the probability of a ZED of size $\ge r$ that includes site $j$ we integrate $\int_{r}^{\infty} dr'\tilde{P}_{r'}$. Thus \footnote{We underscore our underlying assumptions where this critical scaling holds: The system is at a degeneracy point brought about by fine tuning the couplings. The above calculation is valid when the site $i$ is a finite fraction of the ZED linear size away from its center (for simplicity, we assume that $i$ is at the origin for which the coupling is tuned \cite{shen2023universal}) and a disorder average is performed over all such coupling realizations for which degeneracy arises.},
    \begin{eqnarray}
   \Gamma_r \propto r^{-d \kappa_\mathrm{v}}.
\end{eqnarray}
Plugging in the exponents \cite{shen2023universal} $\kappa_\mathrm{v}$, we find that for both the cubic and square lattices the correlations decay with an exponent $d \kappa_d \sim 0.4$.  

Away from couplings at which degeneracy of different GS pairs arises, $\Gamma_{ij} =1$ for all $i$ and $j$. The same also holds true (sans the disorder average) for the GSs of the Ising ferromagnet.

\section{Generalization to Multi-body Interactions}
\label{sec:multibody}

The concept of critical threshold can be generalized to multi-body interactions, including single-body terms (external fields). Consider an interaction term contributing $-J_{i_1i_2\cdots i_m} \sigma_{i_1} \sigma_{i_2} \cdots \sigma_{i_m}$ to the Hamiltonian, involving $m$ spins. The case $m=2$ corresponds to the pairwise interactions discussed in the main text, while $m=1$ corresponds to external field terms.

In the two extreme cases where $J_{i_1i_2\cdots i_m} = \pm \infty$, the spin product $\sigma_{i_1} \sigma_{i_2} \cdots \sigma_{i_m} = \pm 1$ in the ground state. As $J_{i_1i_2\cdots i_m}$ increases continuously from $-\infty$, there exists a unique critical value $J_{\mathrm{c};i_1i_2\cdots i_m}$ at which the system exhibits degenerate ground states. In these degenerate ground states, denoted as $\bm{\sigma}^{(\pm)}$, $\sigma_{i_1} \sigma_{i_2} \cdots \sigma_{i_m}$ can assume both $\pm 1$ values. 

When $J_{i_1i_2\cdots i_m}$ increases beyond $J_{\mathrm{c};i_1i_2\cdots i_m}$ by an amount $\Delta J$, the energy of the $\bm{\sigma}^{(+)}$ configuration decreases by the same amount $\Delta J$. For any system configuration, the energy either decreases or increases by $\Delta J$. This implies that $\bm{\sigma}^{(+)}$ maintains its ground state status until $J_{i_1i_2\cdots i_m} = +\infty$. Similarly, $\bm{\sigma}^{(-)}$ retains its ground state status for $J_{i_1i_2\cdots i_m}$ in the interval $(-\infty, J_{\mathrm{c};i_1i_2\cdots i_m}]$.

In summary, as $J_{i_1i_2\cdots i_m}$ varies from $-\infty$ to $+\infty$, there exists exactly one critical value $J_{\mathrm{c};i_1i_2\cdots i_m}$ at which the ground state changes. 

\section{Contraction Solver Based on LSBS}
\label{sec:contraction_Solver}

In the main text, we discussed using subsystems to solve for bond spin products, or equivalently, relative spin orientations, which we term LSBS (Local Single Bond Solver). A natural question arises: can we leverage LSBS to design a solver capable of determining the complete spin configuration? The answer is affirmative. Here, we present a simple and intuitive approach (though alternative schemes may exist).

\begin{figure*}[htb]
    \centering
      \centering

      \begin{overpic}[width=0.24\textwidth]{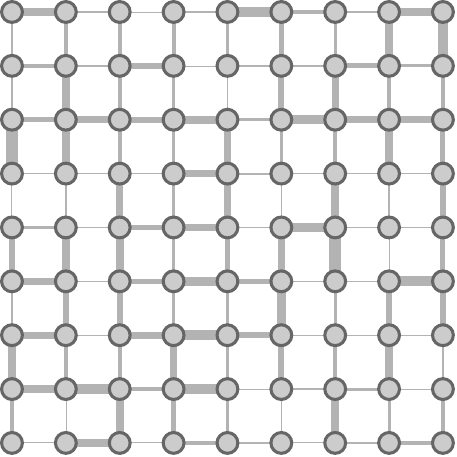}
          \put(-13,0){\fontsize{12}{16}\fontfamily{phv}\selectfont{(a)}}
      \end{overpic}%
      \hspace{0.8cm}
      \begin{overpic}[width=0.24\textwidth]{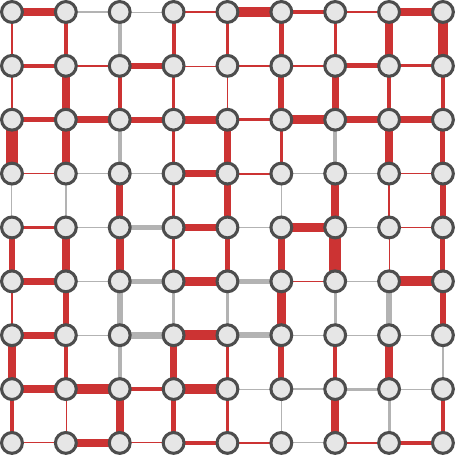}
          \put(-13,0){\fontsize{12}{16}\fontfamily{phv}\selectfont{(b)}}
      \end{overpic}%
      \hspace{0.8cm}
      \begin{overpic}[width=0.24\textwidth]{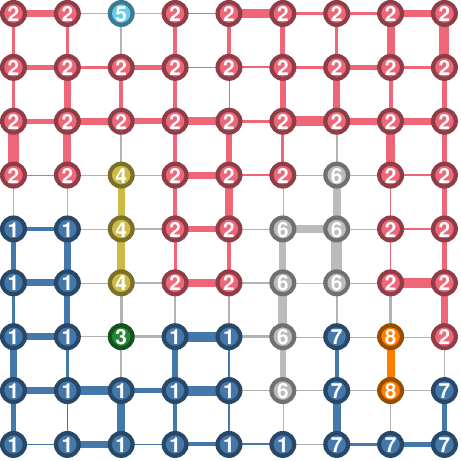}
          \put(-13,0){\fontsize{12}{16}\fontfamily{phv}\selectfont{(c)}}
      \end{overpic}
      
      \vspace{0.06\textwidth}
      
      \begin{overpic}[width=0.24\textwidth]{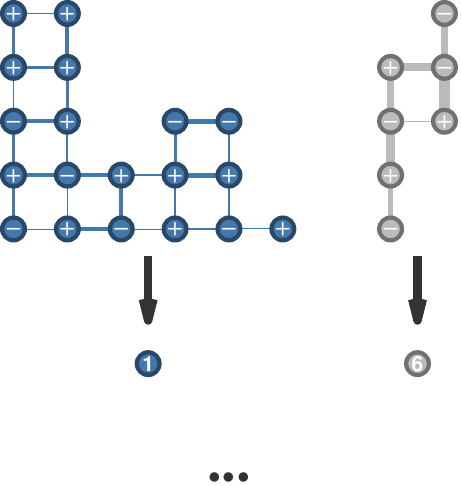}
          \put(-13,0){\fontsize{12}{16}\fontfamily{phv}\selectfont{(d)}}
      \end{overpic}%
      \hspace{0.8cm}
      \raisebox{.7\height}{\begin{overpic}[width=0.24\textwidth]{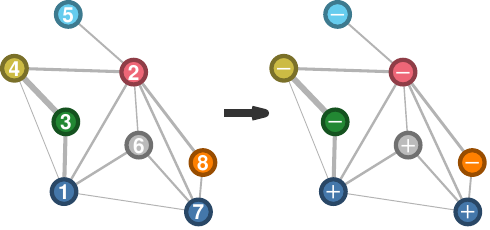}
          \put(-13,-32.5){\fontsize{12}{16}\fontfamily{phv}\selectfont{(e)}}
      \end{overpic}}%
      \hspace{0.8cm}
      \begin{overpic}[width=0.24\textwidth]{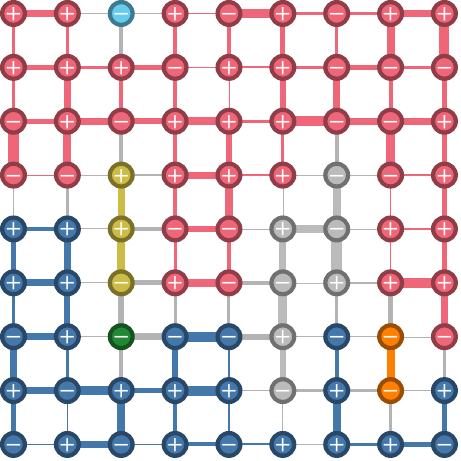}
          \put(-13,0){\fontsize{12}{16}\fontfamily{phv}\selectfont{(f)}}
      \end{overpic}
    \caption{Visualization of the contraction algorithm on a $9 \times 9$ lattice system. (a) \textbf{Vanilla structure}: The original lattice with all bonds shown in uniform gray. (b) \textbf{Well-determined bond emphasis}: Bonds satisfying $|\JC - J_{ij}| > \text{threshold}$ are highlighted in red with increased thickness, while the other bonds remain in light gray. These bonds, being far from their critical thresholds, can be reliably predicted by local solvers. (c) \textbf{Connected components}: The system is partitioned into distinct connected components based on well-determined bonds, with each component colored differently and labeled with component numbers. (d) \textbf{Component contraction}: Within each connected component, since the couplings $J_{ij}$ are far from $\JC$, the local predictions of spin products are reliable. We store the relative spin orientations (shown for components 1 and 6 in the top row with $\pm$ signs) and contract each component into a single site, with this contracted site inheriting all interactions (bonds) between the original component and the rest of the system. (e) \textbf{Ground state on contracted system}: The ground state is computed on the contracted network, with spin assignments shown on the contracted sites. (f) \textbf{Full system reconstruction}: Using the stored relative spin orientation information within each component, the ground state of the entire original system is reconstructed, as shown in the rightmost panel with all components displaying their ground-state spin assignments. The workflow shown in (c)--(f) is also described in the algorithm section.}
    \label{fig:contraction_visualization}
\end{figure*}

We propose an LSBS-based solver, termed the contraction solver. The fundamental idea is as follows: for each bond $\langle i,j \rangle$, we solve for its critical threshold $\JC$ within its corresponding subsystem of size $L_{\rm sub}$ (simultaneously obtaining the relative spin orientation). Subsequently, we consider a special set of bonds that satisfy $|J_{ij} - \JC| \geq \vartheta$, see Fig.~\ref{fig:contraction_visualization}(a-b). As discussed in the main text, these bonds with high stability typically correspond to higher subsystem solving accuracy. We refer to these bonds as \emph{well-determined bonds}.

These well-determined bonds form several connected components, see Fig.~\ref{fig:contraction_visualization}(c). Within each connected component, since the constituent sites are linked together through well-determined bonds, their relative spin orientations from local predictions are reliable. In other words, the relative spin relationships among all sites within each connected component are locked in. Subsequently, we naturally contract each connected component into a single site, see Fig.~\ref{fig:contraction_visualization}(d). The contracted site, serving as a representative of the original connected component within the system, inherits all interactions between that component and all other connected components.
It is worth noting that, after contraction, all the connections between contracted sites are single bonds. This is because, every time two sites sharing common neighbors are contracted (this is the only possibility for the occurrence of multiple bonds), the corresponding bonds are merged (summed together). Of course, the relative spin orientation of the sites to be contracted determines the appropriate sign when adding these bond strengths.

We then solve the contracted system, see Fig.~\ref{fig:contraction_visualization}(e). Finally, utilizing the ground state of the contracted system along with the relative spin orientations within each connected component, we can reconstruct the ground state of the original system, see Fig.~\ref{fig:contraction_visualization}(f).

The complete algorithmic procedure is outlined in Algorithm~\ref{alg:contraction}.

\begin{algorithm}[htb]
\caption[Contraction Solver Algorithm]{Contraction Solver Algorithm}\label{alg:contraction}
\KwIn{Coupling matrix $\mathbf{J}$ of size $N \times N$, site pairs $\langle i,j \rangle$ to be contracted and their spin products $p_{ij} = \sigma_i \sigma_j$}
\KwOut{Contracted matrix $\mathbf{J}^{\text{con}}$, contracted ground state $\boldsymbol{\sigma}^{\text{con}}$ and reconstructed ground state $\boldsymbol{\sigma}$}

\tcp{Initialization:}
\For{$i = 1$ to $N$}{
    $b_i \leftarrow i$ \tcp*{Assign block identifier to each site}
    $\sigma'_i \leftarrow 0$ \tcp*{Initialize relative spin orientation in each block}
}
$\mathbf{J}^{\text{con}} \leftarrow \mathbf{J}$ \tcp*{Initialize contracted matrix}

\tcp{Contraction Loop:}
\For{each site pair $\langle i,j \rangle$ to be contracted}{
    \If{$b_i = b_j$}{
        \textbf{skip} \tcp*{sites already contracted through other pairs}
    }
    
    \If{$\sigma'_i = 0$ and $\sigma'_j = 0$}{
        $\sigma'_i \leftarrow +1$\;
        $\sigma'_j \leftarrow p_{ij}$\;
    }
    \ElseIf{$\sigma'_i \neq 0$}{
        \If{$\sigma'_j \neq \sigma'_i \times p_{ij}$}{
            $\sigma'_j \leftarrow \sigma'_i \times p_{ij}$\;
            \For{$k = 1$ to $N$}{
                \If{$k \neq j$ and $b_k = b_j$}{
                    $\sigma'_k \leftarrow -\sigma'_k$ \tcp*{Flip relative orientations in block $b_j$}
                }
            }
        }
    }
    $J_{b_i b_j} \leftarrow 0$, $J_{b_j b_i} \leftarrow 0$ \tcp*{Contract the blocks}
    Set $b_j$-th row and column of $\mathbf{J}^{\text{con}}$ to zero \tcp*{Clean up the contracted block}
    \For{$k = 1$ to $N$}{
        \If{$b_k = b_j$}{
            $b_k \leftarrow b_i$ \tcp*{Assign all sites in block $b_j$ to block $b_i$}
        }
    }
}

\tcp{Ground State Solution:}
Solve for $\boldsymbol{\sigma}^{\text{con}}$ using $\mathbf{J}^{\text{con}}$

\tcp{Reconstruction:}
\For{$k = 1$ to $N$}{
    \If{$b_k = k$}{
        $\sigma_k \leftarrow \sigma^{\text{con}}_k$\;
    }
    \Else{
        $\sigma_k \leftarrow \sigma^{\text{con}}_k \times \sigma'_k / \sigma'_{b_k}$\;
    }
}
\end{algorithm}

\begin{figure}[htb]
    \centering
    \includegraphics[width=0.48\textwidth]{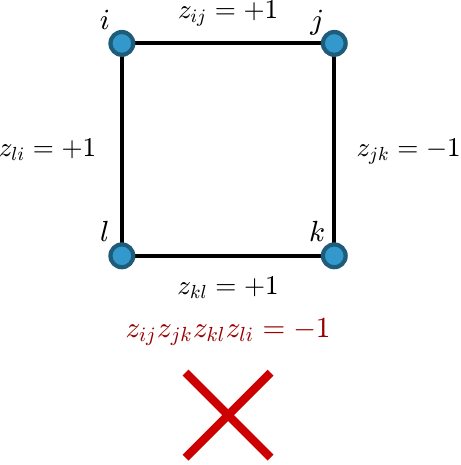}
    \caption{Illustration of a consistency check for improving the accuracy of contraction solvers. A square plaquette with four nodes is shown. For any valid spin configuration, the product $z_{ij} z_{jk} z_{kl} z_{li} = (\sigma_i \sigma_j) \cdot (\sigma_j \sigma_k) \cdot (\sigma_k \sigma_l) \cdot (\sigma_l \sigma_i)$ must equal $+1$. However, if exactly one or three bonds are incorrectly predicted locally, the product $z_{ij} z_{jk} z_{kl} z_{li}$ of the local predictions will equal $-1$, indicating the presence of prediction errors (marked by the red cross).}
    \label{fig:component_refinement}
\end{figure}

To enhance the accuracy of contraction solvers, simple strategies can be employed to identify and exclude incorrectly locally predicted bonds. As illustrated in Fig.~\ref{fig:component_refinement}, we consider elementary square plaquettes containing four nodes. From mathematical consistency, we must have $(\sigma_i \sigma_j) \cdot (\sigma_j \sigma_k) \cdot (\sigma_k \sigma_l) \cdot (\sigma_l \sigma_i) = \sigma_i^2 \sigma_j^2 \sigma_k^2 \sigma_l^2 = +1$ for any valid spin configuration. However, if exactly one or three bonds are incorrectly predicted within such a plaquette, the product $z_{ij} z_{jk} z_{kl} z_{li}$ of the local predictions (where $z_{ij} = \sigma_i \sigma_j$ denotes the predicted spin product) will yield $-1$, signaling the presence of erroneous local predictions. A straightforward strategy to improve solver reliability is to examine all plaquettes and only include bonds from those plaquettes satisfying $z_{ij} z_{jk} z_{kl} z_{li} = +1$ as ``predictions with confidence'' for contraction.

\begin{figure*}
    \centering
    \includegraphics[width=0.49\textwidth]{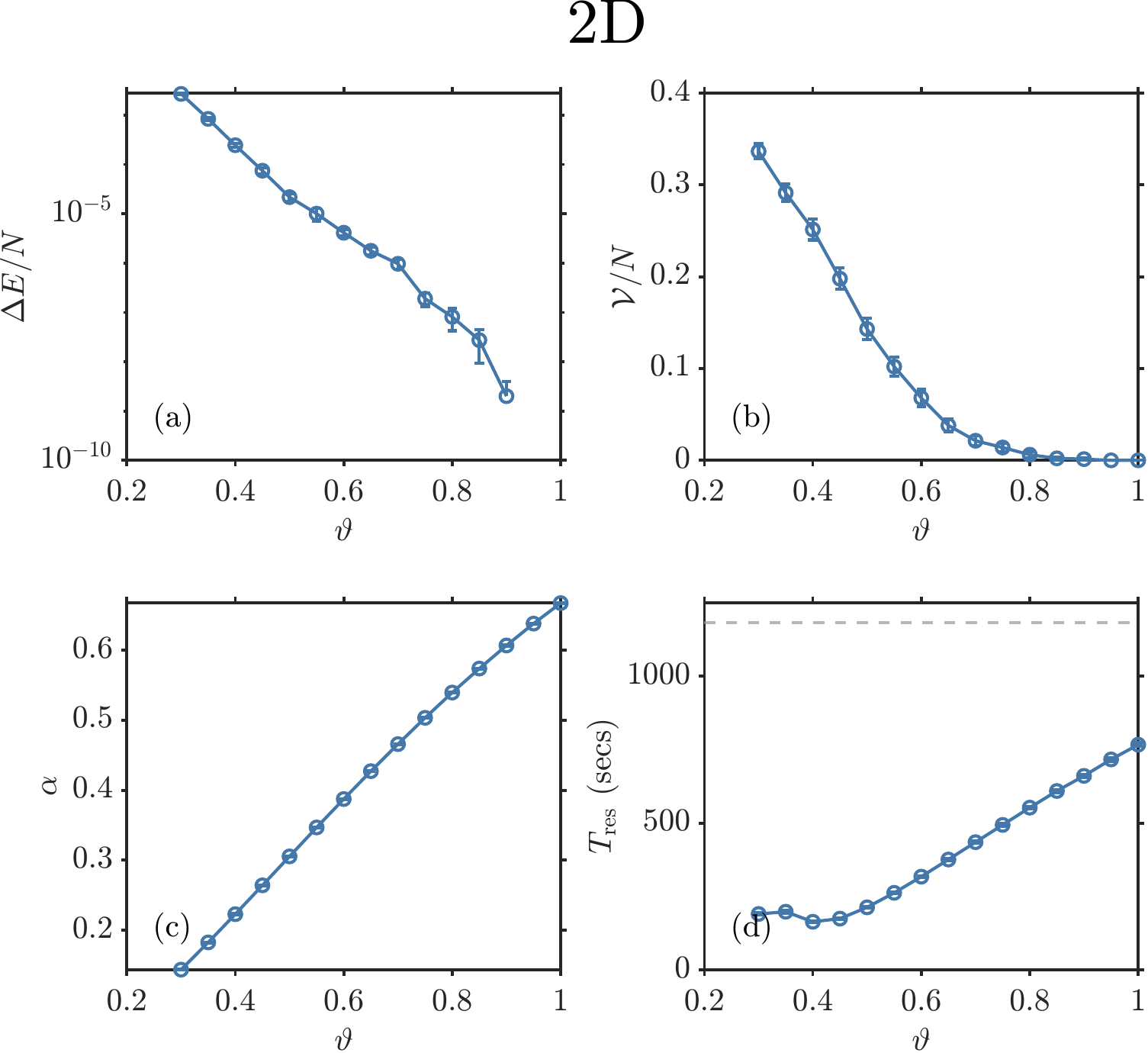}
    \includegraphics[width=0.49\textwidth]{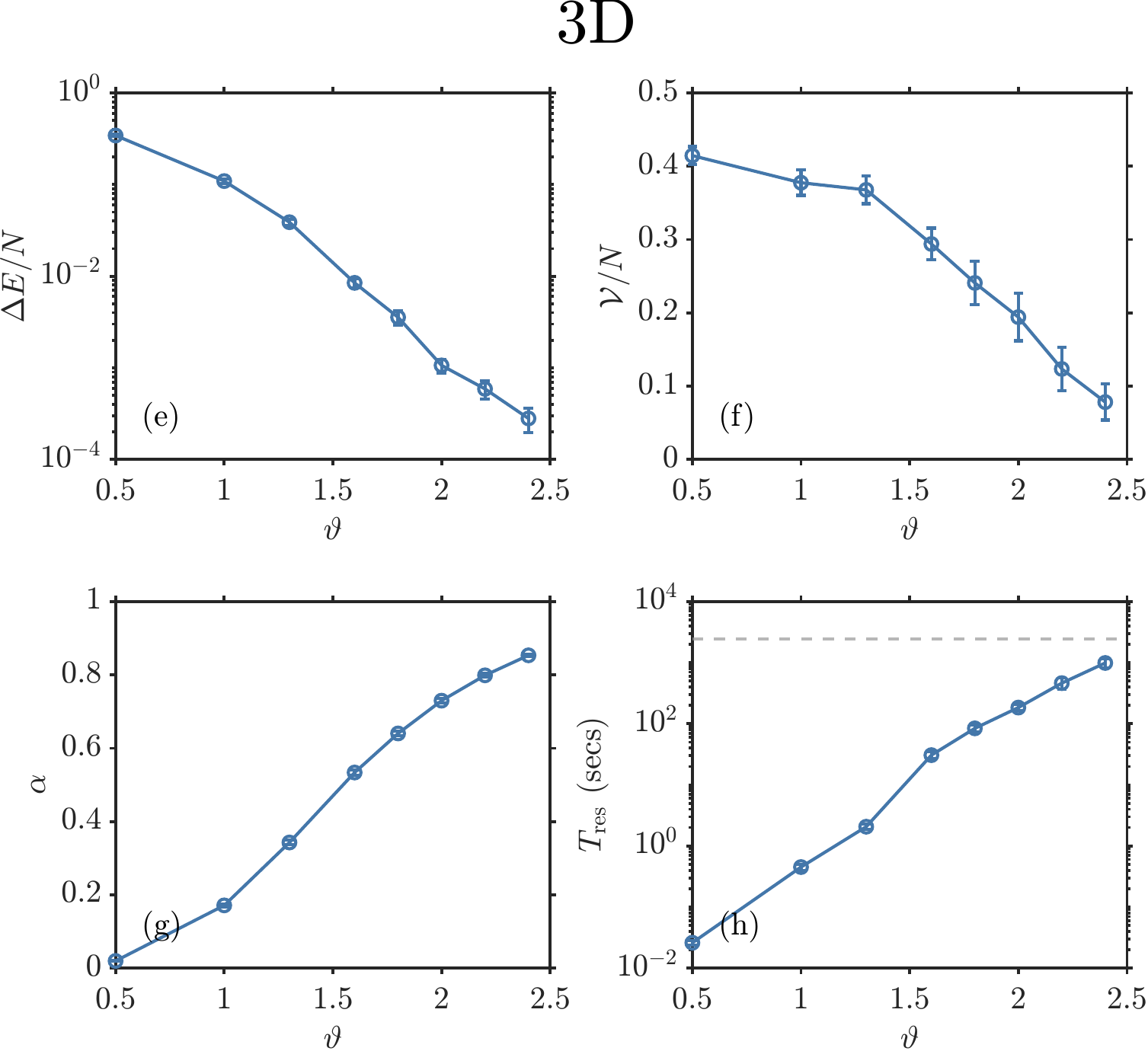}
    \caption{Performance of the contraction solver based on LSBS on 2D (Top) and 3D (Bottom) spin-glass systems, averaged over multiple disorder realizations. For the 2D system, we used subsystems of size $L_{\rm sub}=64$ and a full system of size $L=256$. $\vartheta$ is the threshold for $|J_{ij} - \JC| \geq \vartheta$. For the 3D system, we used subsystems of size $L_{\rm sub}=9$ and a full system of size $L=12$. Panels (a) and (e): Average per-spin energy difference $\Delta E/N$ between the solution obtained by the contraction solver and the true ground state. Due to some energy differences being 0 and the Y-axis being on a logarithmic scale, some points are not visible. Panels (b) and (f): Normalized spin difference $\mathcal{V}/N$ between the solution configuration and the ground-state configuration. Panels (c) and (g): Ratio $\alpha$ of the number of bonds in the contracted system relative to the number of bonds in the original system. Panels (d) and (h): Computation time required to solve the contracted system. The dashed line indicates the time required to solve the full system directly.}
    \label{fig:contraction_solver_performance}
\end{figure*}

We evaluate the performance of the contraction solver on 2D systems with $L=256$ using subsystems of size $L_{\text{sub}}=64$, and on 3D systems with $L=12$ using subsystems of size $L_{\text{sub}}=9$, as shown in Fig.~\ref{fig:contraction_solver_performance}. Across multiple disorder realizations, we assess the following metrics: per-spin energy difference $\Delta E/N$ (Fig.~\ref{fig:contraction_solver_performance}(a)(e)), normalized spin difference $\mathcal{V}/N$ (Fig.~\ref{fig:contraction_solver_performance}(b)(f)), ratio of the number of bonds in the contracted system relative to the number of bonds in the original system (Fig.~\ref{fig:contraction_solver_performance}(c)(g)), and computation time required to solve the contracted system and the original system (Fig.~\ref{fig:contraction_solver_performance}(d)(h)).

The value of $\vartheta$ determines the quality of configurations computed by the contraction solver. Generally, as $\vartheta$ increases, the proportion of well-defined bonds decreases (resulting in larger contracted systems and consequently longer computation time), but simultaneously, these bonds are more likely to be correctly predicted locally. This trend is clearly evident from Fig.~\ref{fig:contraction_visualization}. The quantitative performance of the contraction solver, as shown in Fig.~\ref{fig:contraction_solver_performance}, shows good accuracy. For the 2D case, when $\vartheta > 0.9$, the contraction solver with $L_{\rm sub}=64$ achieves perfect ground-state predictions on all $N_{\rm sample}=176$ disorder realizations of $L=256$ systems. Even when predictions are not perfect, the degradation in solution quality caused by incorrectly predicted bonds remains well controlled. Specifically, the average energy error per spin can be as low as $10^{-8}$. Given that the average ground-state energy per spin is around 1.3, this error is small. For the 3D case, due to computational limitations, we were unable to compute systems with larger $L_{\rm sub}$ or $L$. Even under these constraints, the average energy error per spin remains at the order of $10^{-4}$ for $L_{\rm sub}=9$ and $L=12$. It is reasonable to expect that with more advanced computational resources enabling calculations for larger $L_{\rm sub}$ and $L$, we should be able to achieve similar results as in the 2D case. We emphasize that when a bond is incorrectly predicted, it does not lead to an avalanche-like amplification of energy errors. A heuristic argument for this behavior is that such incorrect predictions likely correspond to large excitations (exceeding the subsystem size) of $O(1)$ magnitude—otherwise, it would be difficult to explain the substantial differences in spin difference counts observed in Fig.~\ref{fig:contraction_visualization}(b)(f) alongside the relatively small energy differences shown in Fig.~\ref{fig:contraction_solver_performance}(a)(e).

\begin{figure}
    \centering
    \includegraphics[width=0.45\textwidth]{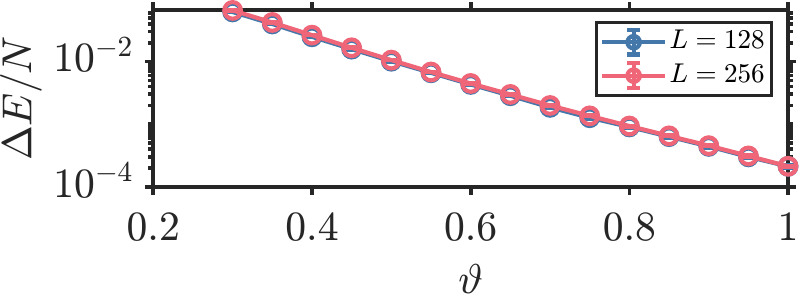}
    \caption{The averaged energy difference per spin, between the acquired solution and the true ground state, for different total system sizes $L=128, 256$ and fixed subsystem size $L_{\text{sub}}=4$ (we set a relatively small subsystem size to reduce the boundary effect). }
    \label{fig:contraction_scalability}
\end{figure}

Regarding the scalability of the contraction solver, we make several important observations. First, as demonstrated in Fig.~\ref{fig:solver_performance} in the main text, the performance of LSBS depends solely on the subsystem size—that is, the error rate does not decay with increasing total system size. This implies that even when maintaining the same subsystem size $L_{\text{sub}}$, we can guarantee that the solution quality will not deteriorate as the total system size $L$ increases for a given value of $\vartheta$. This property is verified in Fig.~\ref{fig:contraction_scalability}, where we solve relatively large systems of $L=128$ and $L=256$ using subsystems of size $L_{\text{sub}}=4$. In this scenario, boundary effects can be almost neglected, demonstrating that the solution quality remains the same for systems with a fourfold difference in the number of nodes ($L=128$ versus $L=256$). Consequently, to maintain the same solution quality, the computational time for solving the $\JC$ determination step scales as $O(L)$, since the time required to compute $\JC$ for each bond is a constant independent of $L$. The $O(L)$ time complexity is evidently lower than that of any spin-glass ground-state solver algorithm and can therefore be neglected in the overall computational cost analysis.

\section{Minimum Vertex Cover as an Ising Problem}
\label{sec:mvc_ising}

Many important NP-hard combinatorial optimization problems, including the Minimum Vertex Cover (MVC), can be encoded as Ising formulations\cite{lucas2014ising}. Here, we take the MVC problem as a representative example to study the effectiveness of local solvers—showing how our locality-based analysis applies beyond traditional spin-glass systems.

The Minimum Vertex Cover (MVC) problem is a classic NP-hard combinatorial optimization problem. Given an undirected graph $G = (V, E)$, a vertex cover is a subset of vertices $C \subseteq V$ such that every edge $e \in E$ has at least one endpoint in $C$. The MVC problem seeks the smallest such subset---that is, it aims to find a vertex cover of minimal size. Formally, the goal is to minimize $|C|$ subject to the condition that for every edge $\langle ij \rangle \in E$, either $i \in C$ or $j \in C$. The MVC problem has wide applications in areas such as network security, resource allocation, and compiler optimization, and serves as an important benchmark for the study of computational complexity and algorithm design\cite{chen2024vertex, chen2016approximation}.

Mapping to Ising spins $\sigma_i \in \{-1, +1\}$ (where $\sigma_i = +1$ indicates vertex $i$ is in the cover), the Hamiltonian becomes:
\begin{equation}
\label{eq:mvc_hamiltonian}
H_{\text{MVC}} = A \sum_{i \in V} \frac{1 + \sigma_i}{2} + B \sum_{\langle ij \rangle \in E} \frac{1 - \sigma_i}{2} \frac{1 - \sigma_j}{2}.
\end{equation}

Expanding to standard Ising form with unified external fields:
\begin{equation}
\label{eq:mvc_ising_expanded}
H_{\text{MVC}} = -\sum_{i \in V} h_i \sigma_i + \frac{B}{4} \sum_{\langle ij \rangle \in E} \sigma_i \sigma_j + \frac{A|V| + B|E|}{4},
\end{equation}
where the unified external field is $h_i = -A/2 + (B/4) \cdot \text{deg}(i)$. Choosing $A = 1$ and $B = 2$ (ensuring $B > A$), this simplifies to:
\begin{equation}
\label{eq:mvc_simplified}
H_{\text{MVC}} = -\sum_{i \in V} \left(-\frac{1}{2} + \frac{\text{deg}(i)}{2}\right) \sigma_i - \frac{1}{2} \sum_{\langle ij \rangle \in E} \sigma_i \sigma_j + \frac{|V| + 2|E|}{4}.
\end{equation}

\begin{figure}[htb]
    \centering
    \includegraphics[width=0.4\textwidth]{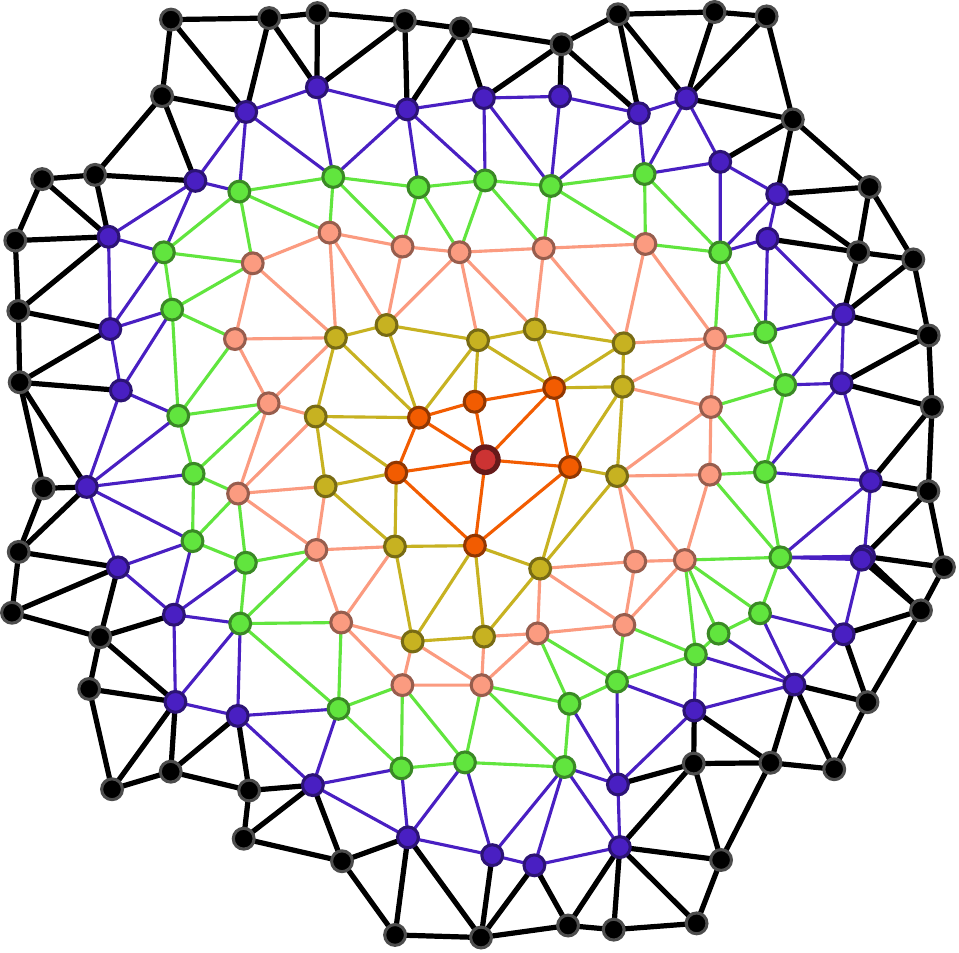}
    \caption{Layered structure of a random planar graph for MVC analysis. The graph contains $n = 10,000$ sites generated via Delaunay triangulation of uniformly randomly distributed points in a unit circle. sites are colored according to their topological distance $r$ from the geometric center: $r = 0$ (center, red), $r = 1-5$ (various colors), and $r = 6$ (black).}
    \label{fig:layered_mvc_graph}
\end{figure}

In the Ising formulation, each edge corresponds to a ferromagnetic bond ($J_{ij} = -1/2$), and each site is associated with an external field $h_i = (\text{deg}(i)-1)/2$. To demonstrate the applicability of our locality analysis to the MVC problem, we examine LSBS on random planar graphs generated using Delaunay triangulation, see Fig.~\ref{fig:layered_mvc_graph}. Our graph generation process begins with $n = 10,000$ (the number is large enough to ensure it the generated subgraph is fully contained in the system) uniformly randomly distributed points in a unit circle. Then, we extract the subgraph corresponding to the sites within topological distance $r$ from the geometric center, see Fig.~\ref{fig:layered_mvc_graph}. Naturally, these subgraphs with different $r$ values form differently sized subsystems.

\begin{figure}[htb]
    \centering
    \includegraphics[width=0.48\textwidth]{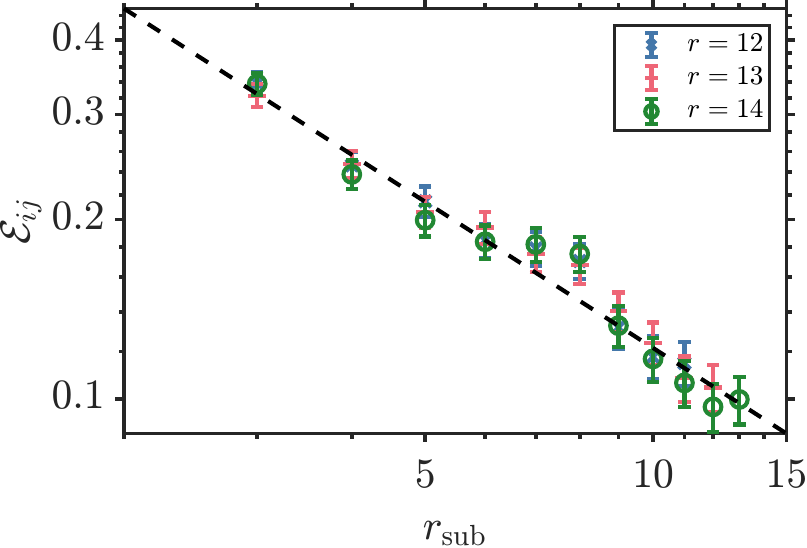}
    \caption{Error rate analysis for MVC on random planar graphs. The LSBS error rate follows a power-law-like decay with subsystem size, similar to the spin-glass case. Fit parameters: $\kappa = 0.80(5)$, $\ell_{\mathcal{E}} = 0.7(1)$ (cf.\ Eq.~\ref{eq:solver}), with $\chi^2/{\rm d.o.f} = 1.76$.}
    \label{fig:mvc_performance}
\end{figure}

Similar to the spin glass case, we can compute the spin value of the central site within a subsystem of relatively small $r$ as well as within a larger system with a greater value of $r$. In fact, due to the presence of external fields, this computation is even more direct than in the zero-field spin glass case, where only spin products of a bond can be determined rather than the value of a particular spin. After obtaining the central spin values from both the local and the larger systems, we can directly compare whether these two computed spin values are the same. The error rate analysis, like the one for spin glass, is shown in Fig.~\ref{fig:mvc_performance}.

\newpage
\bibliographystyle{my}
\bibliography{ref.bib}